\documentclass[useAMS,usenatbib]{mn2e}

\usepackage{gensymb}
\usepackage{graphicx}
\usepackage{cprotect}
\usepackage{amssymb}
\usepackage{afterpage}

\def\ms{\hbox{m\,s$^{-1}$}}         
\def\m2s2{\hbox{\,m$^{2}$\,s$^{-2}$}} 
\def\kms{\hbox{\,km\,s$^{-1}$}}       
\def\Msun{\hbox{$M_{\odot}$}}             
\def\Rsun{\hbox{$R_{\odot}$}}
\def\Mjup{\hbox{$\mathrm{M}_{\rm J}$}}
\def\Rjup{\hbox{$\mathrm{R}_{\rm J}$}}

\def\Kepler{{\it Kepler}}
\def\ten[#1]{$\;\times 10^{#1}$}
\def\teff{$T_{\rm eff}$}
\def\logg{$\log g$}
\newcommand{\e}[1]{{\;\times\; 10^{#1}}}
\title[Absolute masses and radii determination]{Absolute masses and radii determination in multiplanetary systems without stellar models}
\author[J. M. Almenara et al.]
{\parbox{\textwidth}{J. M. Almenara$^{1,2,3}$\thanks{E-mail: \texttt{jose-manuel.almenara-villa@ujf-grenoble.fr} (JMA); \texttt{Rodrigo.Diaz@unige.ch} (RFD)}, R. F. D\'{i}az$^{4}$$^{\star}$, R. Mardling$^{4,5}$, S. C. C. Barros$^{3}$, C. Damiani$^{6}$,\\ G. Bruno$^{3}$, X. Bonfils$^{1,2}$, and M. Deleuil$^{3}$}\vspace{1.0cm}\\
\parbox{\textwidth}{$^{1}$Univ. Grenoble Alpes, IPAG, F-38000 Grenoble, France\\
$^{2}$CNRS, IPAG, F-38000 Grenoble, France\\
$^{3}$Aix Marseille Universit\'e, CNRS, LAM (Laboratoire d'Astrophysique de Marseille) UMR 7326, F-13388, Marseille, France\\
$^{4}$Observatoire Astronomique de l'Universit\'e de Gen\`eve, 51 chemin des Maillettes, CH-1290 Versoix, Switzerland\\
$^{5}$School of Physics and Astronomy, Monash University, Victoria, 3800, Australia\\
$^{6}$Institut d'Astrophysique Spatiale, Universit\'e Paris-Sud \& CNRS, F-91405 Orsay, France}}
\begin{document}

\date{Accepted 2015 July 28. Received 2015 July 20; in original form 2015 April 20}

\pagerange{\pageref{firstpage}--\pageref{lastpage}} \pubyear{2015}

\maketitle

\label{firstpage}

\begin{abstract}
  The masses and radii of extrasolar planets are key observables for understanding their interior, formation and evolution. While transit photometry and Doppler spectroscopy are used to measure the radii and masses respectively of planets relative to those of their host star, estimates for the true values of these quantities rely on theoretical models of the host star which are known to suffer from systematic differences with observations. When a system is composed of more than two bodies, extra information is contained in the transit photometry and radial velocity data. Velocity information (finite speed-of-light, Doppler) is needed to break the Newtonian $MR^{-3}$ degeneracy. We performed a {\it photodynamical modelling} of the two-planet transiting system  Kepler-117 using all photometric and spectroscopic data available. We demonstrate how absolute masses and radii of single-star planetary systems can be obtained without resorting to stellar models. Limited by the precision of available radial velocities (38~\ms), we achieve accuracies of 20 per cent in the radii and 70 per cent in the masses, while simulated 1~\ms\ precision radial velocities lower these to 1 per cent for the radii and 2 per cent for the masses. Since transiting multi-planet systems are common, this technique can be used to measure precisely the mass and radius of a large sample of stars and planets. We anticipate these measurements will become common when the {\it TESS} and {\it PLATO} mission provide high-precision light curves of a large sample of bright stars. These determinations will improve our knowledge about stars and planets, and provide strong constraints on theoretical models.  
\end{abstract}


\section{Introduction}

The mass and radius of extrasolar planets are usually obtained relative to those of the host star. Generally, stellar evolution tracks are used for field stars to infer their physical parameters (mass, radius, age) from the spectroscopic parameters (\teff, \logg, and metallicity) obtained from the modelling of a stellar spectrum. If the star is transited by a planet, it is possible to estimate the stellar density \citep{2007ApJ...664.1190S}, that can replace \logg, typically the most uncertain of the three atmospheric parameters, as input for the stellar models. The typical error is 5 per cent in the mass and radius \citep{2011PASP..123..412W}, but usually these errors do not take into account the systematic errors of the models that can be up to 10 per cent \citep{2012ApJ...757..112B}. Asteroseismology provides stellar radii and masses with a  typical precision of 3 and 7 per cent, respectively \citep{2013ApJ...767..127H}. But this determination uses scaling relations \citep{1986ApJ...306L..37U,1995A&A...293...87K}, that require calibration using stellar models and also depends on the stellar effective temperature.

The {\it Gaia} satellite \citep{2001A&A...369..339P} will provide the stellar radius of a large number of stars with a precision of around 3 per cent, but the measurement will depend on the stellar effective temperature, the extinction, the bolometric correction, and the distance to the star. The best source of direct empirical determination of stellar masses and radii are double-lined eclipsing binaries \citep{2010A&ARv..18...67T}. Equivalent measurements in planetary systems imply the detection of the planet radial velocity \citep{2010Natur.465.1049S}, not achievable for a large number of planets at present.

While the orbital elements of single-planet systems are constant, they are functions of time when more than one planet is present. In particular, planet-planet interactions perturb the timing of transits so that they are no longer strictly periodic. When detectable, the associated {\it transit timing variations} or TTVs and transit shape variations contain valuable information about the system parameters \citep[e.g.][]{2010arXiv1006.3727R}, especially planet-to-star mass ratios which are not available for single-planet systems.

When photometric timing data alone is available, it is not possible to deduce absolute masses and radii using purely Newtonian modelling because of the latter's inherent $M R^{-3}$ degeneracy\footnote{In other words, the model is invariant to scaling the lengths by a factor and the masses by the same factor at cubic exponent.}. This degeneracy can be broken, however, when {\it velocity} information is available \citep{2005MNRAS.359..567A,2013ApJ...762..112M}, either in the form of variations in the time-of-arrival of light signals due to barycentric motion in a triple or higher-order system, or in the form of radial velocity measurements (or both). This information provides an absolute scale of the system. While the timing effect due to the finite speed of light has been used to directly measure the masses and radii of the three stars in the triple system KOI-126 \citep{2011Sci...331..562C}, it is currently undetectable in single and binary stars hosting planets, being at the level of 1~s or less. However, supplementing photometric data with radial velocity measurements has allowed for the direct measurement of the masses and radii of both the stars and planets in the circumbinary planetary systems Kepler-16 \citep{2011Sci...333.1602D}, Kepler-34, and Kepler-35 \citep{2012Natur.481..475W}. Note that the fractional uncertainties in the masses and radii of the three stars in KOI-126 reduced from 10 and 3 per cent, respectively, using photometry alone, to 3 and 0.5 per cent when low-resolution radial velocity information was included.

While the success of the studies described above is in large part due to the exquisite photometric precision of the \Kepler\ telescope \citep{2010Sci...327..977B}, an extremely important aspect was the use of {\it photometricdynamical (or photodynamical)} modelling.

To date, most studies have proceeded in a different manner: the individual transits are fitted separately to obtain the {\it times of mid-transit}, often fixing the remaining transit parameters and thus effectively averaging over (and therefore discarding) valuable information contained in the whole transit. The obtained transit times are used to obtain a mean ephemeris, the departures from which are finally fitted using an $N$-body integrator. This allows computing the $N$-body model at a much sparser resolution that the one needed to model each single photometric observation, reducing drastically the required computing time. In contrast, the photodynamical approach models the whole light curve consistently by coupling $N$-body integrations with a model of the flux variations due to transits and occultations.

The aim of this paper is to demonstrate that the photodynamical approach can be successfully employed to determine accurate masses and radii of both the star and planets in {\it single}-star planetary systems independently of any stellar models when high-precision radial velocity data are available.\footnote{Note that the photodynamical model has been used before to study multi-planetary systems with single host stars but the stellar parameters have been constrained using asteroseismology \citep{2012Sci...337..556C,2013Sci...342..331H}, thus again relying on stellar models.}
Applying this approach to Kepler-117 \citep{2014ApJ...784...45R,2015A&A...573A.124B}, a system composed of an F9-type star, a 0.7~\Rjup\ planet in an 18.8~d orbit (planet~b), and a 1.1~\Rjup\ companion in a 50.8~d orbit (planet~c). \citet{2015A&A...573A.124B} detected TTVs in this system, which allowed them to measure the masses of the system planets using additional information on the stellar parameters based on spectroscopic measurements.

  As the light time travel effect is negligible in this system, we use the {\it SOPHIE} radial velocities reported by \citet{2015A&A...573A.124B} to break the degeneracy mentioned above. In Sect.~\ref{sect.data}, we describe the data employed in the analysis. In Sect.~\ref{sect.pdm}, we present the details of the photodynamical modelling. In Sect.~\ref{sect.results}, we present our results, and finally in Sect.~\ref{sect.discussion}, we discuss them and their importance in the framework of upcoming space missions like {\it TESS} and {\it PLATO}.

\section{Data} \label{sect.data}

\Kepler\ observed 67 transits of the inner planet and 29 transits of the outer one between 2009 May and 2013 May. The \Kepler\ light curves of all Quarters (Q1 - Q17) were retrieved from the Mikulski Archive for Space Telescopes (MAST) archive\footnote{http://archive.stsci.edu/index.html.} We preferred short-cadence data (about one point per minute; quarters Q4-Q17) whenever available. Only the data around transits was modelled, after normalizing it with a parabola, and to account for the integration time of long cadence data (about one point every 30~min; quarters Q1-Q3), the model light curve was oversampled by a factor of 20 and then binned back to match the cadence of the data points. We use the `SAP' light curve, which we corrected for the flux contamination using the `CROWDSAP' value in the fits file header. We do not detect transit depth differences among different seasons. Therefore, the given contamination values seem self-consistent. 

The {\it SOPHIE} \citep{2008SPIE.7014E..0JP,2013A&A...549A..49B} radial velocity observations of Kepler-117 are described in \citet{2015A&A...573A.124B}. Briefly, 14 radial velocity measurements at a resolution power of 40 000 were obtained between 2012 July and 2013 November. 

\section{Photodynamical model} \label{sect.pdm}

The photodynamical model describes the light curve and radial velocity data at any moment in time accounting for the dynamic interactions via an $N$-body simulation. The model parameters are the stellar mass and radius, the coefficients of a quadratic limb-darkening law \citep{1977A&A....61..809M}, and the planetary mass, planet-to-star radius ratio, and the orbital parameters ($a$, $e$, $i$, $\omega$, $n$, and $M$; see Table~\ref{table}) at a fixed reference time ($t_{\mathrm{ref}}$) for each orbiting planetary companion. The system was integrated over the span of \Kepler\ and {\it SOPHIE} observations with a time step of 0.04~d, which produces a maximum error of 2~ppm in the interpolated light curve model. In this way, the instantaneous system parameters, including the sky-projected planet-star separation, are known at each step. This, together with the planet-to-star-radius ratio and the limb darkening coefficients, defines the instantaneous light curve model \citep{2002ApJ...580L.171M}. Additionally, the line-of-sight projected star velocity is computed and compared to the {\it SOPHIE} radial velocities. As $N$-body integrator, we employed the Bulirsch-Stoer algorithm implemented in the \textsc{mercury} code \citep{1999MNRAS.304..793C}.

To sample from the posterior distributions of the parameter models we used the Monte Carlo Markov Chain (MCMC) code implemented in the \textsc{pastis} package \citep{2014MNRAS.441..983D}. We use the \citet{2013Sci...342..331H} parametrization to minimize the correlation between the model parameters, which can reduce the efficiency of the MCMC algorithm, and impede a proper exploration of the entire joint posterior distribution (see Table~\ref{table}). 

We included a number of additional parameters in the model: a radial velocity linear drift, a global light curve normalization factor for the short- and long-cadence \Kepler\ data, and a multiplicative jitter parameter for each data set. The longitude of the ascending node at $t_{\mathrm{ref}}$ of only one of the two planets is explored, while the other is kept fixed (this is equivalent to fitting the difference of longitudes of the ascending nodes at $t_{\mathrm{ref}}$). For a spherical star, the model does not depend on the values of the individual longitudes of the ascending node.

\begin{figure}
\includegraphics[height=7cm]{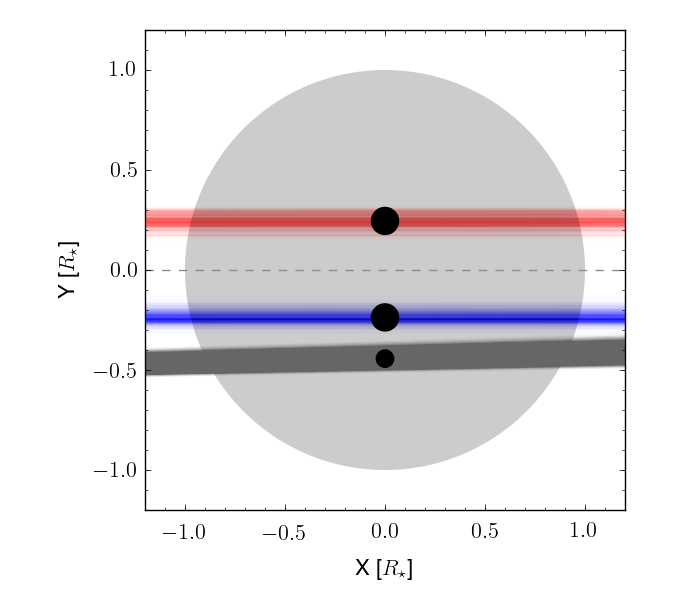}
\caption{Two possible configurations for orbital inclinations. Stellar crossings paths seeing by \Kepler\ of 1000 random MCMC steps from the photodynamical model fit are plotted. Planet~c: red (inclination planet~c $>$ 90), and blue (inclination planet~c $<$ 90), planet~b: grey. Stellar disc in light grey.}
\label{hemispheres}
\end{figure}

Finally, we considered the two possible configurations for orbital inclinations: both planets transit the same or different stellar hemispheres (Fig.~\ref{hemispheres}). We set the planet~b to transit one hemisphere ($i_b < 90\degree$) and leave planet~c free to be on any hemisphere ($ 0\degree < i_c < 180\degree$). Uniform priors were used for all the parameters. The starting point of the MCMC algorithm in parameter space was the previous solution found for this system \citep{2015A&A...573A.124B}. We ran 40 chains of 50 000 steps each. The chains were thinned using their autocorrelation length and merged together for a total of 3817 independent samples from the posterior. The mode and the posterior 68.3 per cent credible interval of the system parameters are given in Table~\ref{table}.

\begin{table*}
\renewcommand{\arraystretch}{0.7}
\centering
\cprotect\caption{Model parameters. Posterior mode and 68.3 \% credible intervals. The orbital elements have the origin at the star (Asteroidal parameters in the \textsc{mercury} code) and are given for the reference time $t_{\mathrm{ref}} = 2 454 967.63$~BJD$_{\mathrm{TDB}}$.}\label{table}
\begin{tabular}{lcc}
\hline
Parameter & \multicolumn{2}{c}{Mode and 68.3 \% credible interval} \\
\hline
Stellar mass, $M_\star$ [\Msun]                                 & 0.40$^{+0.34}_{-0.20}$  & [0; 1.557]$^{a}$\\
Stellar radius, $R_\star$ [\Rsun]$^{b}$                     & 1.12 $\pm$ 0.23       & [0.490; 1.884]$^{a}$\\
Stellar density, $\rho_{\star}$ [$\rho_\odot$]$^{b}$         & 0.2886 $\pm$ 0.0065     & \\
Surface gravity, \logg\ [cgs]                                   & 3.973$^{+0.057}_{-0.11}$ & \\
Linear limb darkening coefficient, $u_{\mathrm{a}}$$^{b}$           & 0.420 $\pm$ 0.026         & \\
Quadratic limb darkening coefficient, $u_{\mathrm{b}}$$^{b}$           & 0.130 $\pm$ 0.042         & \\
Systemic velocity (at BJD$_{\mathrm{TDB}}$ 2,456,355), $\gamma_0$ [\kms]$^{b}$ & -12.9506 $\pm$ 0.0097         & \\
Linear radial velocity drift, $\gamma_1$ [m s$^{-1}$ yr$^{-1}$]$^{b}$        & 3 $\pm$ 20         & \medskip\\

\multicolumn{1}{l}{} & \emph{Kepler-117b} & \emph{Kepler-117c} \smallskip\\
Semi-major axis, $a$ [au]                                 & 0.101 $\pm$ 0.023           & 0.197 $\pm$ 0.044 \\
Eccentricity, $e$                                         & 0.05257$^{+0.00084}_{-0.0018}$ & 0.03085$^{+0.0013}_{-0.00071}$ \\
Inclination, $i$ [\degree]$^{b}$                     & 88.667 $\pm$ 0.042           & 89.644 $\pm$ 0.043$^{c}$\\
Argument of pericentre, $\omega$ [\degree]                & 256.6$^{+2.4}_{-1.7}$         & 300.1$^{+3.3}_{-1.5}$\\
Longitude of the ascending node, $n$ [\degree]$^{b}$ & 181.18 $\pm$ 0.20            & 180 (fixed) \\
Mean anomaly, $M$ [\degree]                               & 340.5$^{+1.5}_{-3.0}$         & 141.01$^{+0.93}_{-4.1}$ \\
Radius ratio, $R_{\mathrm{p}}/R_\star$$^{b}$                     & 0.04689 $\pm$ 0.00014        & 0.07105 $\pm$ 0.00017 \\
Planet mass, $M_{\mathrm{p}}$ [$\Mjup$]                             & 0.031$^{+0.031}_{-0.014}$     & 0.65$^{+0.56}_{-0.32}$  \\
Planet radius, $R_{\mathrm{p}}$[$\Rjup$]                            & 0.51 $\pm$ 0.12             & 0.77 $\pm$ 0.18   \\
Planet density, $\rho_{\mathrm{p}}$ [$\mathrm{g\;cm^{-3}}$]                    & 0.3218$^{+0.0055}_{-0.015}$    & 1.779$^{+0.055}_{-0.037}$  \\
Planet surface gravity, $\log$\,$g_{\mathrm{p}}$ [cgs]             & 2.535$^{+0.059}_{-0.12}$      & 3.467$^{+0.057}_{-0.12}$\\
$\alpha_1$ [BJD$_{\mathrm{TDB}}$-2,450,000]$^{d, b}$            & 4978.81123 $\pm$ 0.00058   & 4968.63162 $\pm$ 0.00038\\
$\alpha_2$ [d]$^{d, b}$                   & 18.774480$^{+7.7\e{-5}}_{-1.4\e{-4}}$ & 50.77830$^{+2.7\e{-4}}_{-1.2\e{-4}}$ \medskip\\

\Kepler\ long-cadence jitter$^{b}$ & 1.069 $\pm$ 0.036     & \\
\Kepler\ short-cadence jitter$^{b}$ & 0.9962 $\pm$ 0.0024  & \\
{\it SOPHIE} jitter$^{b}$               & 0.98$^{+0.26}_{-0.14}$  &\medskip\\

$\frac{M_{\mathrm{p,b}}+M_{\mathrm{p,c}}}{M_\star}\,\,$$^{b}$ & 0.0016503$^{+8.1\e{-6}}_{-4.8\e{-6}}$ & \\
$\frac{M_{\mathrm{p,b}}}{M_{\mathrm{p,c}}}\,\,$$^{b}$ & 0.05151$^{+4.8\e{-4}}_{-1.3\e{-3}}$ & \\
$e_{\mathrm{b}} \cos \omega_{\mathrm{b}} - \frac{a_{\mathrm{c}}}{a_{\mathrm{b}}} e_{\mathrm{c}} \cos \omega_{\mathrm{c}} \,\,$$^{b}$ & -0.0425$^{+0.0011}_{-0.0033}$ &\\
$e_{\mathrm{b}} \cos \omega_{\mathrm{b}} + \frac{a_{\mathrm{c}}}{a_{\mathrm{b}}} e_{\mathrm{c}} \cos \omega_{\mathrm{c}} \,\,$$^{b}$ & 0.0176$^{+0.0063}_{-0.0025}$ &\\
$e_{\mathrm{b}} \sin \omega_{\mathrm{b}} - \frac{a_{\mathrm{c}}}{a_{\mathrm{b}}} e_{\mathrm{c}} \sin \omega_{\mathrm{c}} \,\,$$^{b}$ & 0.0006 $\pm$ 0.0019 &\\
$e_{\mathrm{b}} \sin \omega_{\mathrm{b}} + \frac{a_{\mathrm{c}}}{a_{\mathrm{b}}} e_{\mathrm{c}} \sin \omega_{\mathrm{c}} \,\,$$^{b}$ & -0.10341$^{+2.2\e{-3}}_{-9.6\e{-4}}$ &\\

\hline
\end{tabular}
\begin{list}{}{}
\item $^{a}$ 99 per cent Highest Density Interval (HDI).  
\item $^{b}$ MCMC jump parameter. 
\item $^{c}$ reflected with respect to $i = 90^\circ$, the supplementary angle is equally probable.
\item $^{d}$ $\alpha_1 \equiv t_{\mathrm{ref}} - \frac{\alpha_2}{2\pi}\left(M-E+e\sin{E}\right)$ with $E=2\arctan{\left\{\sqrt{\frac{1-e}{1+e}}\tan{\left[\frac{1}{2}\left(\frac{\pi}{2}-\omega\right)\right]}\right\}}$; $\alpha_2 \equiv \sqrt{\frac{4\pi^2a^{3}}{G \left( M_{\star} + M_{\mathrm{b}} + M_{\mathrm{c}} \right) }}$; $\alpha_2$ is not representative of the orbital periods of the planets, that moreover is not constant. $\alpha_1$ and $\alpha_2$ should not be used to predict transit times. Instead, the orbital parameters at $t_{\mathrm{ref}}$ should be used in a $N$-body integration. 
\item $\Msun$ = 1.98842\ten[30]~kg, \Rsun = 6.95508\ten[8]~m, $\Mjup$ = 1.89852\ten[27]~kg, $\Rjup$ = 7.1492\ten[7]~m
\end{list}
\end{table*}

\section{Results} \label{sect.results}

\begin{figure*}
\centering
\hspace{-2cm}\includegraphics[width=18cm]{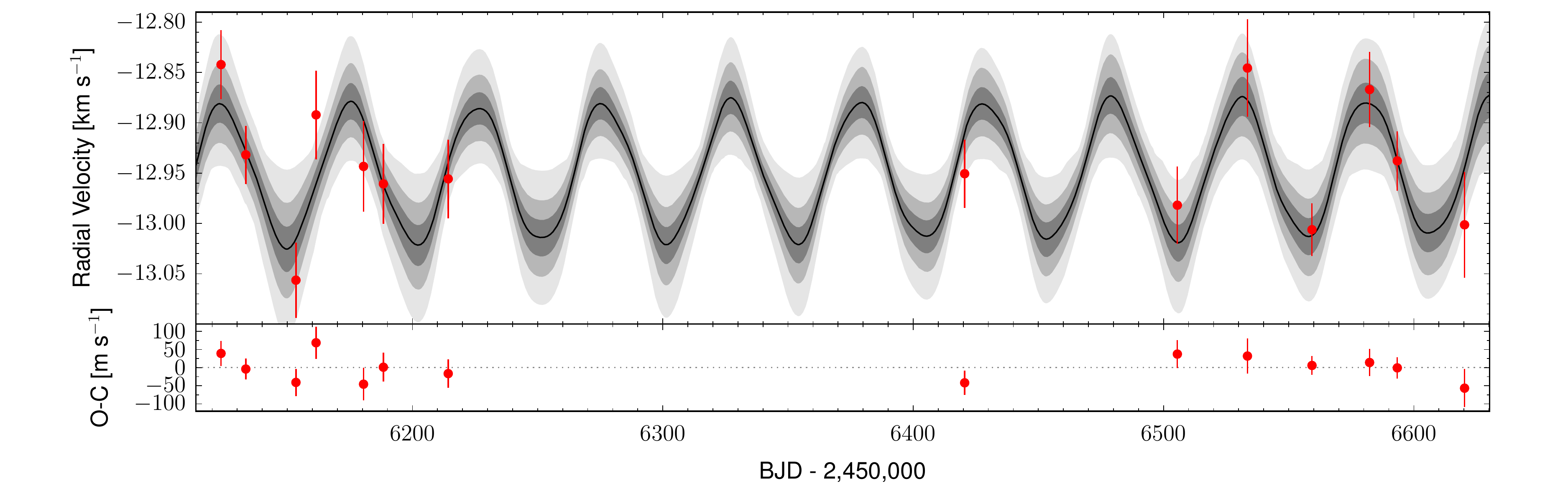}\hspace{-2cm} 
\caption{Radial velocities of Kepler-117 observed by {\it SOPHIE} and photodynamical model fit. The 68.3, 95.5, and 99.7 per cent Bayesian confidence intervals are plotted in different intensities of grey.}
\label{rv}
\end{figure*}

Figs~\ref{transitb} and \ref{transitc} show the photodynamical transit model. Note that our model naturally reproduces simultaneous transits, as in transit no. 15 of planet~c (no. 40 of planet~b). In Fig.~\ref{rv} we show the {\it SOPHIE} radial velocities and the model 1, 2, and 3~$\sigma$ credible intervals for each time. Fig.~\ref{pyramid} shows the parameter posterior distributions and correlations.

The model parameters are in agreement with the values reported by \citet{2015A&A...573A.124B}, but the uncertainties are significantly reduced: the uncertainties of the unitless parameters are between 2 and 28 times smaller. As we discuss below, this is one of the advantages of employing the full photodynamical model instead of computing the central times of transit separately.

\begin{figure}
\includegraphics[width=8.5cm]{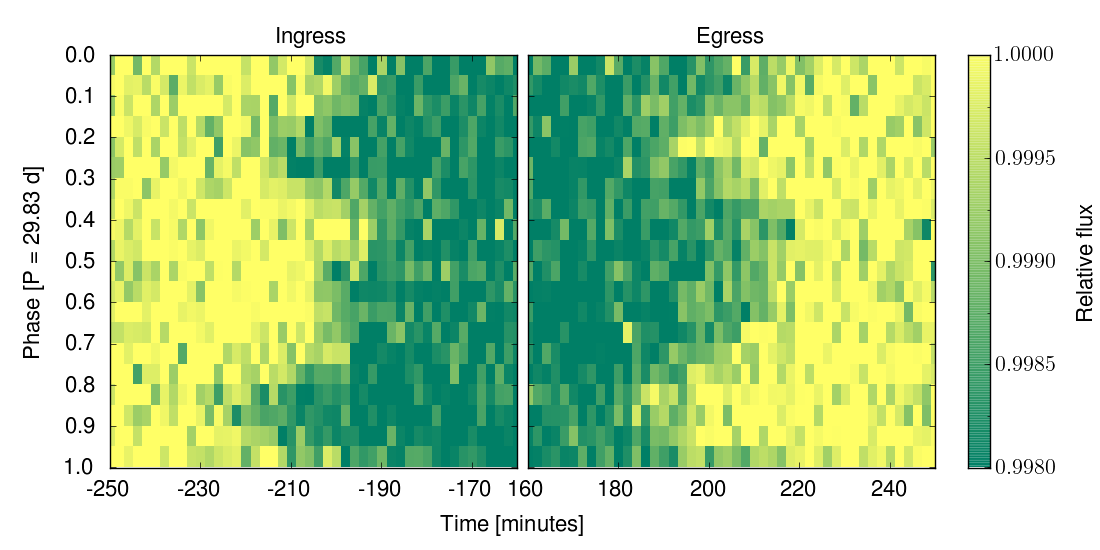} 
\caption{TTV of planet~b. Transit ingress and egress observed by \Kepler, with relative flux colour coded (the so-called river or lava plot). The times are measured from the expected transit time for a linear ephemeris. The transits are sorted according to the orbital phase of the 1:1 resonance period.}
\label{ttv}
\end{figure}

\begin{figure}
  \hspace{1cm}
\includegraphics[width=12cm]{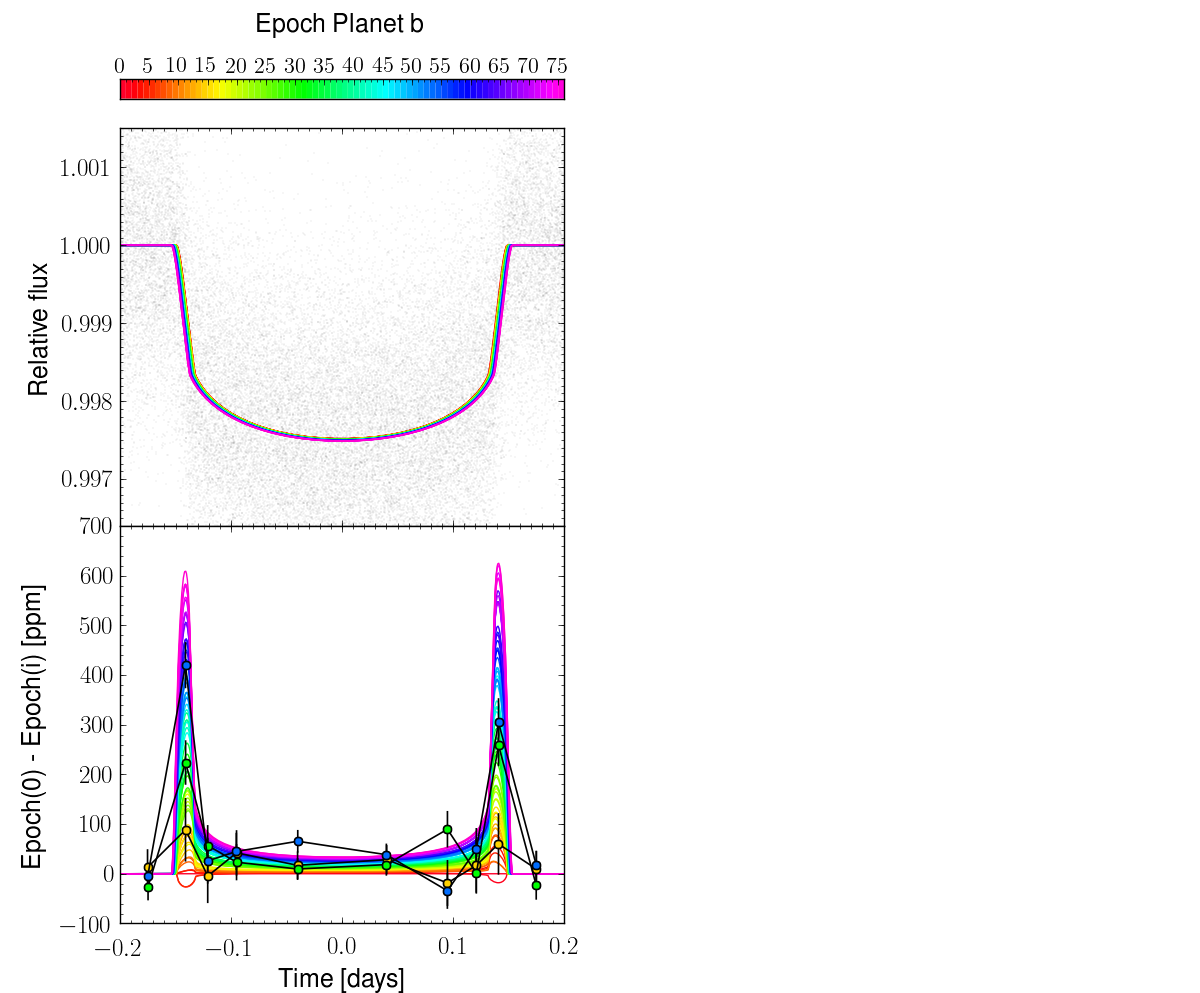} 
\caption{Transit shape evolution of planet~b during \Kepler\ observations. Upper panel: Phase-folded \Kepler\ photometry and maximum-likelihood transit model corrected for TTV for planet~b. The transit models of all epochs are plotted but are hardly distinguishable. Lower panel: difference between the transits of different epochs (colour coded) and the model of the first transit. The effect is observed in the data points by combining one third of the transits (black lines and circles with 3 different colours). Note that the difference between different epochs reaches 600~ppm.}
\label{shapeevolution}
\end{figure}

\begin{figure*}
\centering
\includegraphics[width=8.5cm]{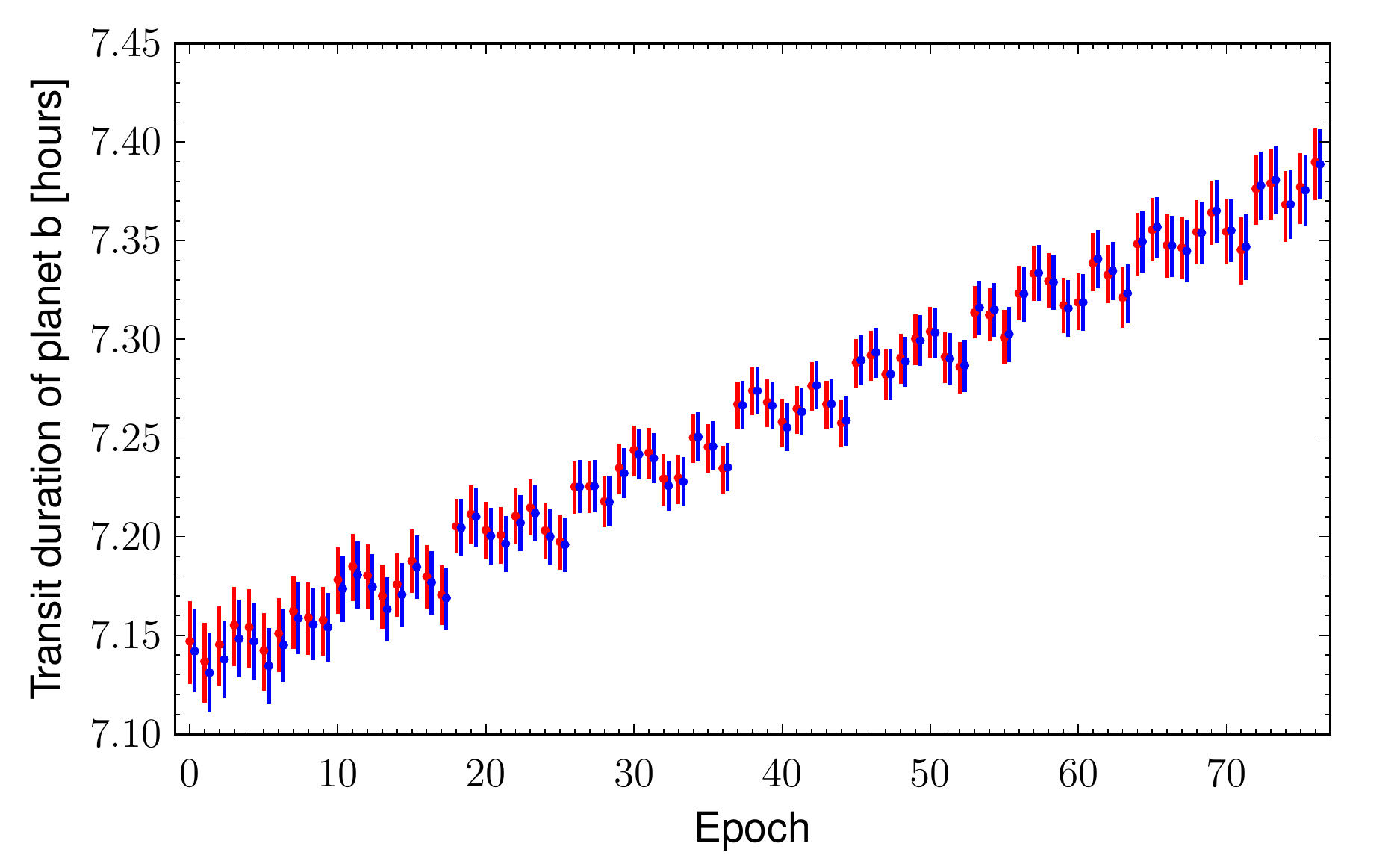}
\includegraphics[width=8.5cm]{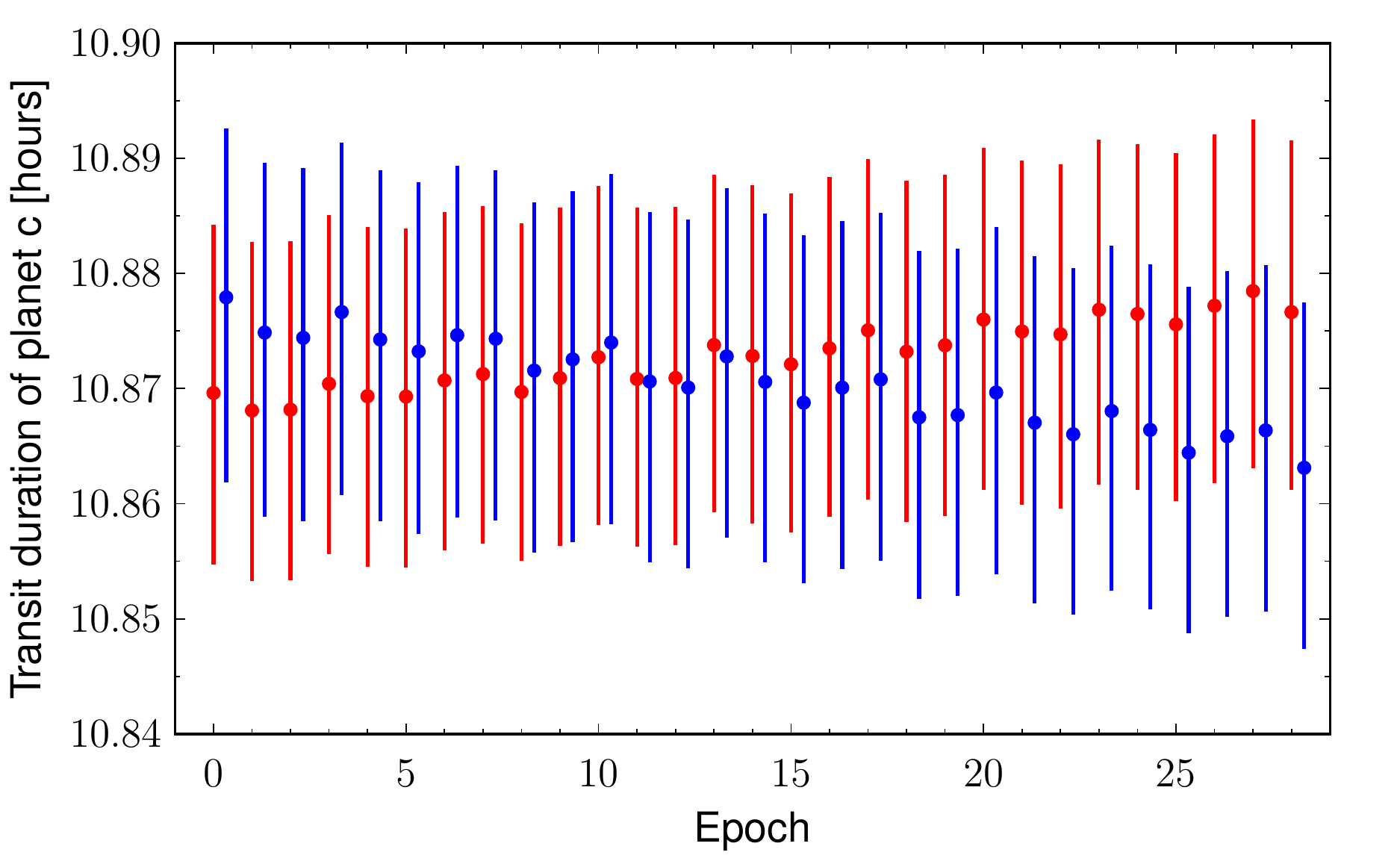} 
\caption{TDVs of planet~b and c during \Kepler\ observations. The two possible solutions for the inclination of planet~c are plotted in different colours, $i_{\mathrm{c}}>$90$\degree$ is shown in red and $i_{\mathrm{c}}<$90$\degree$ is shown in blue (slightly shifted in epoch for clarity). The duration change during \Kepler\ observations (last - first transit) is: planet~b = $14.6\pm1.7\ (i_{\mathrm{c}}>90)$, $14.8\pm1.7\ (i_{\mathrm{c}}<90)$ minutes; planet~c = $25\pm76\ (i_{\mathrm{c}}>90)$, $-53\pm77\ (i_{\mathrm{c}}<90)$ seconds. The mean error in the transit times of planet~c during \Kepler\ observations is 20~s, so it is not possible to differentiate between the two configurations.}
\label{TDV}
\end{figure*}

With our model the information in the TTVs and the transit shape is fully exploited, which explains the improvement with respect to the \citet{2015A&A...573A.124B} parameter determination. Figs~\ref{ttv} and \ref{shapeevolution} show the evolution of transit times and shapes, respectively. As can be seen, changes in transit shape are clearly detected for planet~b. On the other hand, the transit shape variations predicted by the model for planet~c are too small to be detectable with the available data. The transit duration of the inner planet increases by $14.7 \pm 1.7$~min during the timespan of the \Kepler\ observations (Fig.~\ref{TDV}). The duration and depth of the transits of planet~b are increasing as the transits become progressively central (Fig.~\ref{LongCen}). Eventually the transit paths along the stellar disc will cross, and mutual eclipses between planets could occur. This effect can aid to distinguish between the two possible configurations of the orbital inclinations, either both planets in the same or different halves of the stellar disc. With the available data both configurations are equally likely. Another possibility would be to precisely measure the transit duration of planet~b in a few years from now. The model predicts that by 2020 the transit duration of planet~b will differ by $\sim 3.6$~min between both configurations.

\begin{figure*}
\includegraphics[width=5.8cm]{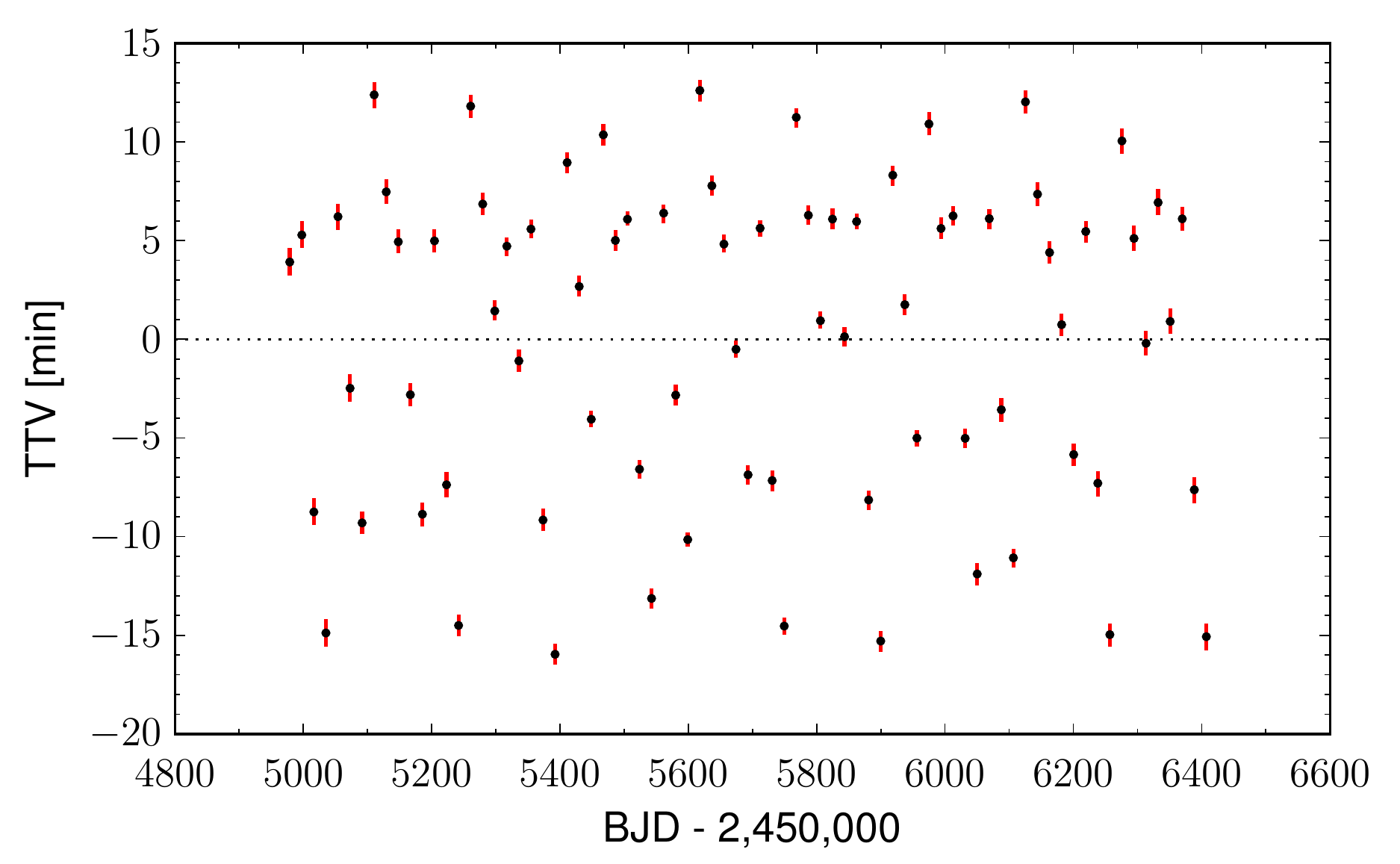} 
\includegraphics[width=5.8cm]{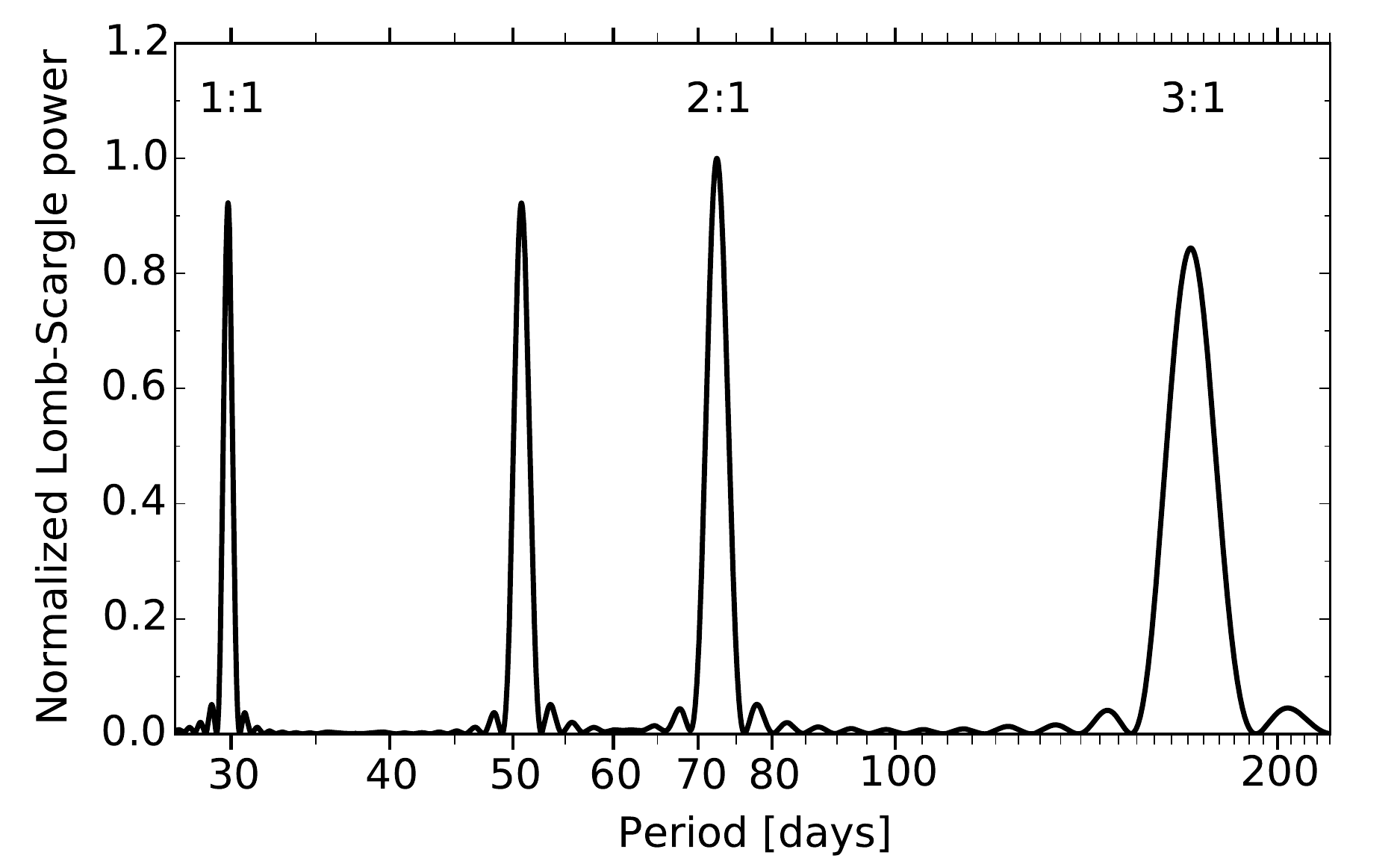} 
\includegraphics[width=5.8cm]{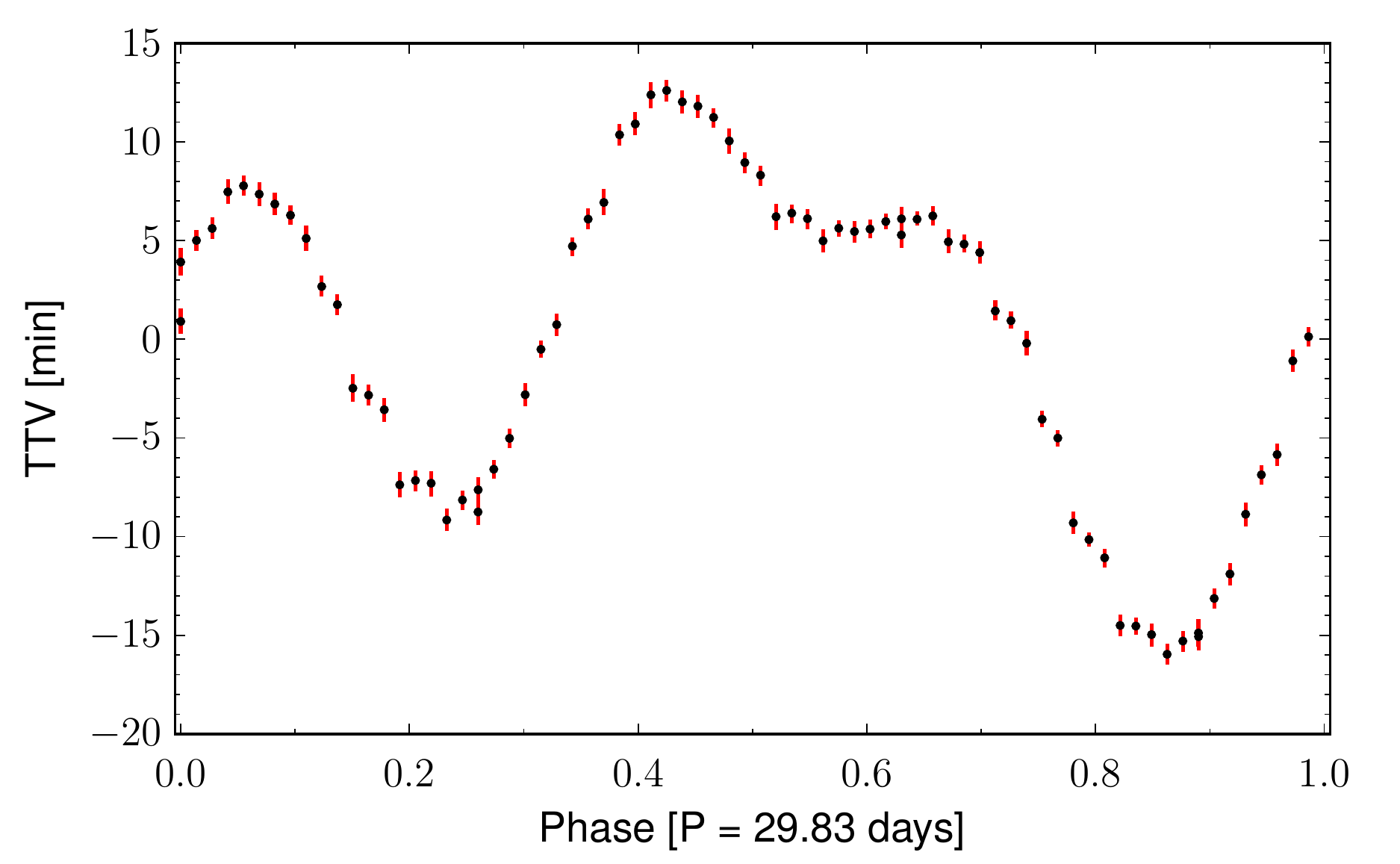}\\ \vspace{0.5cm}
\includegraphics[width=5.8cm]{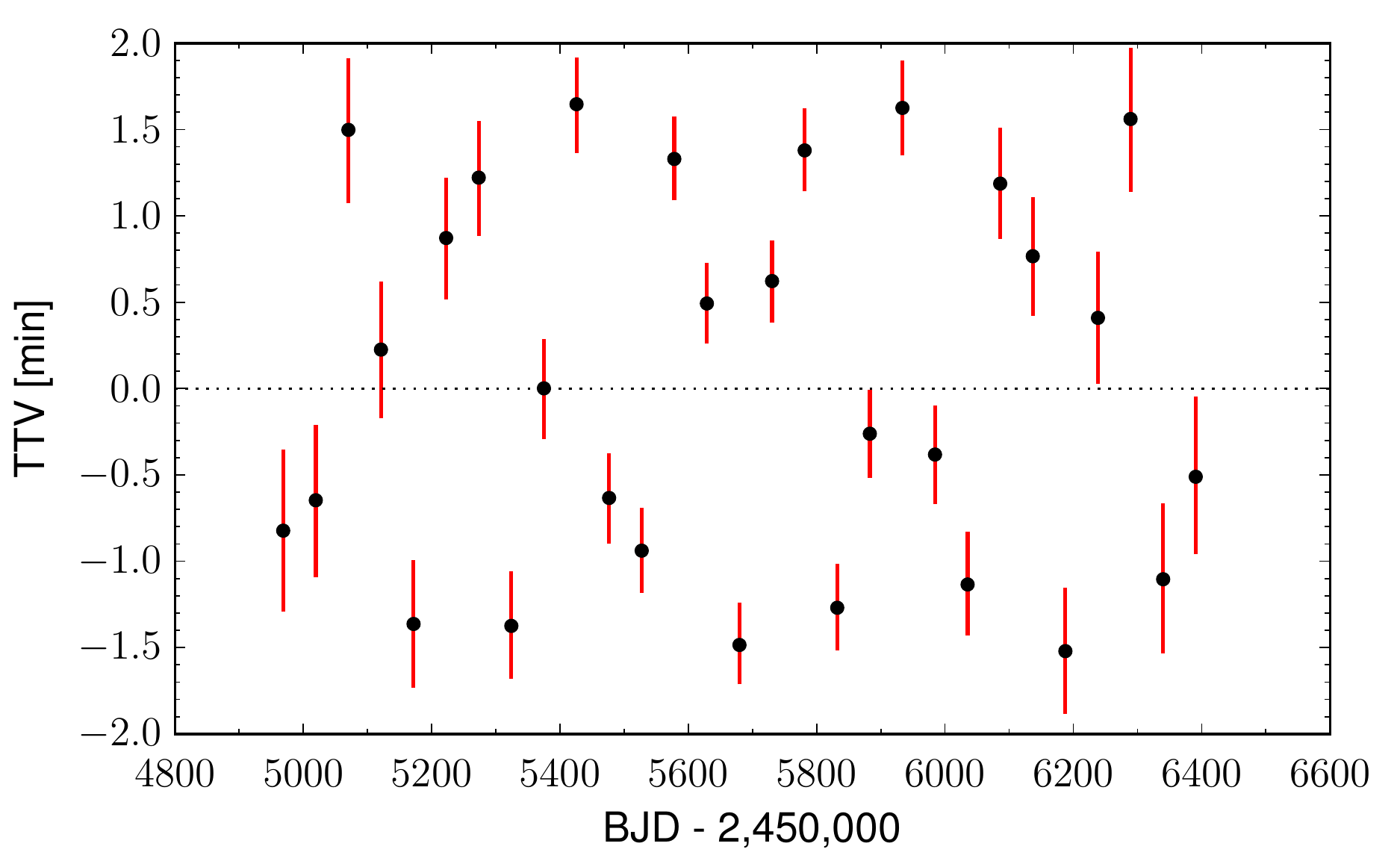} 
\includegraphics[width=5.8cm]{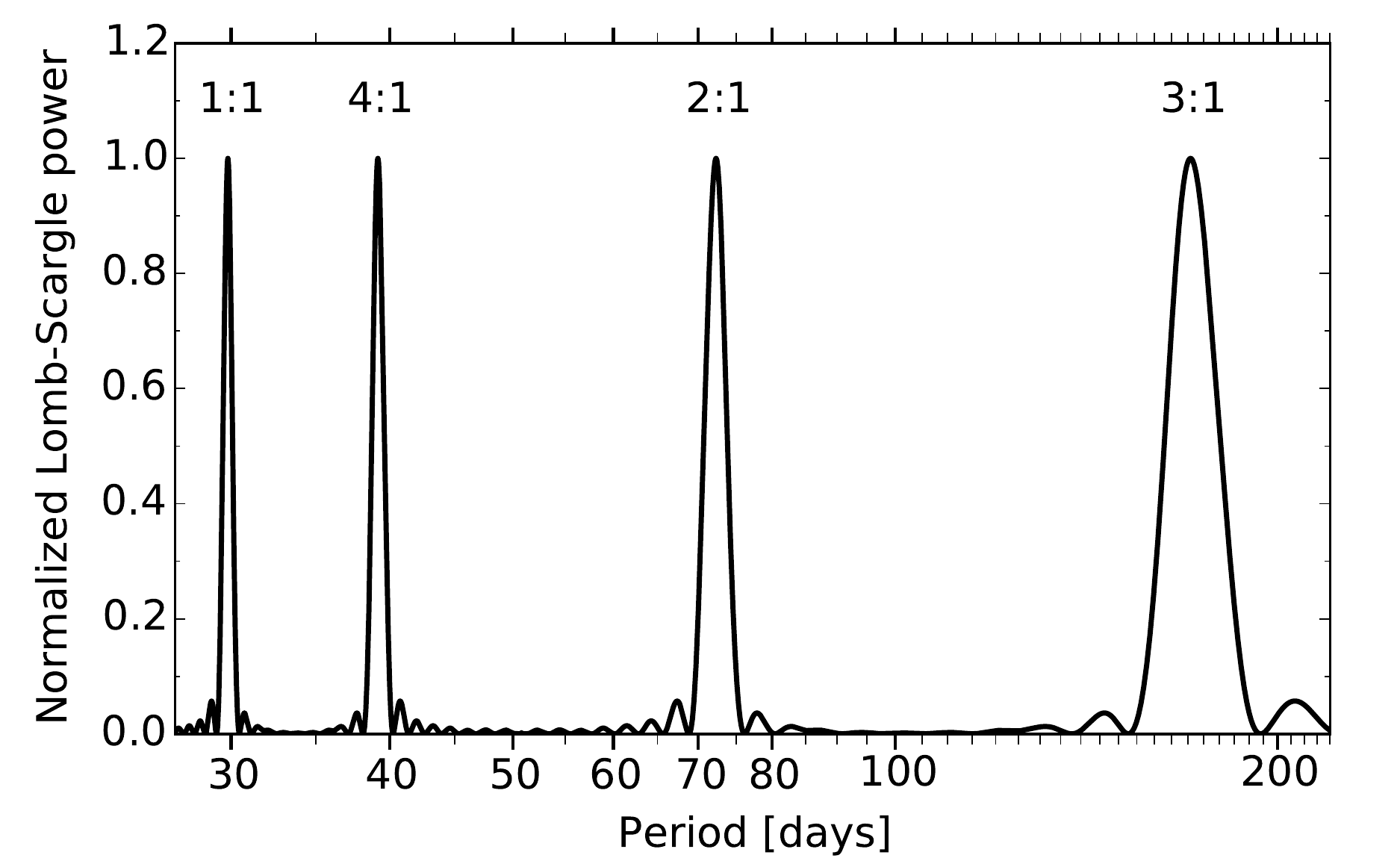} 
\includegraphics[width=5.8cm]{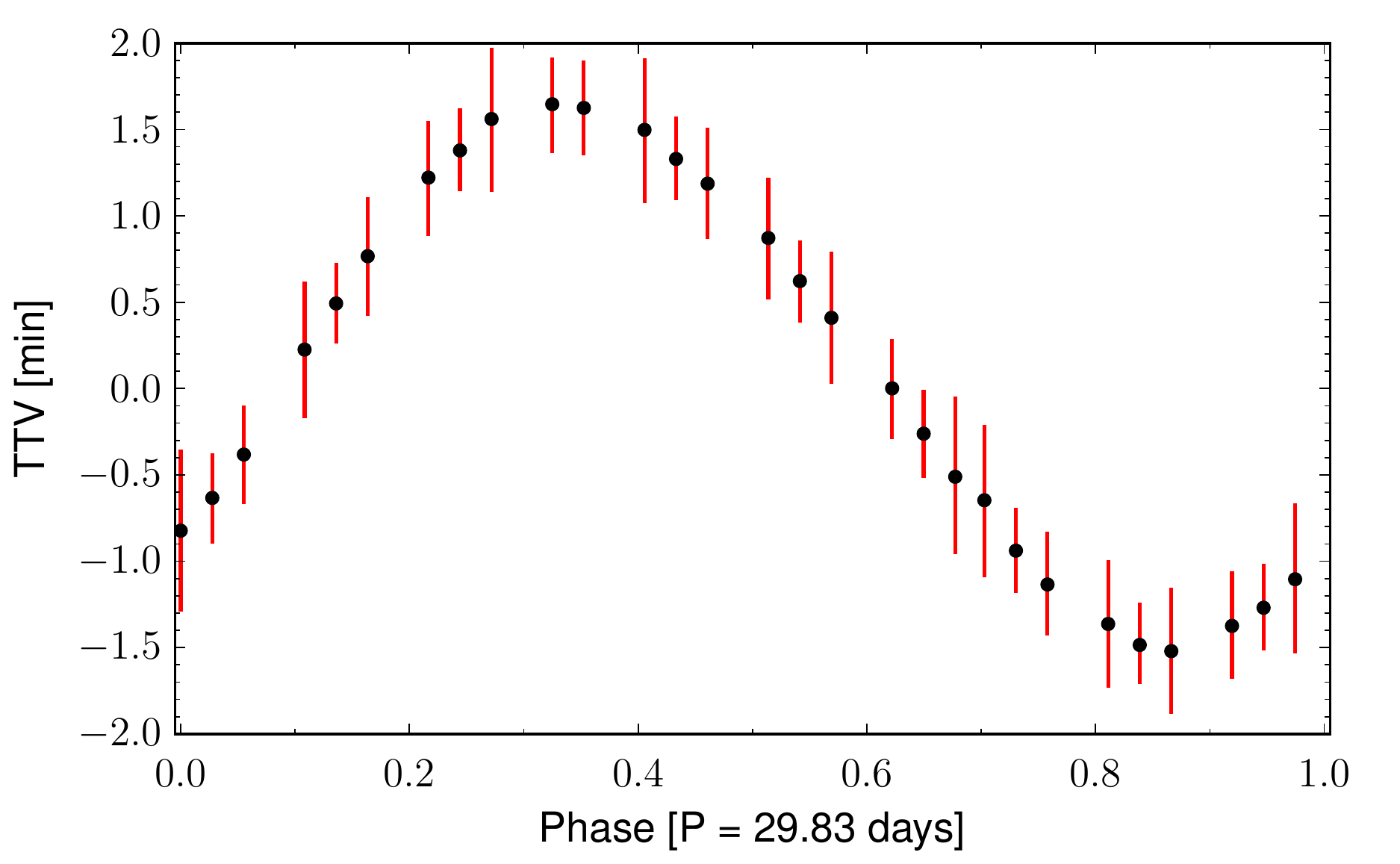} 
\caption{TTVs of planet~b (upper panels) and planet~c (lower panels) obtained with the photodynamical model fit. From left to right: TTVs (calculated respect to a linear ephemeris) against mid-transit time. Lomb-Scargle periodogram \citep{1989ApJ...338..277P} of the TTVs. The peaks corresponding to the modulation periods of different resonances are noted ($P_{1:1}$ = 29.83$\pm$0.23~d, $P_{2:1}$ = 72.3$\pm$1.3~d, $P_{3:1}$ = 170.9$\pm$7.4~d, and $P_{4:1}$ = 39.15$\pm$0.39~d). The remaining peaks correspond to aliases of the sampling frequencies \citep[see][]{1987AJ.....93..968R}. TTVs folded at period of the 1:1 resonance (29.83~d).}
\label{TTV}
\end{figure*}

The central times of the transit obtained from the photodynamical model exhibit a 29 min amplitude variation with respect to a linear ephemeris\footnote{The linear ephemeris named trough the paper refers to a linear fit to the mid-transit times (of the transits observed by \Kepler) obtained with the photodynamical model fit (Sect.~\ref{sect.pdm}). The linear ephemeris for planet~b is: BJD$_{\mathrm{TDB}}$ = 2 454 978.8214(12) + 18.795931(27) $\times$ Epoch, and for planet~c: BJD$_{\mathrm{TDB}}$ = 2 454 968.63220(25) + 50.790374(15) $\times$ Epoch, the errors of the last digit are indicated in parenthesis.} for the inner planet. We also detect clear TTVs of the outer planet of 3~min amplitude (Fig.~\ref{TTV}). 

With a period ratio of around 2.7, the Kepler-117 system is many resonance widths away from both the 2:1 and 3:1 resonances, the widths of which are around 0.06 and 0.02, respectively, for this system\footnote{The width of a resonance, $\Delta\sigma$, is defined to be such that the associated harmonic angle librates (rather than circulates) when $\sigma-\Delta\sigma<\sigma<\sigma+\Delta\sigma$, where $\sigma$ is the period ratio. Physically, a librating harmonic angle allows for substantial energy and angular momentum to be exchanged between the orbits over many orbital periods, as is evidenced by the significant TTVs one associates with resonant systems.} \citep{2013MNRAS.435.2187M}. None the less, these two strong harmonics of the disturbing function are responsible for significant perturbations to the orbital elements (Fig.~\ref{ShortCen}), which in turn, manifest themselves in detectable TTVs, as well as significant Fourier power in the associated circulation periods of the harmonic angles. For a system with orbital periods $P_{\mathrm{b}}$ and $P_{\mathrm{c}}$ and period ratio $\sigma=P_{\mathrm{c}}/P_{\mathrm{b}}$, the circulation period of an $n'\!:\!n$ harmonic is $P_{n':n}=P_{\mathrm{c}}/|n\,\sigma-n'|$, which for the 1:1, 2:1, 3:1, and 4:1 harmonics of the Kepler-117 system are 29.83~d, 72.27~d, 170.86~d, and 39.15~d respectively\footnote{using as periods the median time between consecutive transits, see Fig~\ref{periods}.}, matching the periods detected in the Lomb-Scargle analysis of the TTVs (Fig.~\ref{TTV}).
In spite of being far from resonance, the TTVs of systems like Kepler-117 are rich with information about their architecture and mass distribution; this will be explored elsewhere (Mardling, in preparation).

We emphasize that the observed TTVs and transit duration variations (TDVs) are based on transit times and durations that are obtained as a by-product of the photodynamical code (obtained, respectively, as the mean and the difference of the first and fourth contacts computed from the sky-projected planet-star separation). These measurements are based on the data but are assisted by the assumption that they are produced by the gravitational interactions between the system bodies. Thus the achieved precision in the determination of transit times are improved by a factor of 4 with respect to a classical determination using only the individual transit light curves.

Finally, the absolute masses and radii of the star and both orbiting planets are constrained without using stellar evolutionary models, mainly because of the effect in the transit times. The precision achieved for the star radius and mass is 20, and 70 per cent, respectively, limited by the available radial velocity precision\footnote{The mean radial velocity error is 38~\ms, to be compared with the radial velocity amplitudes reported by \citealt{2015A&A...573A.124B}: $\sim$7~\ms\ for planet~b and $\sim$90~\ms\ due to planet~c.}. \citet{2015A&A...573A.124B} resort to stellar models and obtain a much better precision. The discrepancy between  \citet{2015A&A...573A.124B} and the values reported here is significant to 95 per cent. As the available radial velocity data are not selective enough the posterior distributions are influenced by the chosen priors. In this case -- as the stellar density and radius prior distributions are flat -- the mass prior probability increases towards lower values, which explains, at least partially, the bias for less massive stars. The radius is in turn affected via the correlation with the stellar density. On the other hand, the parameters that are precisely constrained by the available data are in agreement with \citet{2015A&A...573A.124B}. In the next section we discuss the potential of this technique for stars for which 1~\ms\ precision is achievable. In this case, we will show that the mass posterior distribution is completely dominated by the synthetic data, and that the obtained mass is in agreement with the simulated value.

\subsection{Adding high-precision synthetic radial velocities}
\begin{figure*}
\centering
\hspace{-2cm}\includegraphics[width=18cm]{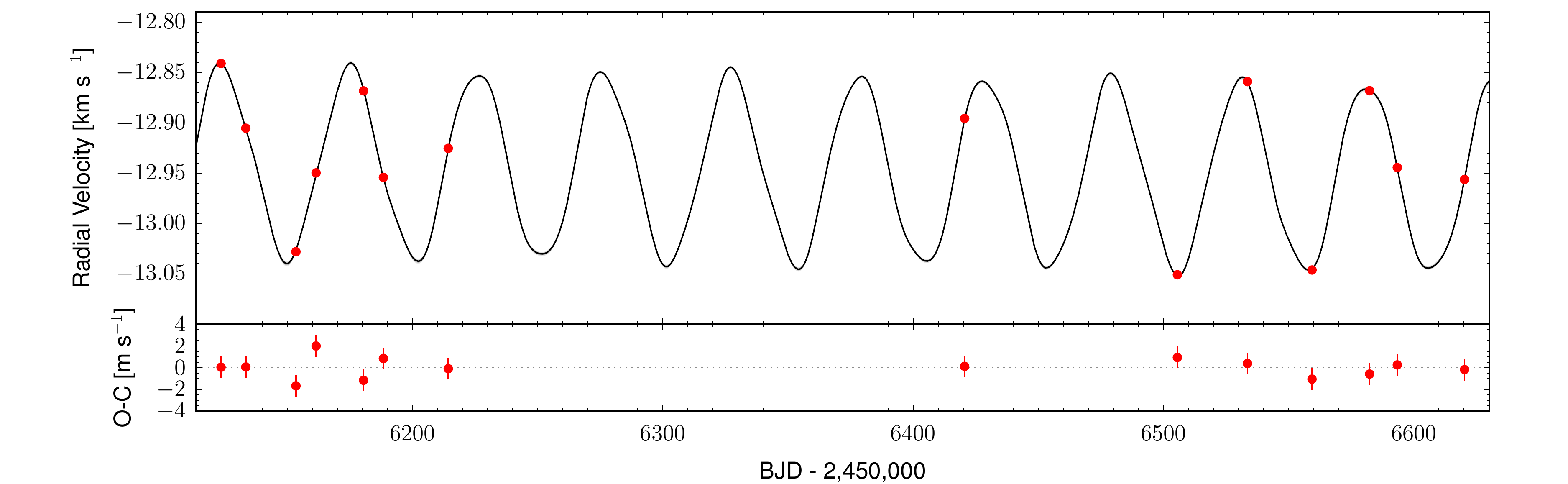}\hspace{-2cm} 
\caption{Idem Fig.~\ref{rv} for the photodynamical model fitted with simulated radial velocities of 1~\ms\ precision.}
\label{rvFake1ms}
\end{figure*}

To probe the capability of the method to characterize targets coming from next generation space-based transit surveys, we replaced the {\it SOPHIE} observations by simulated radial velocity data with a precision of 1~\ms (see Fig.~\ref{rvFake1ms}). The parameters used for the simulation are listed in Table~\ref{simutable}. The stellar mass was chosen close to the one presented by \citet{2015A&A...573A.124B}, but this is irrelevant for the comparison below.

\begin{table*}
\renewcommand{\arraystretch}{0.7}
\centering
\cprotect\caption{Model parameters for the simulated run with 1\ms-precision radial velocities. Simulated value, posterior median and 68.3 \% credible intervals. The orbital elements have the origin at the star (Asteroidal parameters in the \textsc{mercury} code) and are given for the reference time $t_{\mathrm{ref}} = 2 454 967.63$~BJD$_{\mathrm{TDB}}$.}\label{simulated}
\begin{tabular}{lcc}
\hline
Parameter & Simulated value & Median and 68.3 \% credible interval \\
\hline
Stellar mass, $M_\star$ [\Msun]                                & 1.105 & 1.119 $\pm$ 0.017 [1.0736, 1.1575]$^{a}$\\
Stellar radius, $R_\star$ [\Rsun]$^{b}$                   &  & 1.561 $\pm$ 0.012 [1.5321, 1.5865]$^{a}$\\
Stellar density, $\rho_{\star}$ [$\rho_\odot$]$^{b}$       &  & 0.2948 $\pm$ 0.0048 \\
Surface gravity, \logg\ [cgs]                                 &  & 4.1010 $\pm$ 0.0057 \\
Linear limb darkening coefficient, $u_{\mathrm{a}}$$^{b}$           &  & 0.404$^{+0.024}_ {-0.014}$ \\
Quadratic limb darkening coefficient, $u_{\mathrm{b}}$$^{b}$         &  & 0.158$^{+0.025}_ {-0.044}$ \\
Systemic velocity (at BJD$_{\mathrm{TDB}}$ 2,456,355), $\gamma_0$ [\kms]$^{b}$ & -12.94774 & -12.94779  $\pm$ 0.00029 \\
Linear radial velocity drift, $\gamma_1$ [m s$^{-1}$ yr$^{-1}$]$^{b}$ & -12.62 & -12.25 $\pm$ 0.59 \bigskip\\

\emph{Kepler-117b} & & \smallskip\\
Semi-major axis, $a$ [au]                                 & 0.14302      & 0.14362 $\pm$ 0.00074 \\
Eccentricity, $e$                                         & 0.05263     & 0.05251 $\pm$ 0.00085 \\
Inclination, $i$ [\degree]$^{b}$                     & 88.700       & 88.708 $\pm$ 0.040 \\
Argument of pericentre, $\omega$ [\degree]                & 256.5       & 256.3 $\pm$ 1.9 \\
Longitude of the ascending node, $n$ [\degree]$^{b}$ & 181.31       & 181.22 $\pm$ 0.18 \\
Mean anomaly, $M$ [\degree]                               & 340.5        & 340.7 $\pm$ 2.2 \\
Radius ratio, $R_{\mathrm{p}}/R_\star$$^{b}$                     &      & 0.04679 $\pm$ 0.00012 \\
Planet mass, $M_{\mathrm{p}}$ [$\Mjup$]                             & 0.0932    & 0.0946 $\pm$ 0.0016 \\
Planet radius, $R_{\mathrm{p}}$[$\Rjup$]                            &      & 0.7103 $\pm$ 0.0063 \\
Planet density, $\rho_{\mathrm{p}}$ [$\mathrm{g\;cm^{-3}}$]                    &      & 0.3275 $\pm$ 0.0073 \\
Planet surface gravity, $\log$\,$g_{p}$ [cgs]              &       & 2.6671 $\pm$ 0.0073 \\
$\alpha_1$ [BJD$_{\mathrm{TDB}}$-2,450,000]$^{d, b}$              & 4978.81123 & 4978.81127 $\pm$ 0.00048 \\
$\alpha_2$ [d]$^{d, b}$                       & 18.774380     & 18.774386 $\pm$ 4.3$\e{-5}$ \bigskip\\

\emph{Kepler-117c} & & \smallskip\\
Semi-major axis, $a$ [au]                                 & 0.2776  & 0.2788 $\pm$ 0.0014 \\
Eccentricity, $e$                                         & 0.03099 & 0.03121 $\pm$ 0.00073 \\
Inclination, $i$ [\degree]$^{b}$                     & 89.685   & 89.685 $\pm$ 0.034 \\
Argument of pericentre, $\omega$ [\degree]                & 300.2   & 300.0 $\pm$ 1.4 \\
Longitude of the ascending node, $n$ [\degree]            & 180             & 180 \\
Mean anomaly, $M$ [\degree]                               & 140.9   & 141.1 $\pm$ 1.5 \\
Radius ratio, $R_{\mathrm{p}}/R_\star$$^{b}$                     &  & 0.07095 $\pm$ 0.00014 \\
Planet mass, $M_{\mathrm{p}}$ [$\Mjup$]                             & 1.828   & 1.849 $\pm$ 0.027 \\
Planet radius, $R_{\mathrm{p}}$[$\Rjup$]                            &    & 1.0771 $\pm$ 0.0091 \\
Planet density, $\rho_{\mathrm{p}}$ [$\mathrm{g\;cm^{-3}}$]                    &    & 1.838 $\pm$ 0.035 \\
Planet surface gravity, $\log$\,$g_{p}$ [cgs]             &     & 3.5973 $\pm$ 0.0067 \\
$\alpha_1$ [BJD$_{\mathrm{TDB}}$-2,450,000]$^{d, b}$             & 4968.63151 & 4968.63152 $\pm$ 0.00031 \\
$\alpha_2$ [d]$^{d, b}$                      & 50.778361 & 50.778345 $\pm$ 8.0$\e{-5}$ \bigskip\\

\Kepler\ long-cadence jitter$^{b}$ & &  1.067 $\pm$ 0.039 \\
\Kepler\ short-cadence jitter$^{b}$ &  & 0.9963 $\pm$ 0.0024 \\
Radial velocity jitter$^{b}$              &  & 1.07 $\pm$ 0.25 \medskip\\

$\frac{M_{\mathrm{p,b}}+M_{\mathrm{p,c}}}{M_\star}\,\,$$^{b}$ & 0.0016591 & 0.0016584 $\pm$ 2.9$\e{-6}$ \\
$\frac{M_{\mathrm{p,b}}}{M_{\mathrm{p,c}}}\,\,$$^{b}$        & 0.05099   & 0.05112 $\pm$ 0.00030 \\
$e_{\mathrm{b}} \cos \omega_{\mathrm{b}} - \frac{a_{\mathrm{c}}}{a_{\mathrm{b}}} e_{\mathrm{c}} \cos \omega_{\mathrm{c}} \,\,$$^{b}$ & -0.0425  & -0.0425 $\pm$ 0.0017 \\
$e_{\mathrm{b}} \cos \omega_{\mathrm{b}} + \frac{a_{\mathrm{c}}}{a_{\mathrm{b}}} e_{\mathrm{c}} \cos \omega_{\mathrm{c}} \,\,$$^{b}$ & 0.0180   & 0.0178 $\pm$ 0.0028 \\
$e_{\mathrm{b}} \sin \omega_{\mathrm{b}} - \frac{a_{\mathrm{c}}}{a_{\mathrm{b}}} e_{\mathrm{c}} \sin \omega_{\mathrm{c}} \,\,$$^{b}$ & 0.0008 & 0.0013 $\pm$ 0.0016 \\
$e_{\mathrm{b}} \sin \omega_{\mathrm{b}} + \frac{a_{\mathrm{c}}}{a_{\mathrm{b}}} e_{\mathrm{c}} \sin \omega_{\mathrm{c}} \,\,$$^{b}$ & -0.1032   & -0.1036 $\pm$ 0.0015 \\

\hline
\end{tabular}
\label{simutable}
\end{table*}

The photodynamical modelling was repeated identically and the results are shown in Table~\ref{simutable}. With this precision in the radial velocities, the stellar radius and mass are measured with a precision of 1 and 2 per cent, respectively. As the planet-to-star mass and radius ratios are known to a precision better than 2 per cent, independently of the radial velocities, the absolute radii and masses of the planets are also known to 1 and 2 per cent precision.

\section{Discussion} \label{sect.discussion}

We have presented the analysis of Kepler-117 modelling the dynamical evolution of the system during the timespan of the \Kepler\ observations using an $N$-body simulation. Usually, the works in the literature studying the dynamical interactions in multiplanet systems compute first the transit times assuming a fixed transit shape at each epoch and then model the deviation of the thus measured transit times from a linear ephemeris (the TTVs). However, a number of advantages exist in employing a complete dynamical model of the system over this two-step method.
\begin{itemize}
  \setlength{\itemindent}{0.4in}
\item[(a)] Transit times are obtained consistently with the interactions and dynamical evolution of the system. This leads to a much better precision in the transit times. We refer to this determination as gravitationally-assisted transit times (see Fig.~\ref{Tmiderr}). In the case studied here, the transit times are determined with uncertainties four times smaller than the ones reported in \citet{2015A&A...573A.124B}.

\begin{figure}
\includegraphics[width=8.5cm]{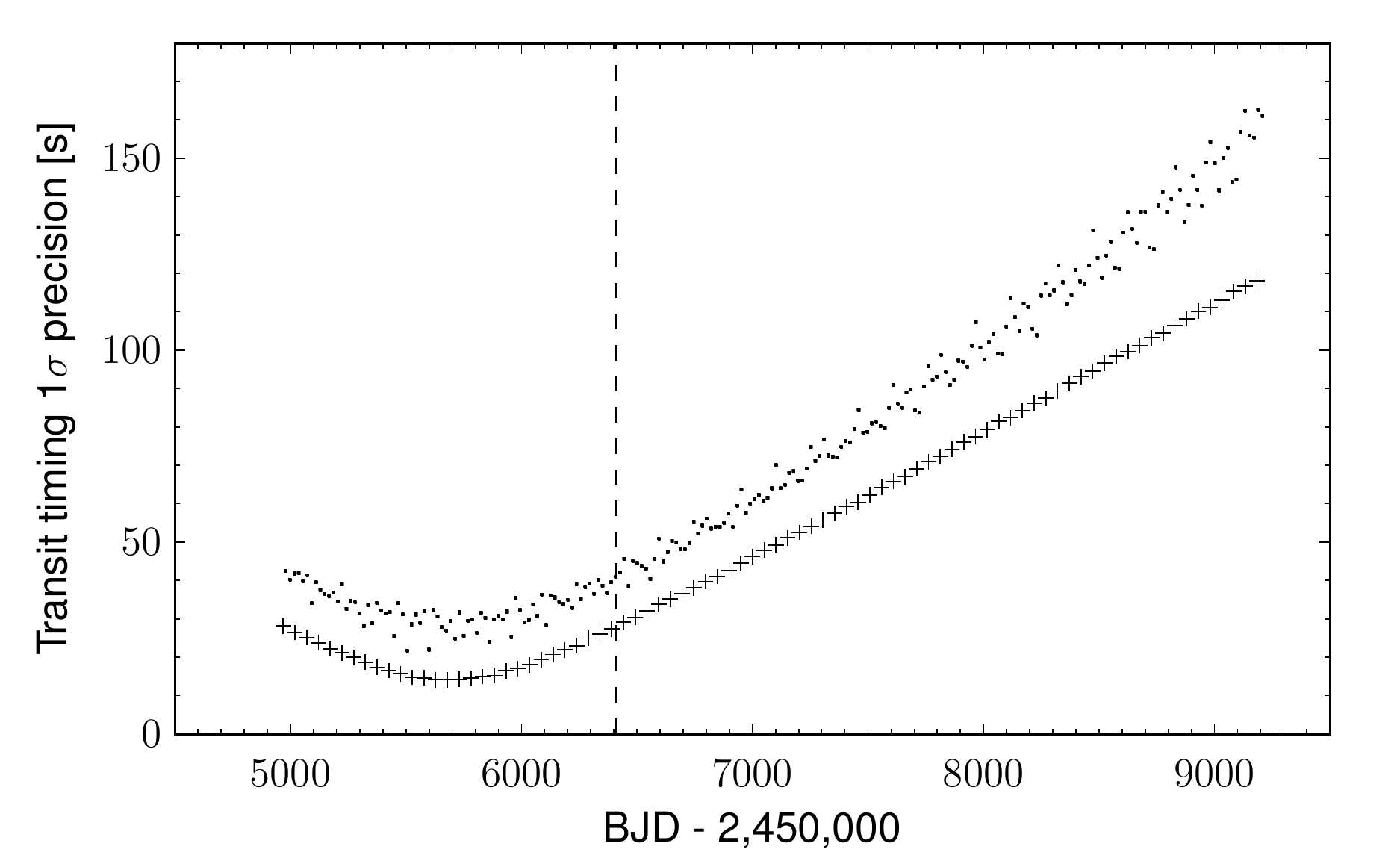}
\caption{Mid-transit time error of planet~b (dots) and c (plus markers) until 2020 from the photodynamical model fit. The vertical dashed line mark the end of \Kepler\ observations. Note that the precision in transit timing is minimum at the middle of \Kepler\ observations. So, precision in transit timing depends on a combination of photometric precision, photometry sampling, and dynamics (through gravity).}
\label{Tmiderr}
\end{figure}

\item[(b)] As a consequence of (a), the precision in transit parameters is improved. Including the shape changes in the model leads to an improved determination of the transit parameters and the derived quantities. With respect to the analysis of \citet{2015A&A...573A.124B} the stellar density is determined with twice the precision, the densities of planets~b and c are known 11 and 4 times (respectively) better, and the difference of the longitude of the ascending node has an uncertainty 28 times smaller. 
  
\item[(c)] The masses and radii of the objects are obtained independently of stellar models if the system scale can be determined. To do this, we have resorted to the {\it SOPHIE} radial velocities, but detecting the light time travel effect, which is negligible for Kepler-117 is another possibility. The independence from stellar evolutionary models make these determinations as valuable as those obtained in double-lined eclipsing binaries.
\end{itemize}

In practice, the TTVs and the changes in the transit shape cannot be detected for all multi-planet systems. However, we note that the TTVs of Kepler-117 were not considered significant by other authors \citep{2013ApJS..208...16M,2015arXiv150305555K}. Other similar systems might have been overlooked as well. Even if computationally expensive, the photodynamical model, as the one described here, permits to fully exploit the observations of multi-transiting planets such as those obtained by the \Kepler\ mission.

The only assumptions of the model are the Newtonian Law of gravitation and the geometry of an opaque disc occulting a bright one that follows a limb-darkening law. However, we have identified five neglected effects: (a) the uncertainty in the contamination of the \Kepler\ photometric mask (it is provided without error), (b) stellar activity, (c) relativistic effects, (d) the light-time effect, and (e) the non-sphericity of the objects. We have repeated the analysis adding a flux contamination (a) as a free parameter. We found an additional contamination factor of -0.7$\pm$1.5 per cent with respect to the one reported in the \Kepler\ archive. The planet-to-star radius ratios seem to be the only parameters affected, with distributions 3-4 times wider compared to the analysis with fixed contamination. In any case, it seems that the contamination of the photometric aperture can be directly measured in this kind of analysis. About the activity (b), there is a 0.1 per cent amplitude variability in the light curve plausibly caused by stellar spots \citep{2015A&A...573A.124B}, that can affect the planet-to-star radius ratios and the radial velocity measurements. In general, stellar activity can have an effect on the parameters obtained by the photodynamical modelling, but probably less than if transit times are measured individually. Here, instead, gravitationally-assisted transit times are constrained by the entire light curve. General relativity (c), and the finite speed of light (d) are negligible in this case. Nor do we expect a significant deviation from sphericity (e) as the star is rotating slowly \citep{2015A&A...573A.124B}, and the planets are in distant orbits, so tidal effects are negligible. An additional source of error are other bodies in the system not taken into account. Their influence depends mainly on their mass. 

In our analysis, the precision achieved for the mass and radius with the available {\it SOPHIE} radial velocities is very poor, specially the mass. Current high-precision spectrographs like {\it HARPS-N} can achieve $\sim$4~\ms\ precision radial velocities on this star. Future space transit search missions will focus on brighter targets for which radial velocity precision better than 1~\ms\ will be achieved. Thus, we have simulated radial velocity measurements with a precision of 1~\ms\ at the times of {\it SOPHIE} observations. We have repeated the analysis with these data obtaining a precision of 1 per cent for the radii and 2 per cent for the masses (both stellar and planetary). As the ratios planet-star radius and masses are determined very accurately ($<$2 per cent), the precision in the masses and radii of the star and planets is similar. This simulation shows the potential of this technique. We have shown that it is possible to achieve a precision comparable with the also direct empirical determinations in double-lined spectroscopic binaries \citep{2010A&ARv..18...67T}. Besides, in binary systems the stars can be affected by interactions between components, especially in low-mass stars.

When a planetary system exhibits detectable gravitational interactions, the orbital parameters of the system, the mean densities of all bodies \footnote{The planet density can be written as $\rho_{\mathrm{p}} = \rho_\star \left(\frac{M_{\mathrm{p}}}{M_\star}\right)\left(\frac{R_{\mathrm{p}}}{R_\star}\right)^{-3}$. When dynamical interactions are detected all terms are known precisely and therefore the planet density can be obtained without further measurements (see Table~\ref{table}).}, the planet-to-star radius and mass ratios, and planet mass ratios are determined by the photometry alone, without resorting to stellar models or to further measurements. If the effect of the finite speed of light is not detected, there are two possibilities to obtain absolute masses and radii: (a) impose priors on the stellar radius or mass (e.g. based on spectroscopic measurements, asteroseismology or stellar radius estimated with {\it Gaia}), or (b) break the Newtonian degeneracy using additional measurements (e.g. radial velocities). In this paper we took the second approach. The shape of the radial velocity curve, except its amplitude, can be inferred from the analysis of the photometry. Therefore, in principle, only two radial velocity observations (ideally at amplitude extrema) are needed to obtain a measurement of the system scale, and determine the masses and radii of the system bodies independently of stellar evolution models. However, more than a couple of radial velocity measurements is desirable, as the presence of drifts (in the case of Kepler-117 there is no detectable drift) or spectrograph systematics can bias the measurement. Stellar jitter due to spots can affect and limit the precision in radial velocity measurements, but its effect can be reduced observing with an infrared spectropolarimeter like SPIRou \citep{2013sf2a.conf..497D}. Given that the full characterization of the system with radial velocities is not needed, this can reduce dramatically the amount of follow-up observations needed for future space missions like {\it TESS} and {\it PLATO}.

In principle the same technique presented here is applicable to all transiting multiplanet systems, limited by the amplitude of the radial velocities, the amplitude of the TTVs, the photometric precision and time sampling. In practice the main limitation comes from the Doppler precision obtained on the relative faint transits host stars. In the future, missions like {\it TESS} \citep{2014SPIE.9143E..20R} and {\it PLATO} \citep{2014ExA....38..249R} will provide a large sample of multiplanet systems around bright stars. For quiet bright stars a precision better than 1~\ms\ is already achievable \citep{2011A&A...534A..58P}. However, multiplanet systems are mostly composed of small planets \citep{2015ApJS..217...31M}. Therefore, even for bright transiting hosts, radial velocities will limit the applicability of the method or the precision at which absolutes masses and radii are determined. Other limiting factor is the time sampling of the light curve. The 25~s observing cadence of {\it PLATO} will be an improvement with respect to previous space missions, and furthermore the fast cameras, with 2.5~s cadence, can improve the transit timing precision for the brightest targets.

In the context of {\it PLATO}, the photodynamical modelling presented here could be complementary to asteroseismology in determining the physical parameters of target stars, as not all stars with multiplanet systems will have detected pulsations. This mass and radius determination is only limited by the precision in photometry and radial velocity measurements, as opposed to the ones determined using stellar models, that depend on the understanding of the involved physical processes taking place in the stars.

\section*{Acknowledgements}
This paper is partially based on observations made with {\it SOPHIE} on the 1.93-m telescope at the Observatoire de Haute-Provence (CNRS), France. 
This paper includes data collected by the \Kepler\ mission. Funding for the \Kepler\ mission is provided by the NASA Science Mission directorate. All of the data presented in this paper were obtained from the MAST. STScI is operated by the Association of Universities for Research in Astronomy, Inc., under NASA contract NAS5-26555. Support for MAST for non-{\it HST} data is provided by the NASA Office of Space Science via grant NNX09AF08G and by other grants and contracts. The team at LAM acknowledge support by CNES grants 98761 (SCCB), 426808 (CD), and 251091 (JMA). JMA and XB acknowledge funding from the European Research Council under the ERC Grant Agreement n. 337591-ExTrA. XB acknowledge the support of the French Agence Nationale de la Recherche (ANR), under the program ANR-12-BS05-0012 Exo-atmos. RFD acknowledge funding from the European Union Seventh Framework Programme (FP7/2007-2013) under Grant agreement no. 313014 (ETAEARTH), and the financial support of the SNSF in the frame of the National Centre for Competence in Research `PlanetS'. We thank S. Udry for discussions on the dynamics of this system, L. Kreidberg for her Mandel \& Agol code, and T. Fenouillet for his assistance with the LAM cluster. 

\bibliographystyle{mn2e}
\bibliography{k117}

\newpage
\appendix

\section[]{Other figures}


\begin{figure*}
\includegraphics[width=17cm]{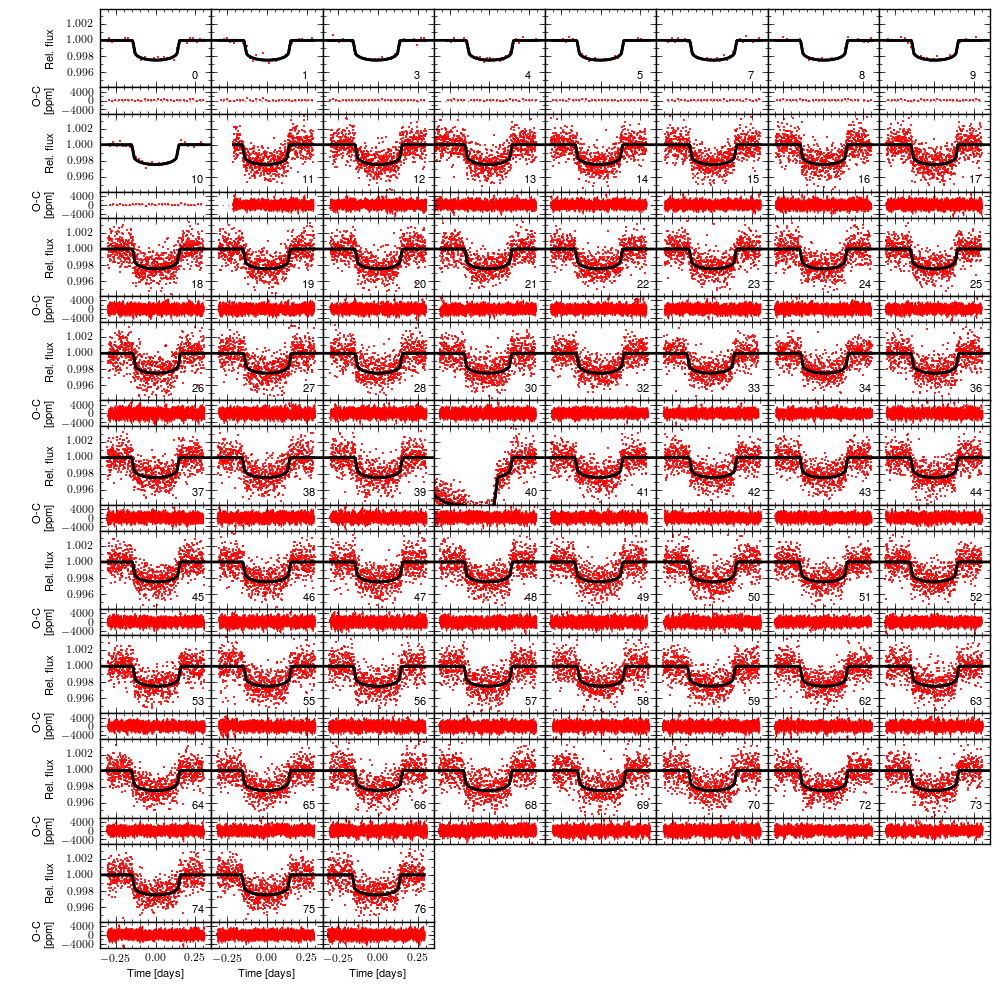} 
\caption{Transits of Kepler-117b observed by \Kepler. Each
  panel shows the photometric time series of the transits observed
  by \Kepler\ and modelled in this work. In the lower part of each panel
  the residuals after subtracting the model to the observed data are
  shown. Each panel is labelled with the transit epoch. Each transit
  is centred relative to a linear ephemeris.}
\label{transitb}
\end{figure*}

\begin{figure*}
\includegraphics[width=17cm]{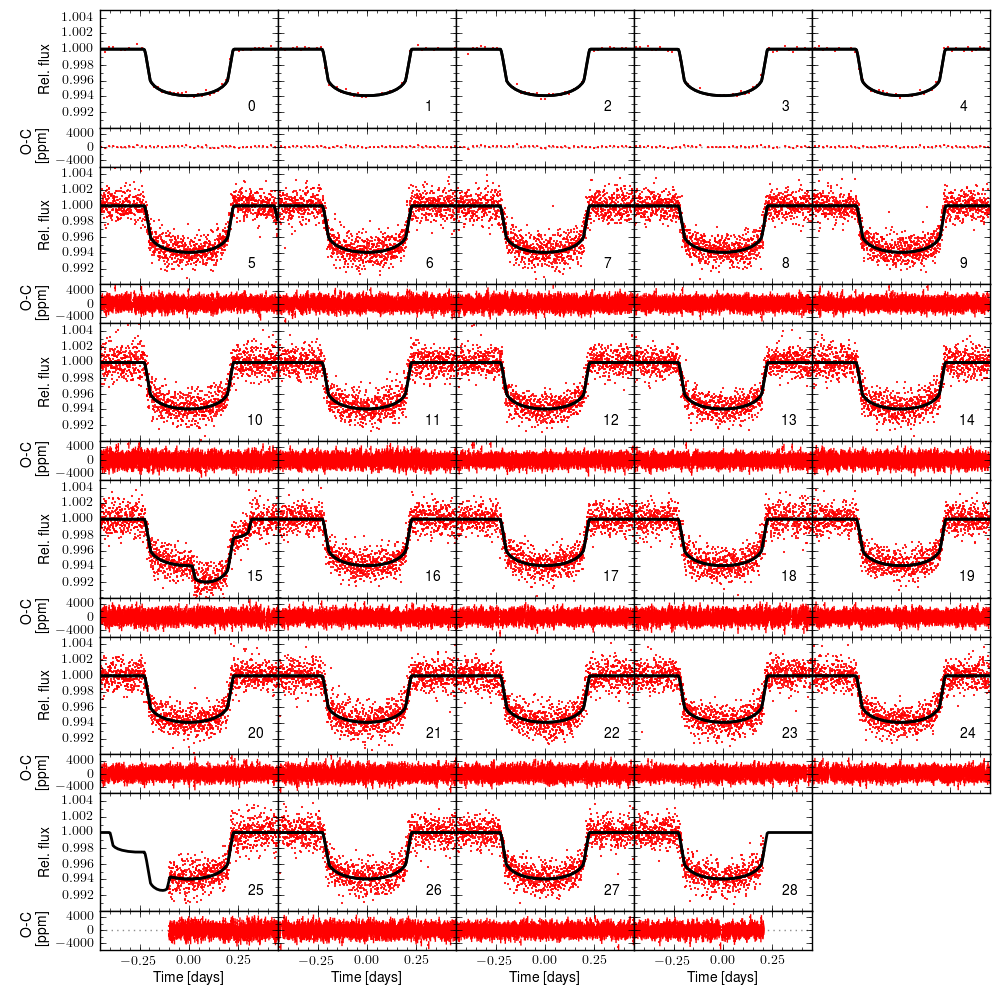} 
\caption{Idem Fig.~\ref{transitb} but for planet~c. }
\label{transitc}
\end{figure*}

\begin{figure*}
  \hspace{-2.3cm}\includegraphics[width=20cm]{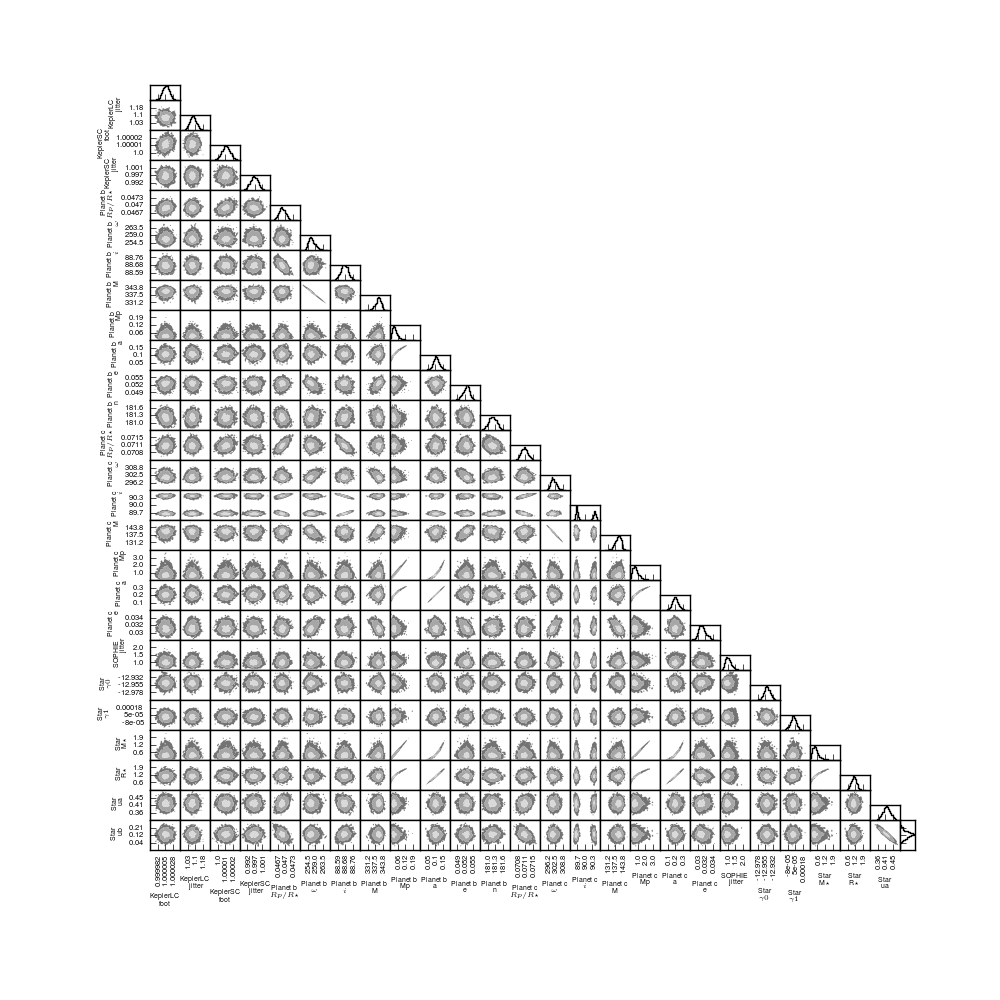} 
  \vspace{-2cm}
  \caption{Two-parameter joint posterior distributions of the MCMC model parameters. The 39.3, 86.5, and 98.9 per cent two-variable joint confidence regions (in the case of a Gaussian posterior, these regions project on to the one-dimensional 1, 2, and 3 $\sigma$ intervals) are denoted by three different grey levels. The histogram of each parameter is shown at the top of each column, except for the parameter on the last line that is shown at the end of the line. Units are the same as in Table~\ref{table}.}
  \label{pyramid}
\end{figure*}


\begin{figure*}
\includegraphics[width=8.5cm]{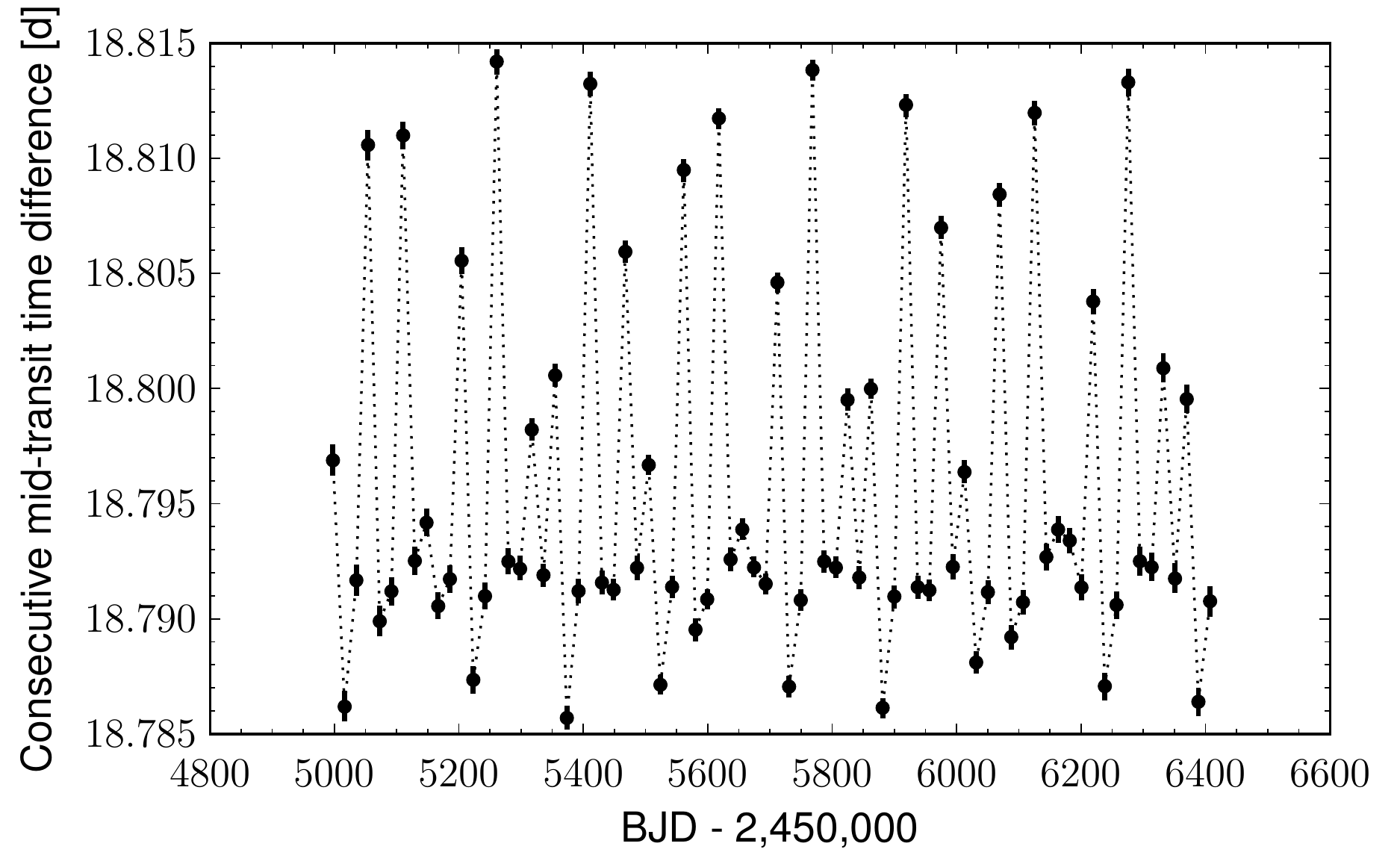}
\includegraphics[width=8.5cm]{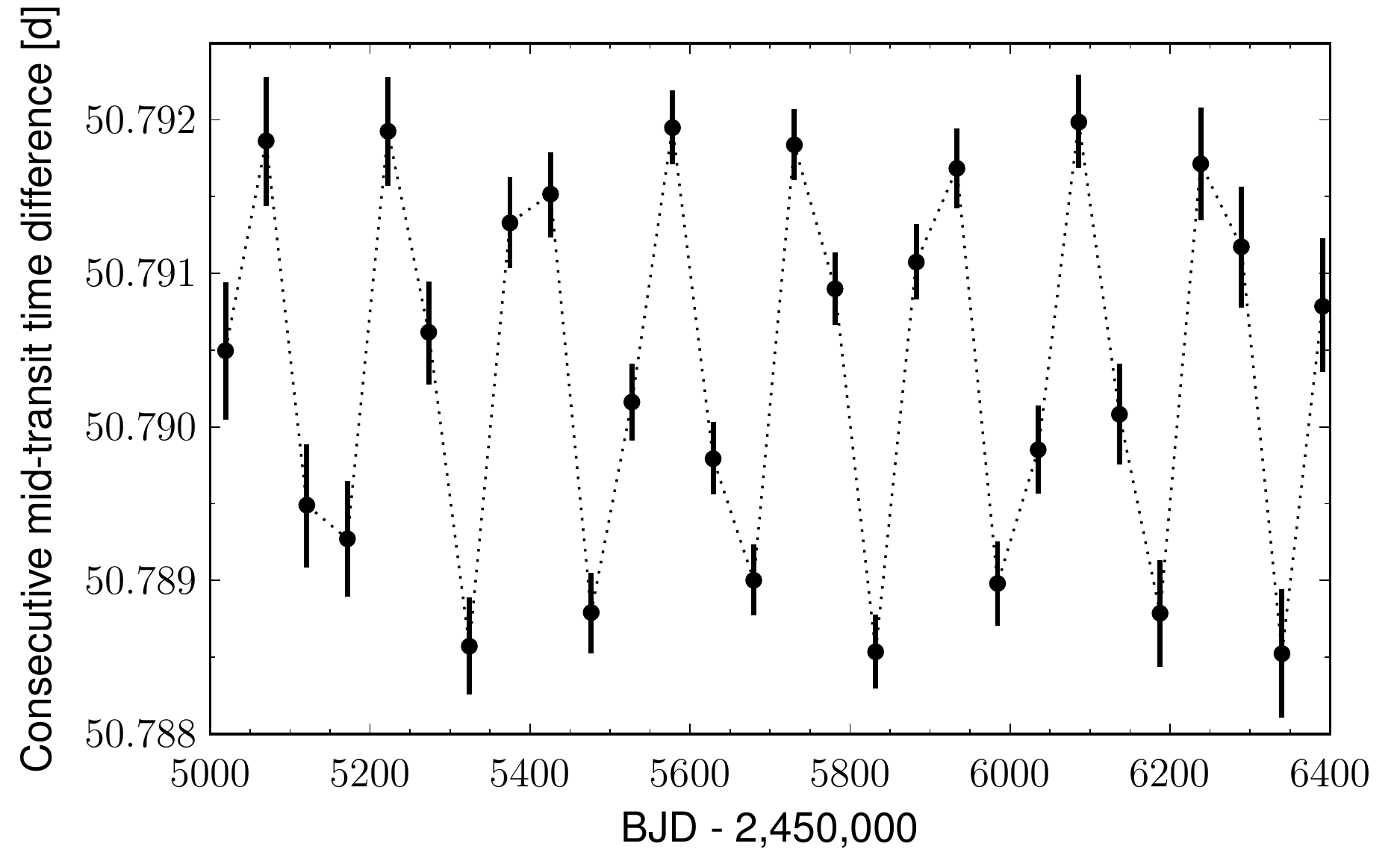}\\
\caption{Time difference between consecutive transits, `period', during \Kepler\ observations for planet~b (left) and planet~c (right). The median value for planet~b is 18.792 275~d, and for planet~c is 50.790 557~d.}
\label{periods}
\end{figure*}

\begin{figure*}
\includegraphics[width=8.5cm]{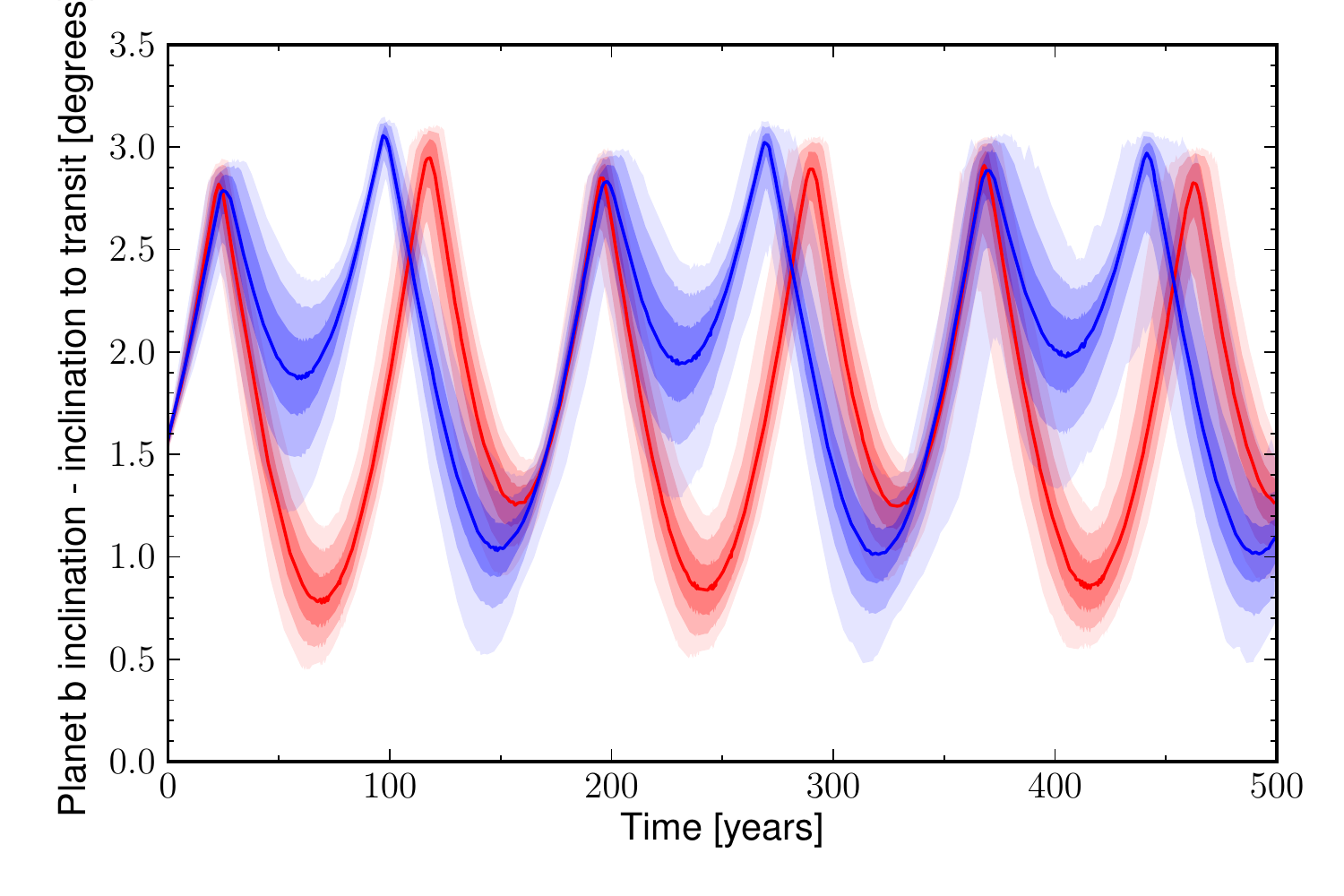} 
\includegraphics[width=8.5cm]{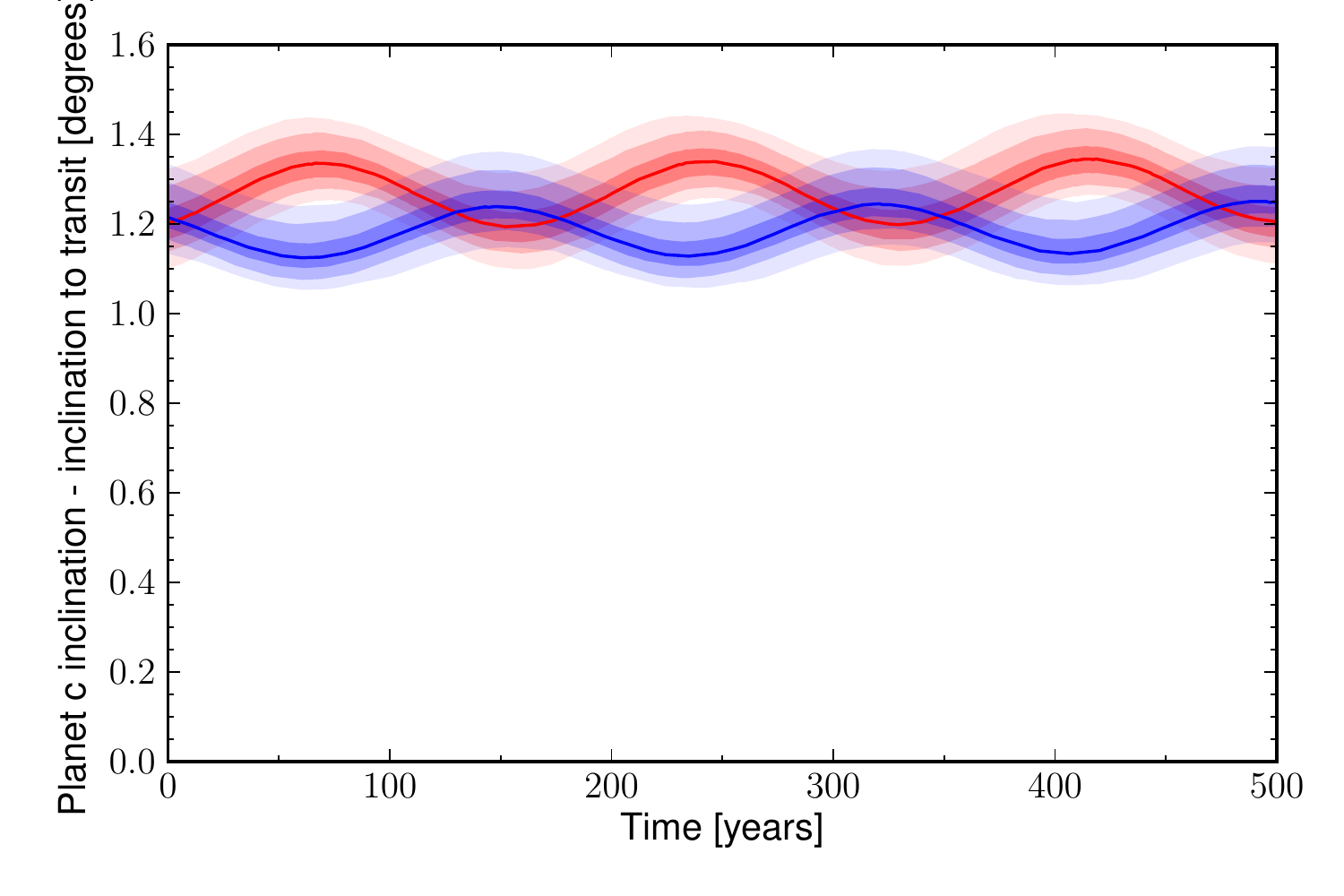} 
\caption{Transit window of planet~b (left), and planet~b (right) for 500 years after \Kepler\ observations. The 68.3, 95.5, and 99.7 per cent Bayesian confidence intervals are plotted in different intensities of red (inclination planet~c $>$ 90), and blue (inclination planet~c $<$ 90). The solid line mark the median of the posterior distribution. $Y$-axis positive means transit. Both planets will transit for the next 500 years.}
\label{TransitWindow}
\end{figure*}


\begin{figure*}
\includegraphics[height=3.2cm]{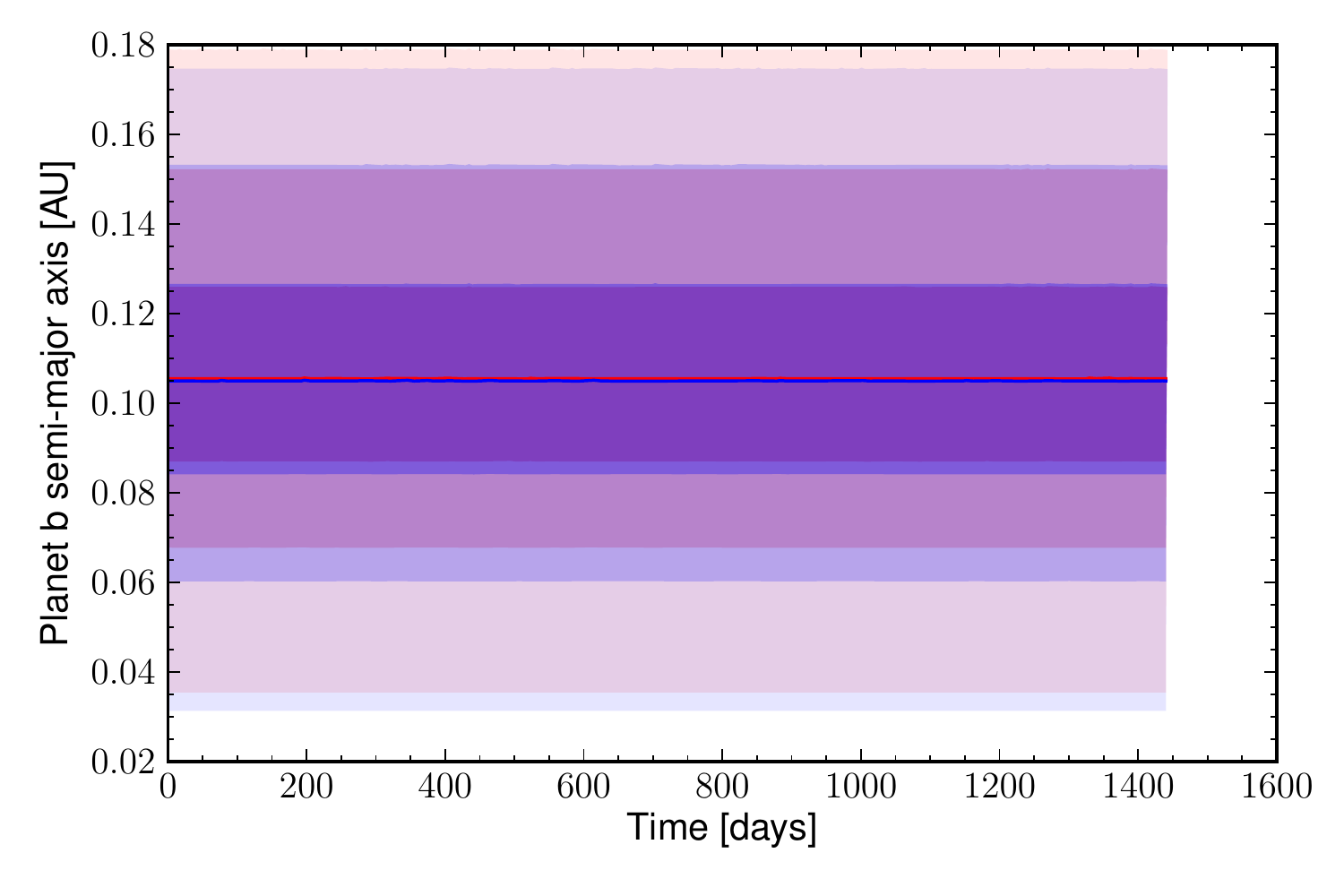}\includegraphics[height=3.2cm]{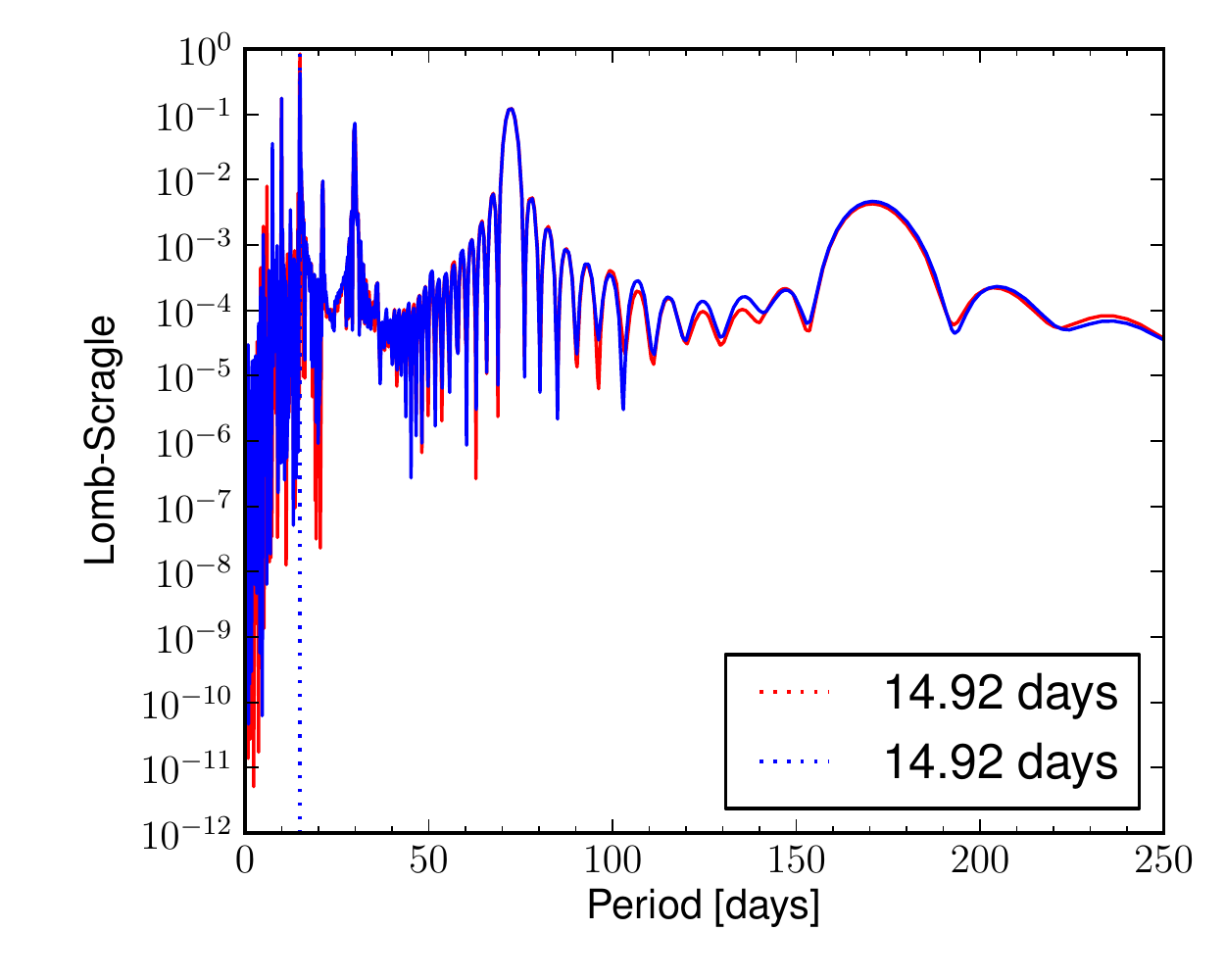} 
\includegraphics[height=3.2cm]{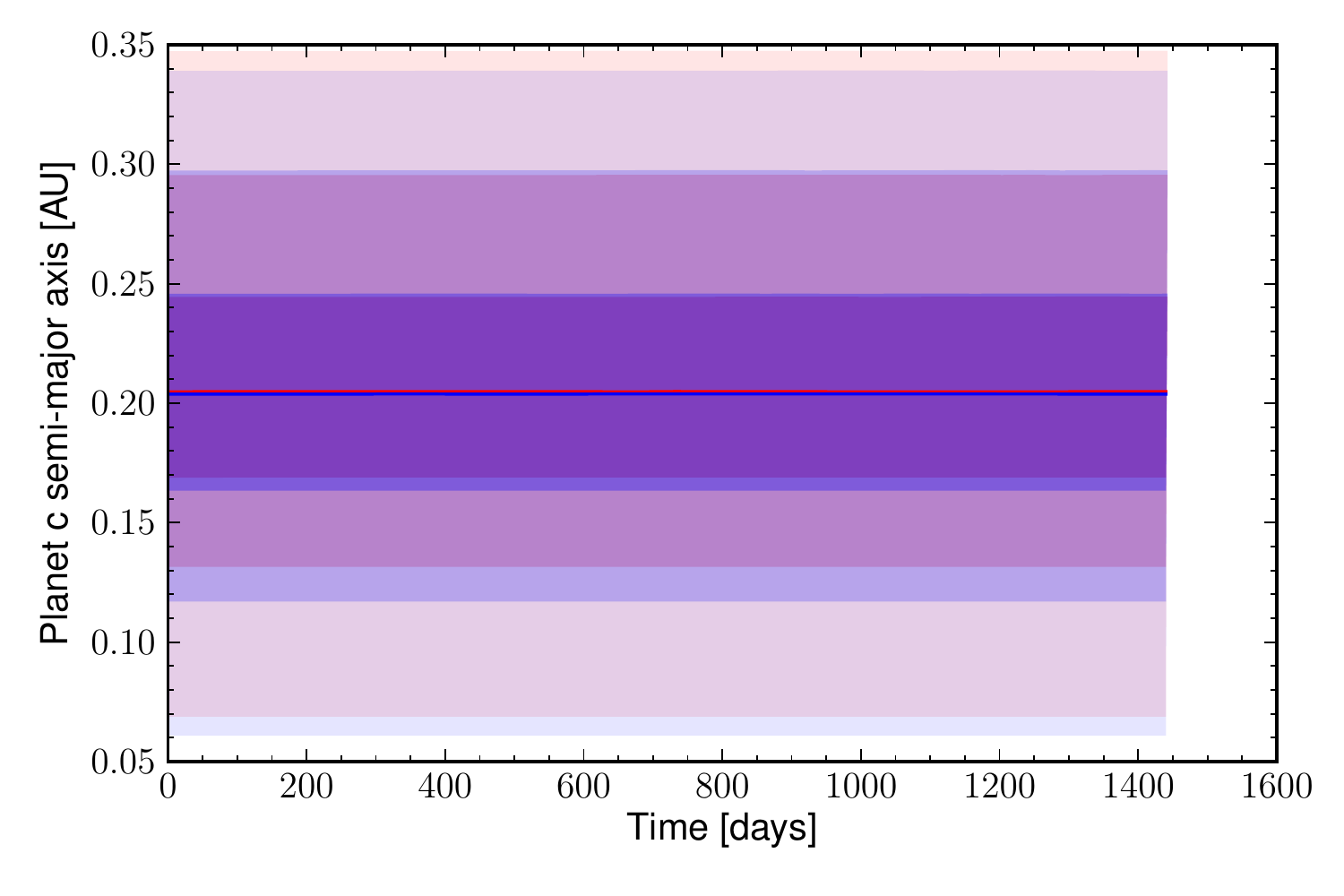}\includegraphics[height=3.2cm]{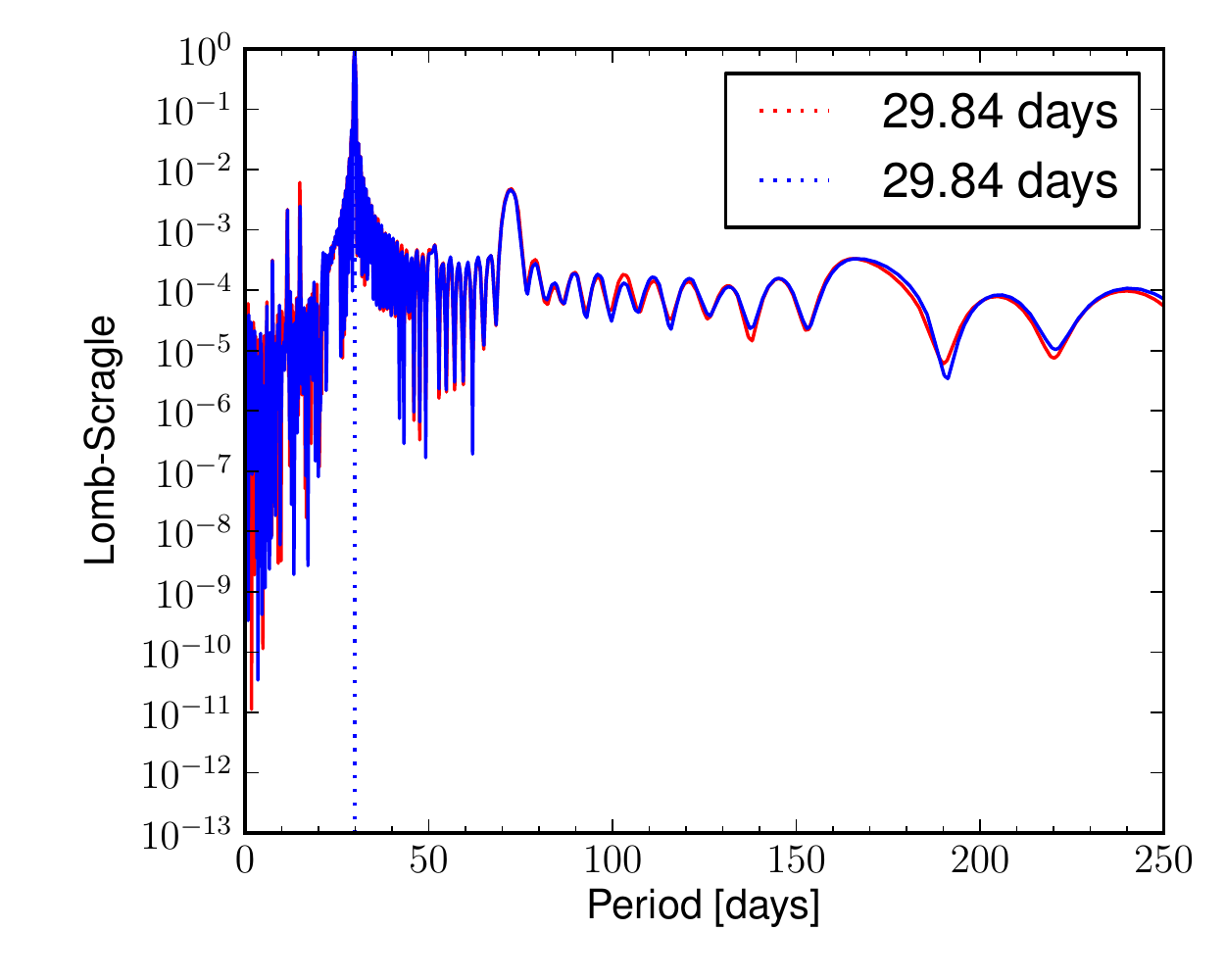}\\ 
\includegraphics[height=3.2cm]{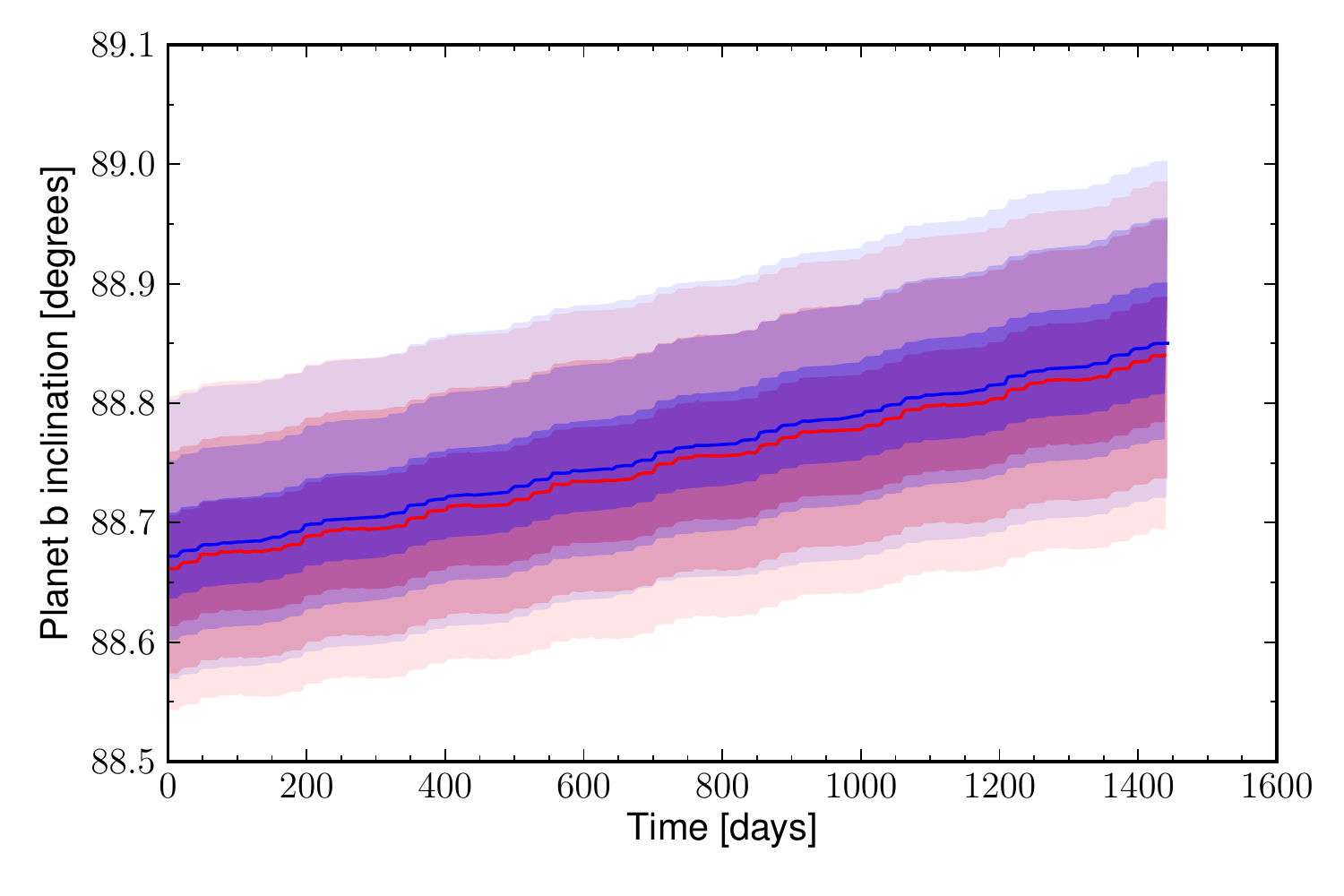}\includegraphics[height=3.2cm]{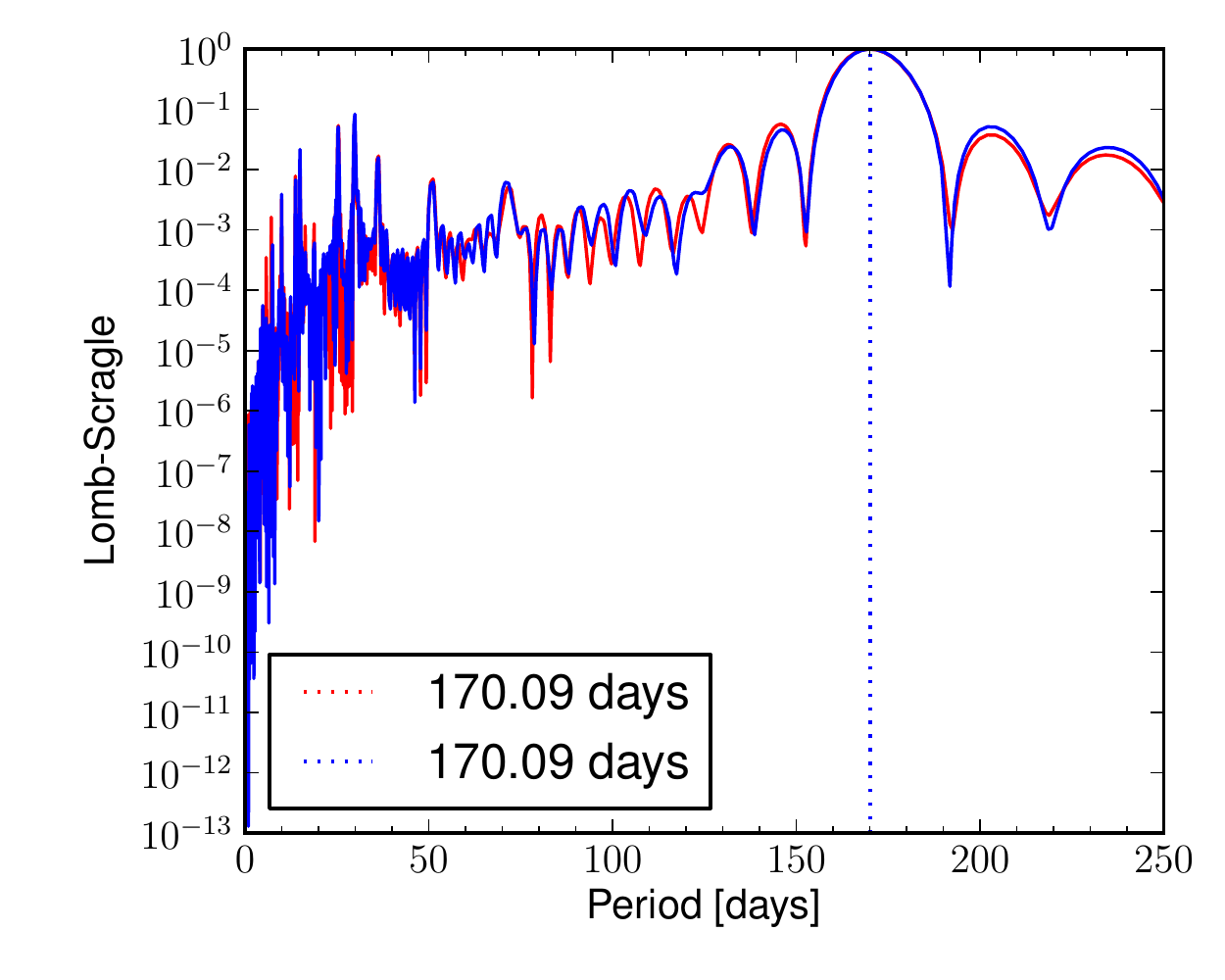}
\includegraphics[height=3.2cm]{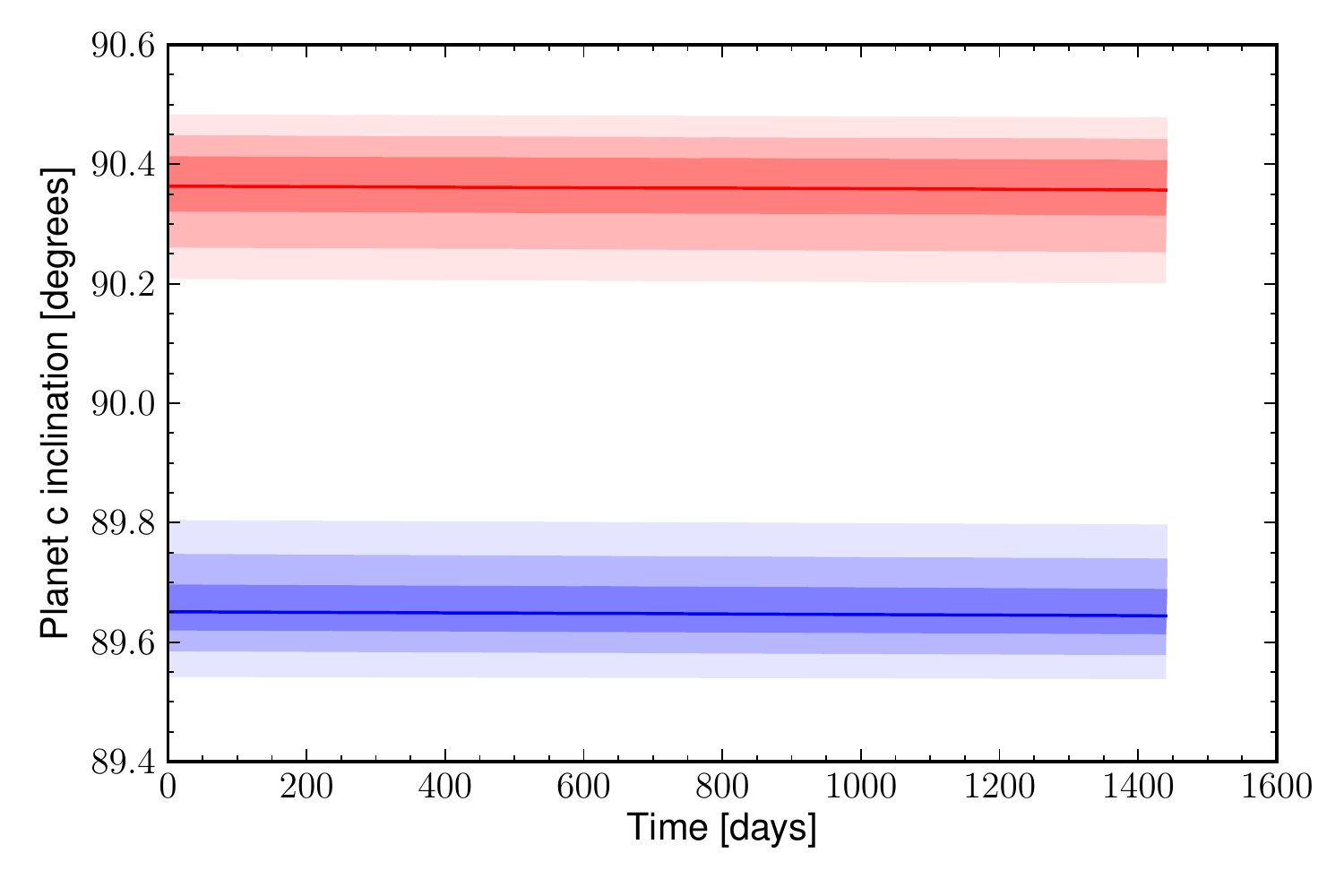}\includegraphics[height=3.2cm]{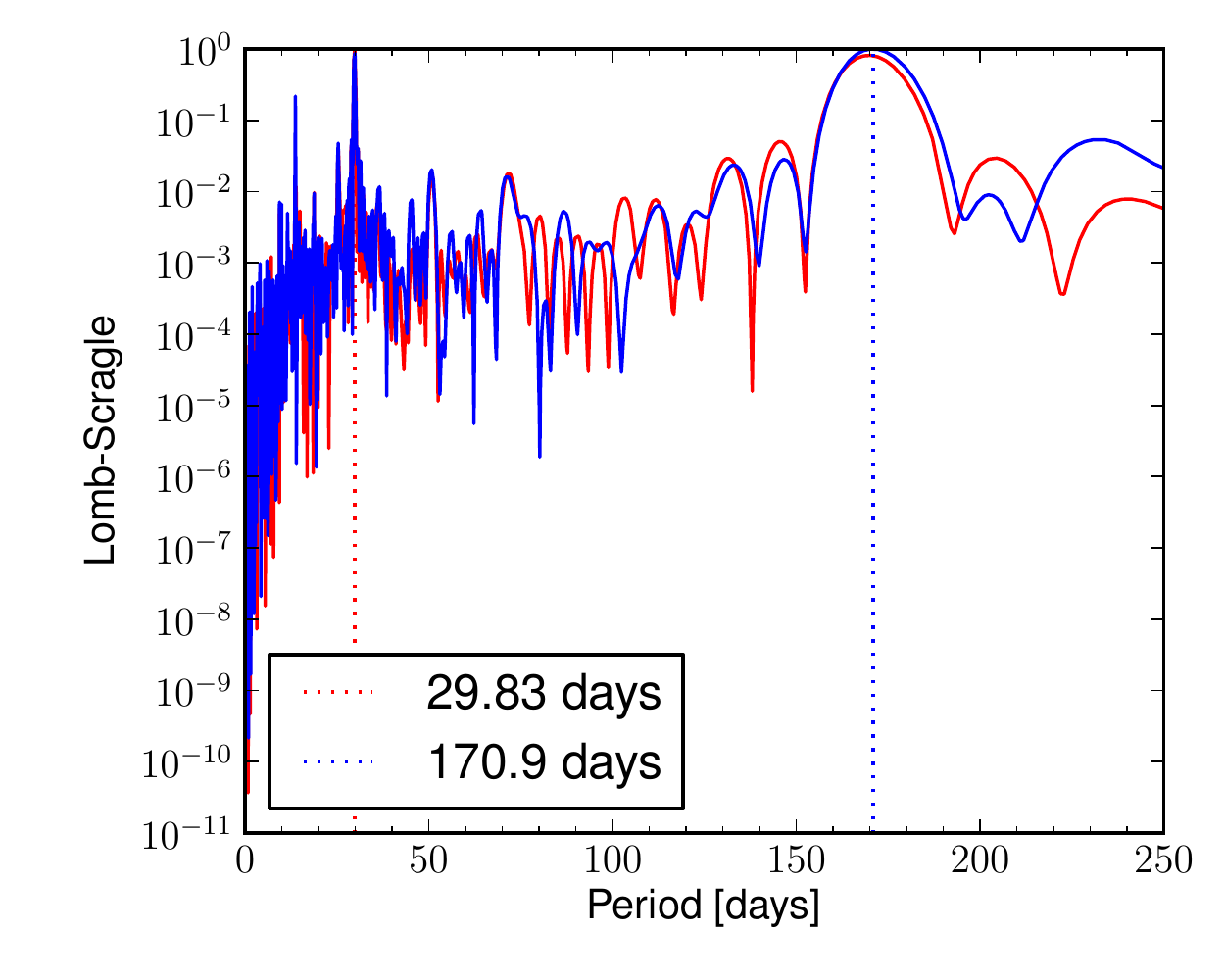}\\
\includegraphics[height=3.2cm]{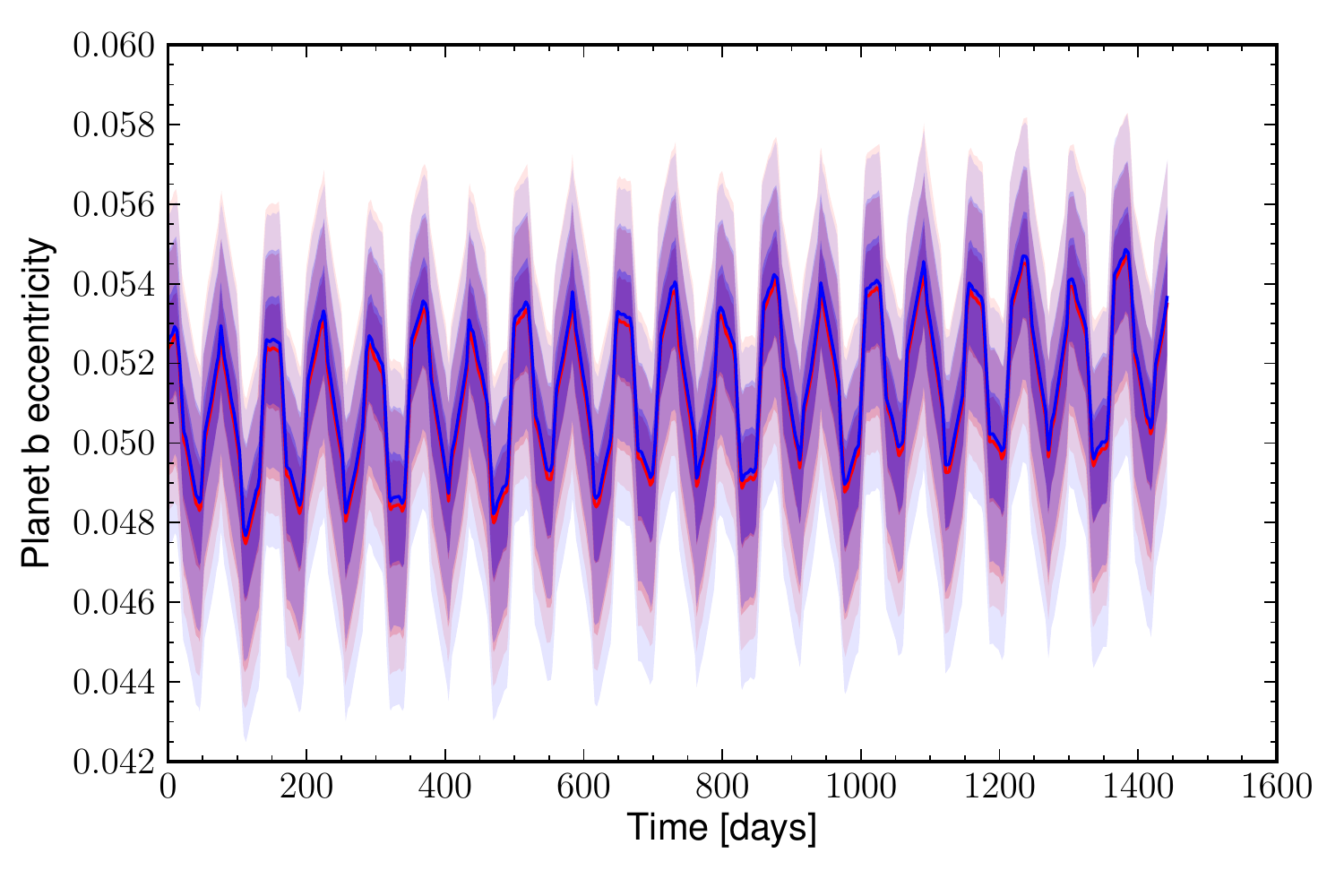}\includegraphics[height=3.2cm]{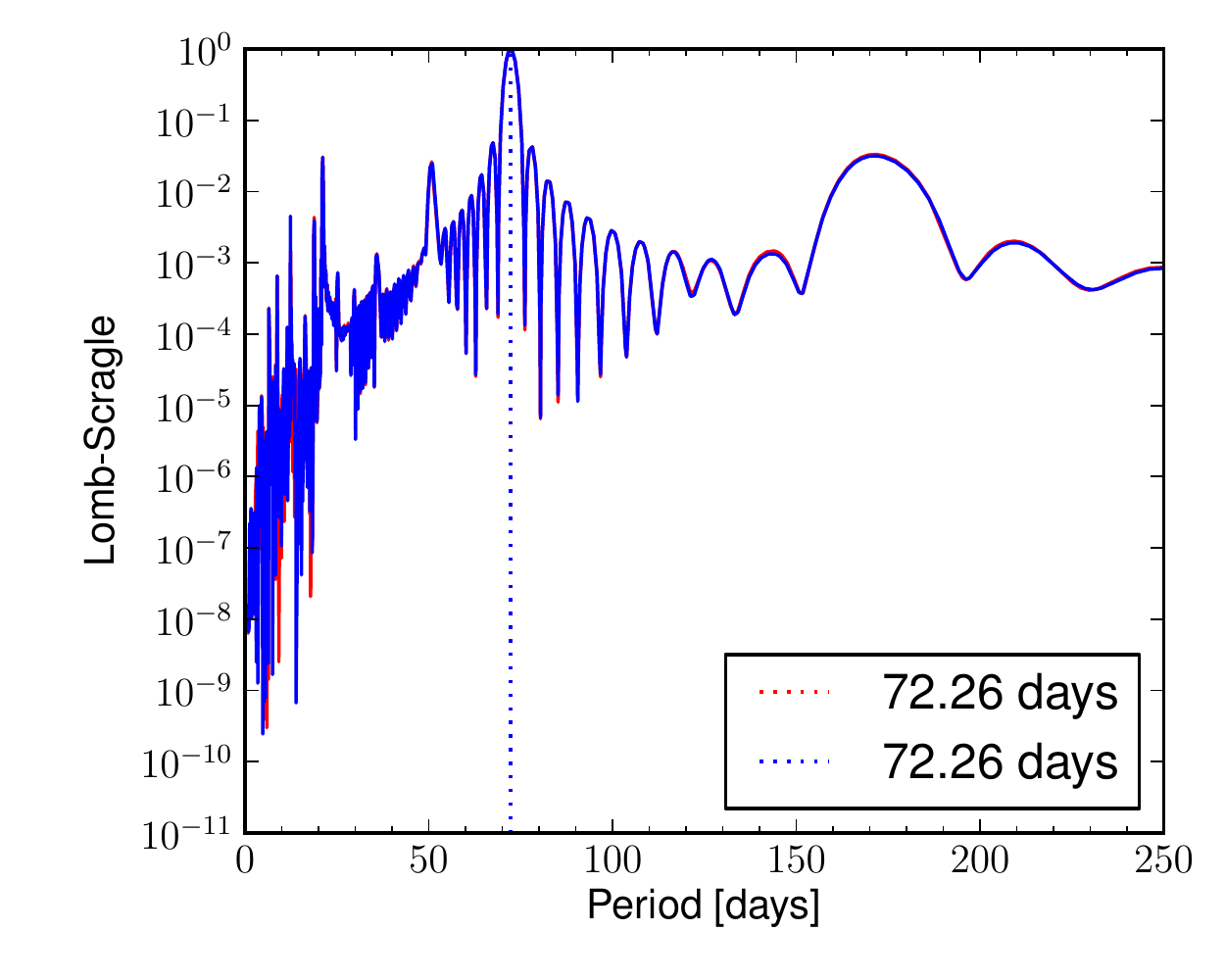}
\includegraphics[height=3.2cm]{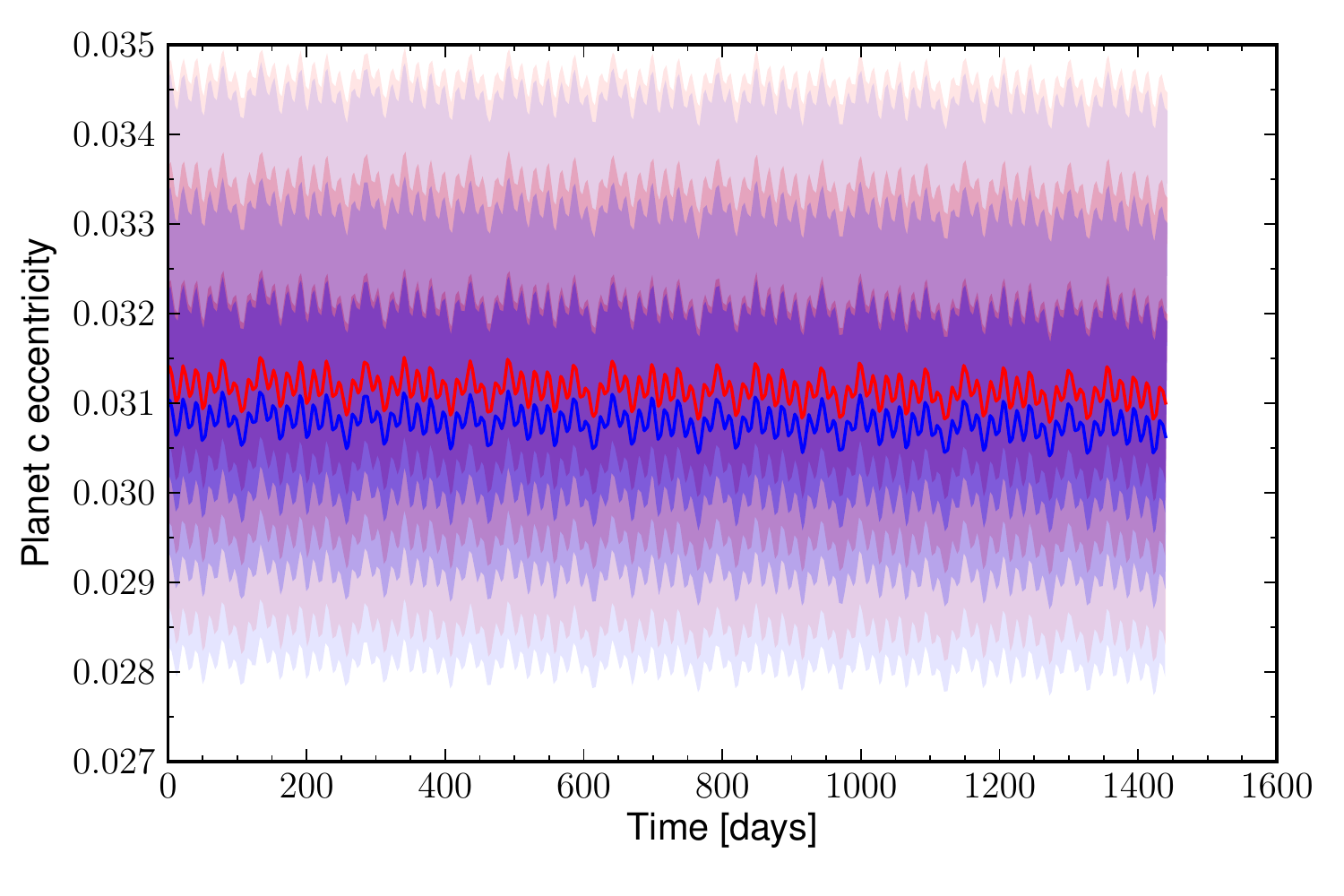}\includegraphics[height=3.2cm]{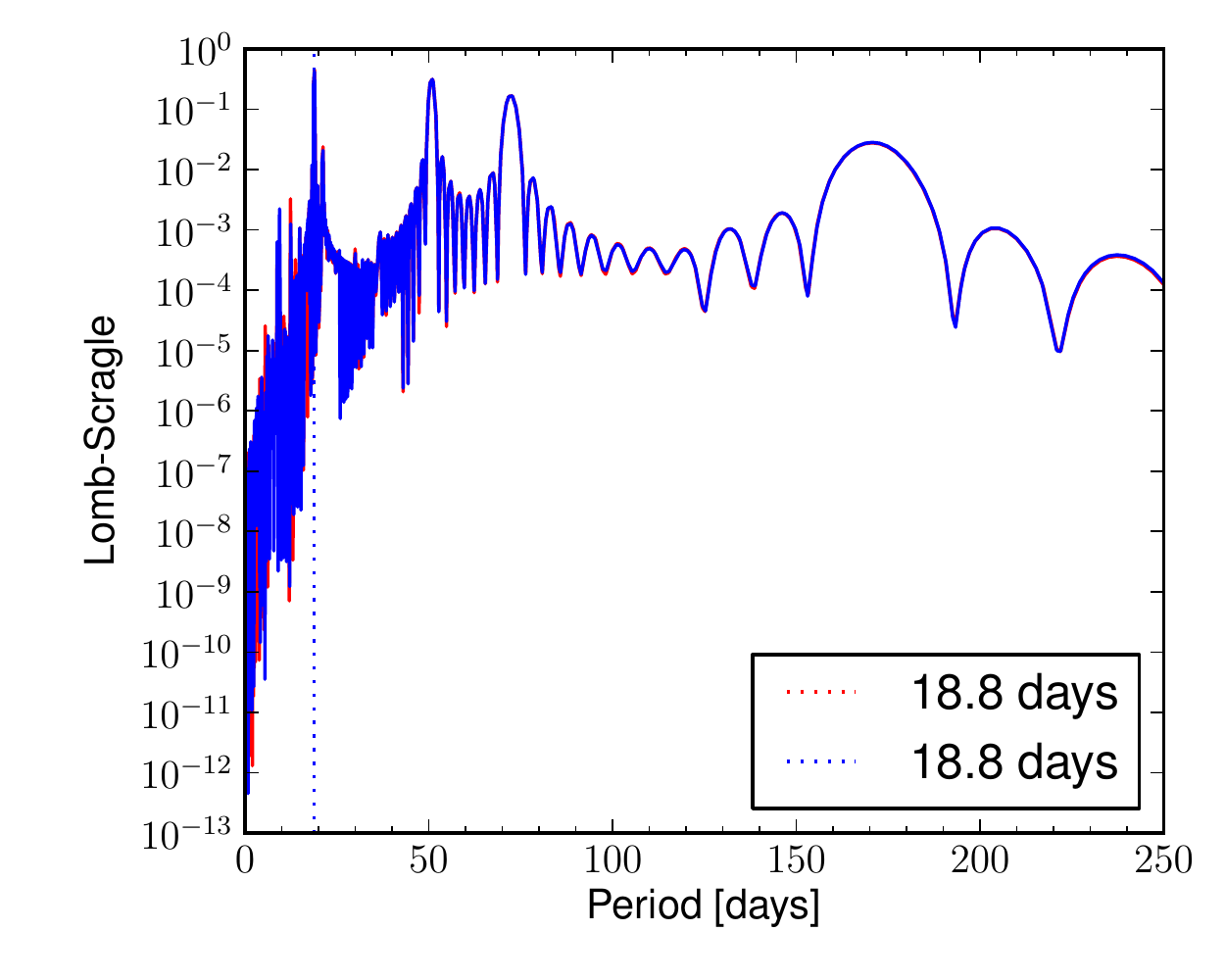}\\
\includegraphics[height=3.2cm]{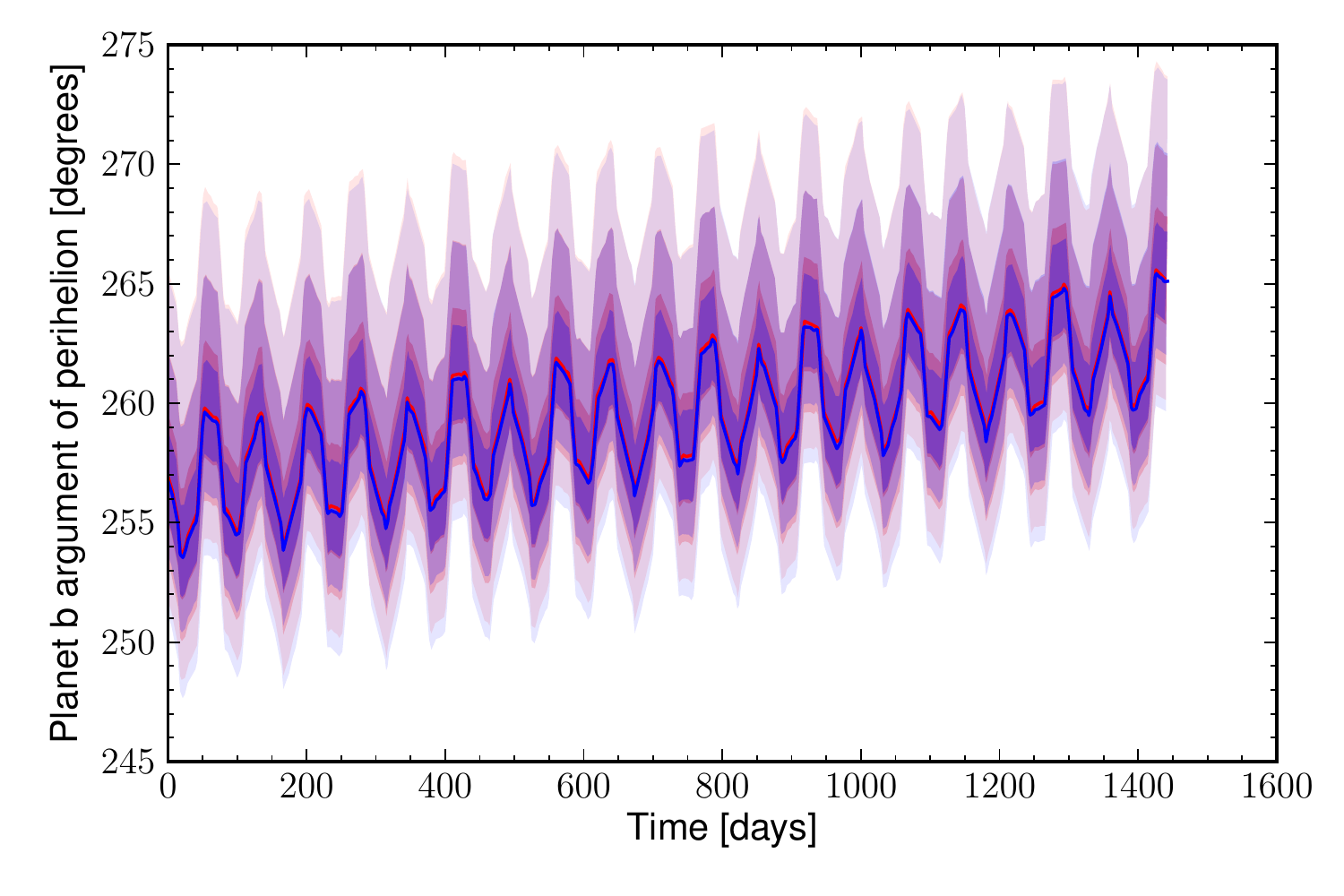}\includegraphics[height=3.2cm]{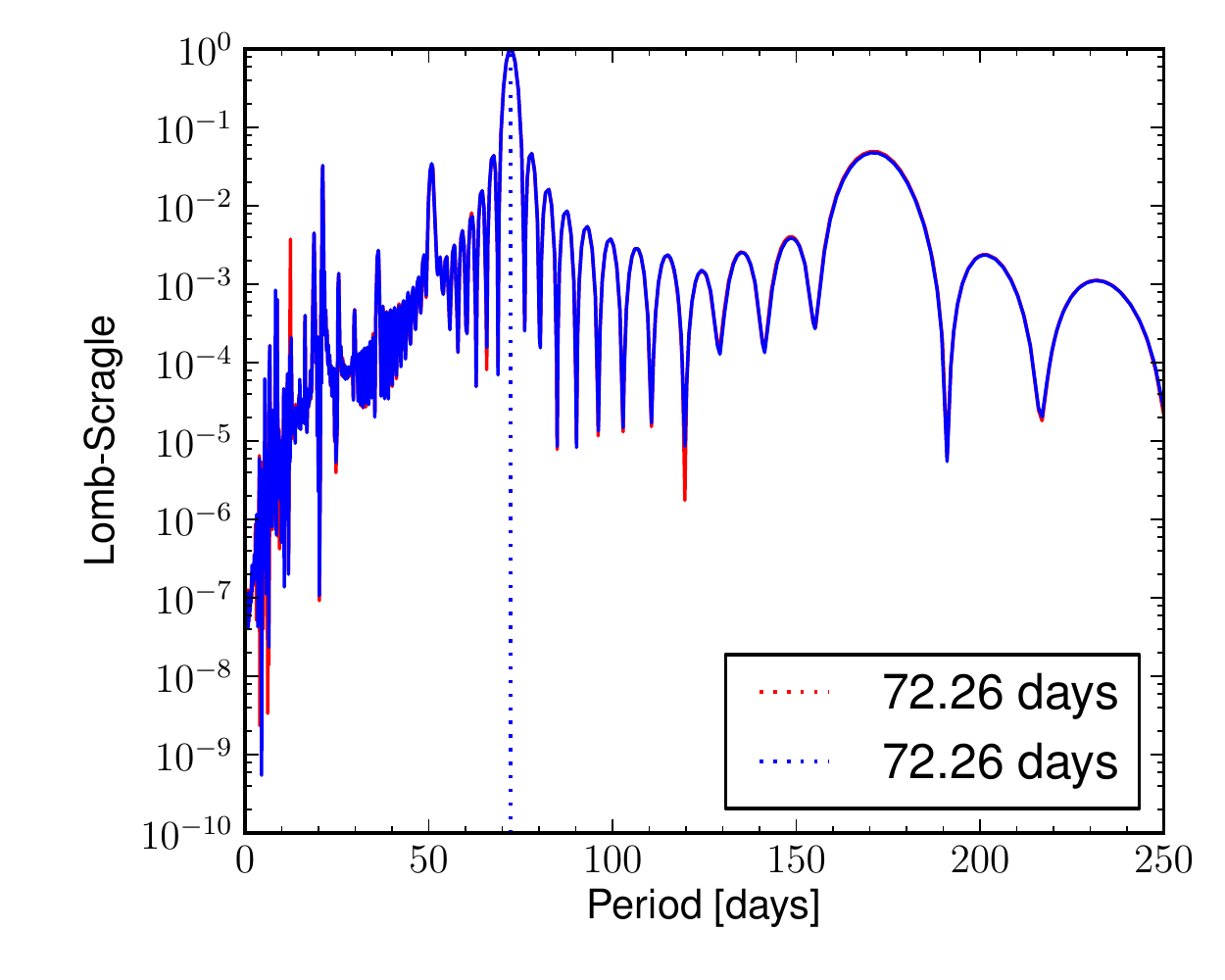}
\includegraphics[height=3.2cm]{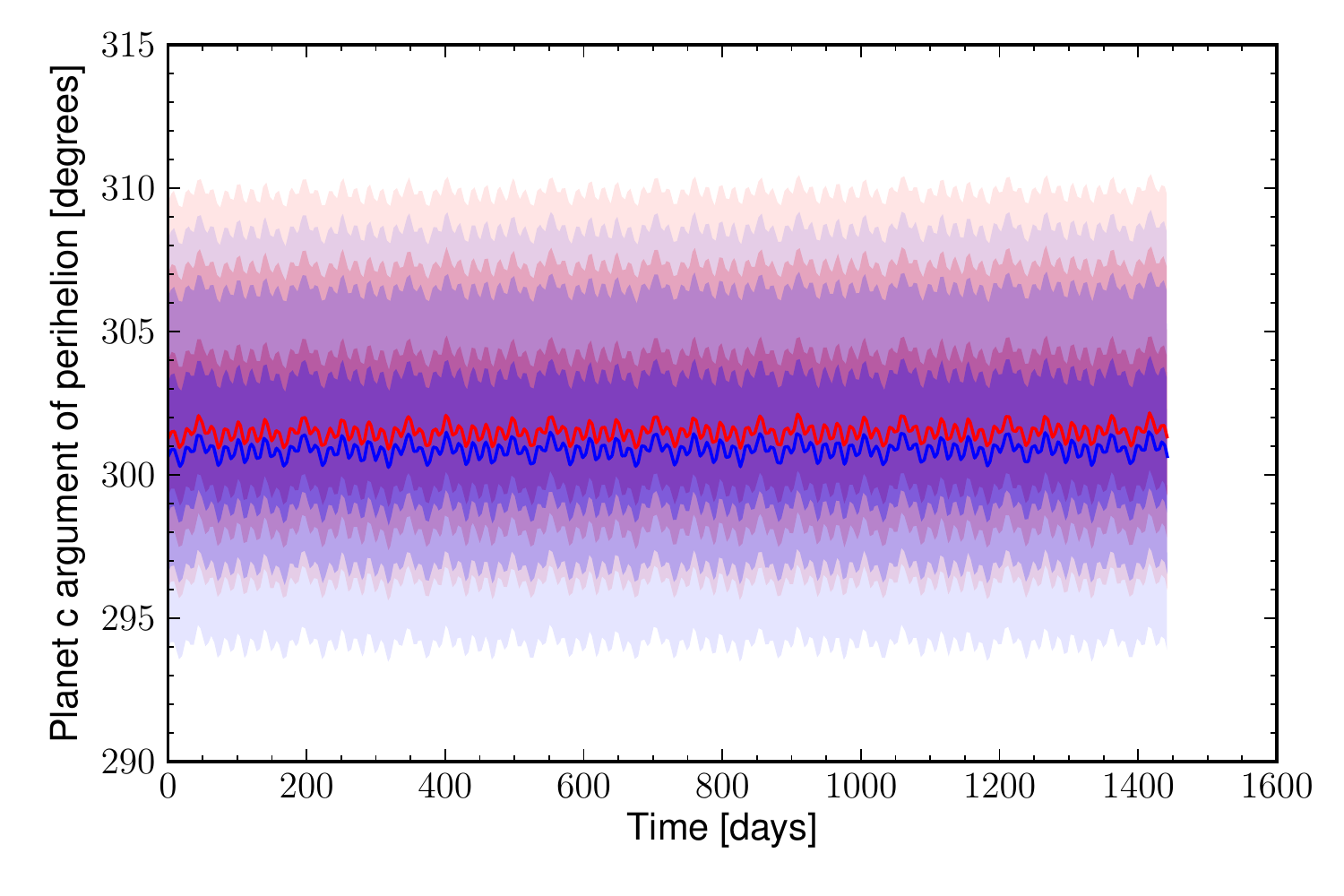}\includegraphics[height=3.2cm]{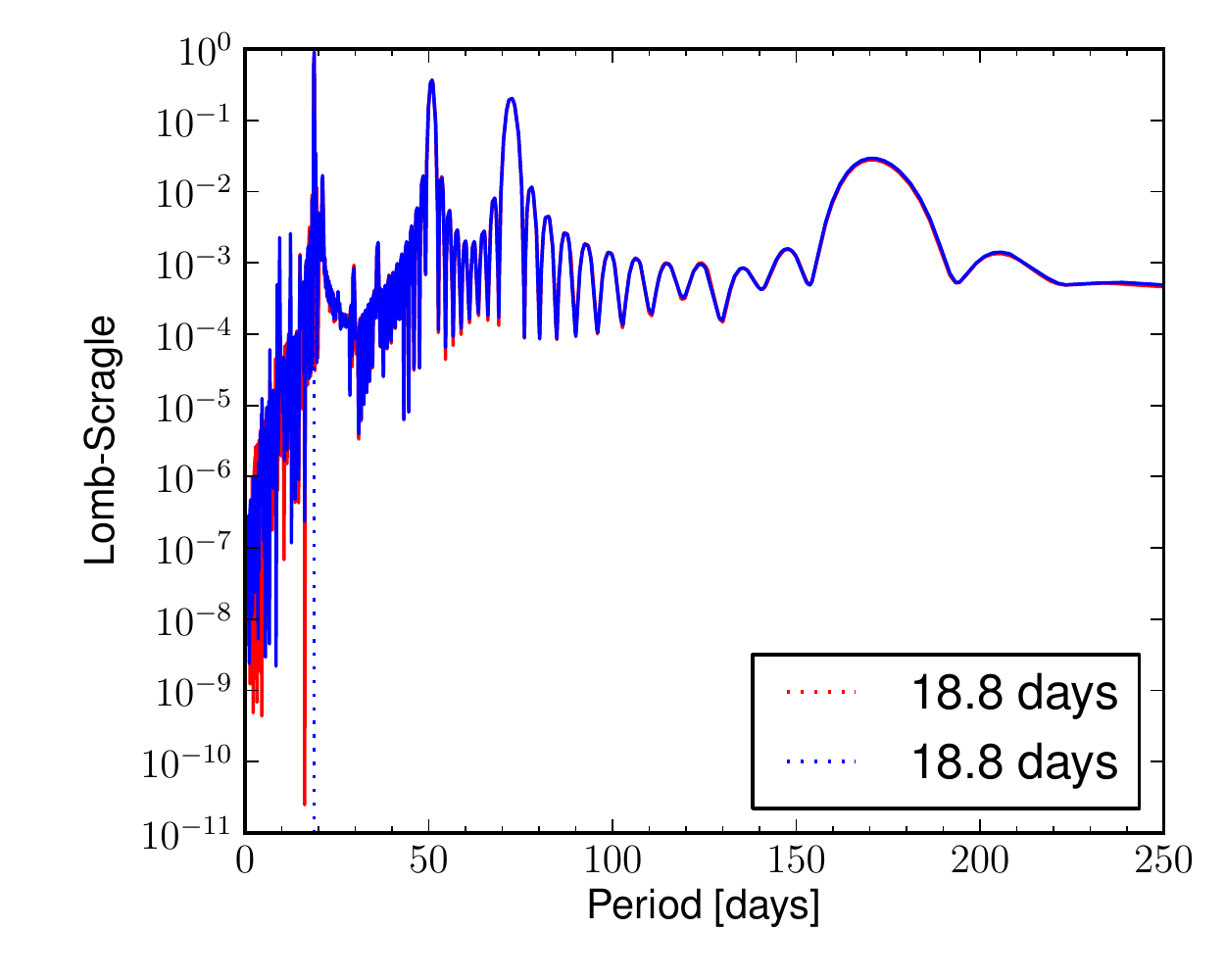}\\
\includegraphics[height=3.2cm]{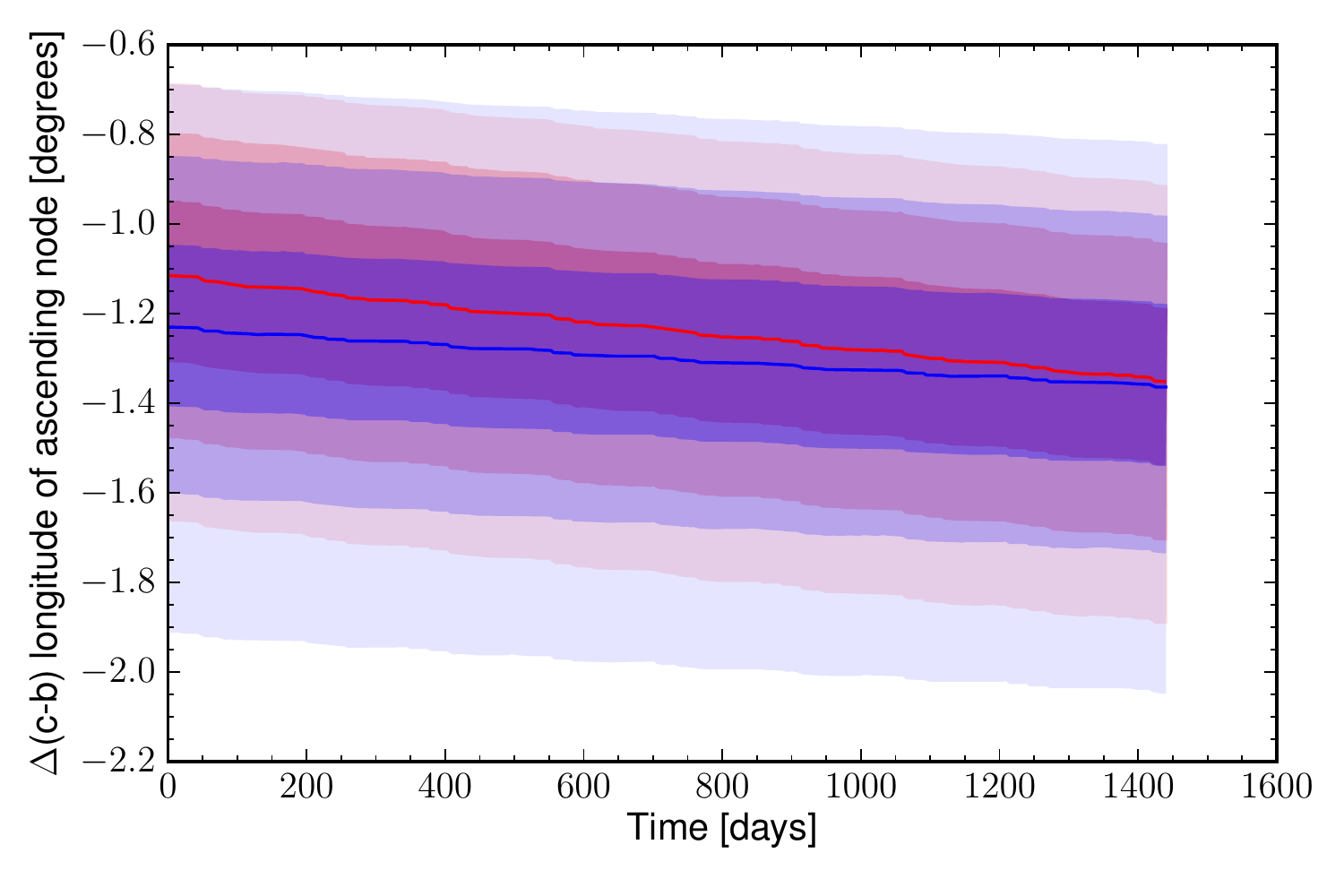}\includegraphics[height=3.2cm]{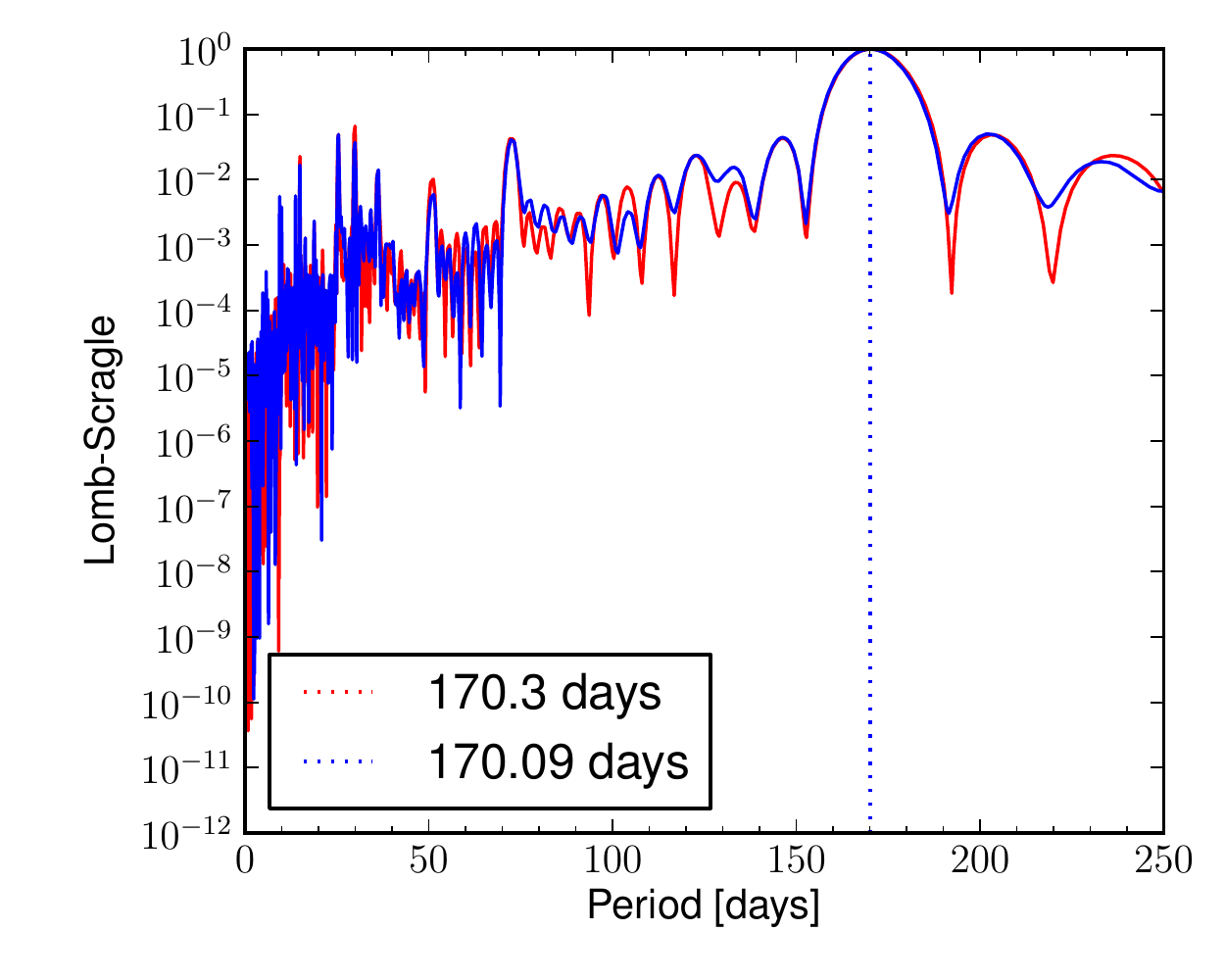}\\
\caption{Evolution of the planets orbital parameters during \Kepler\ observations from the photodynamical model fit (central body origin in the \textsc{mercury} code). The 68.3, 95.5, and 99.7 per cent Bayesian confidence intervals are plotted in different intensities of red (inclination planet~c $>$ 90), and blue (inclination planet~c $<$ 90). The solid line mark the median of the posterior distribution. Next to each plot, the Lomb-Scargle periodogram is shown.}
\label{ShortCen}
\end{figure*}


\begin{figure*}
\includegraphics[height=3.2cm]{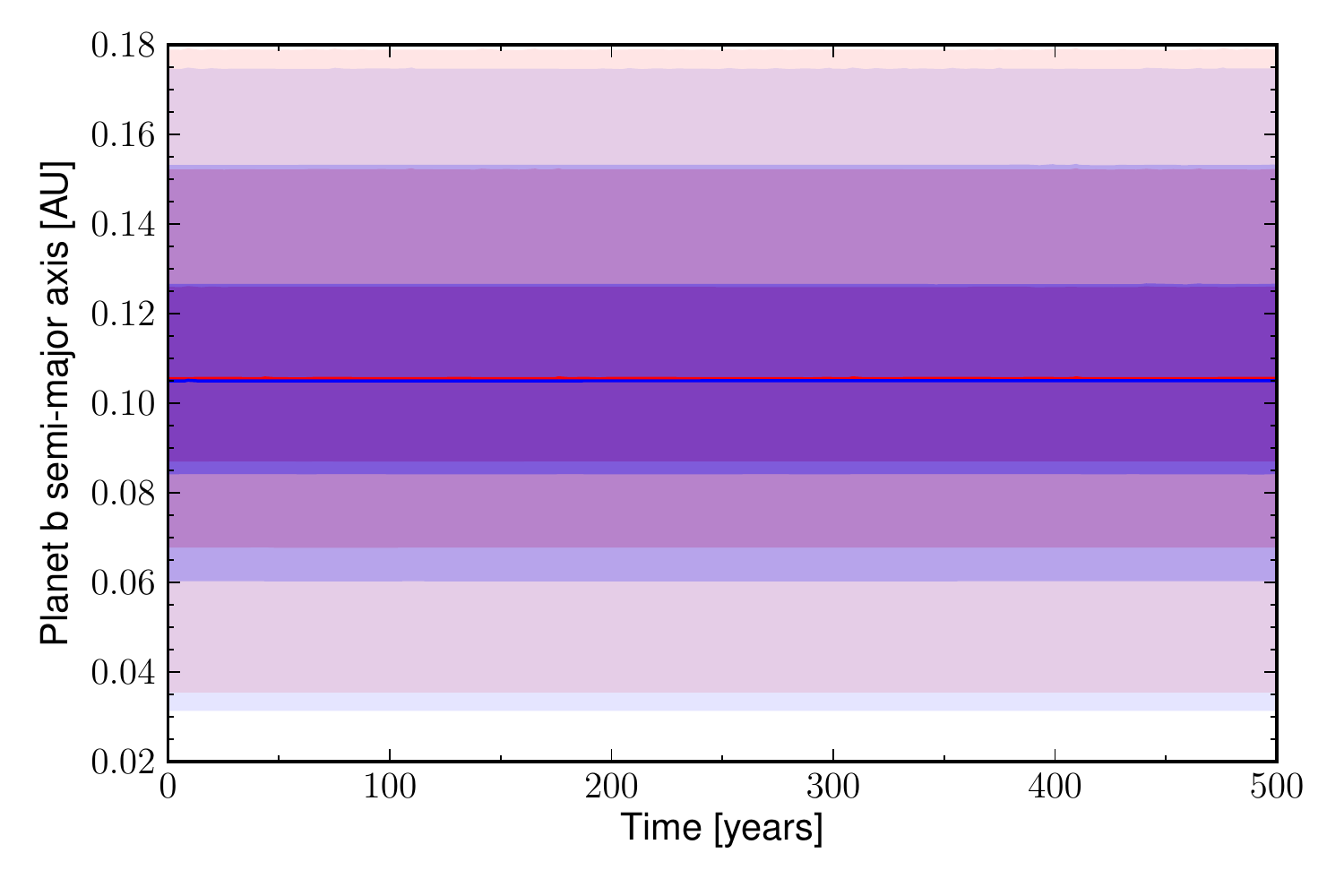}\includegraphics[height=3.2cm]{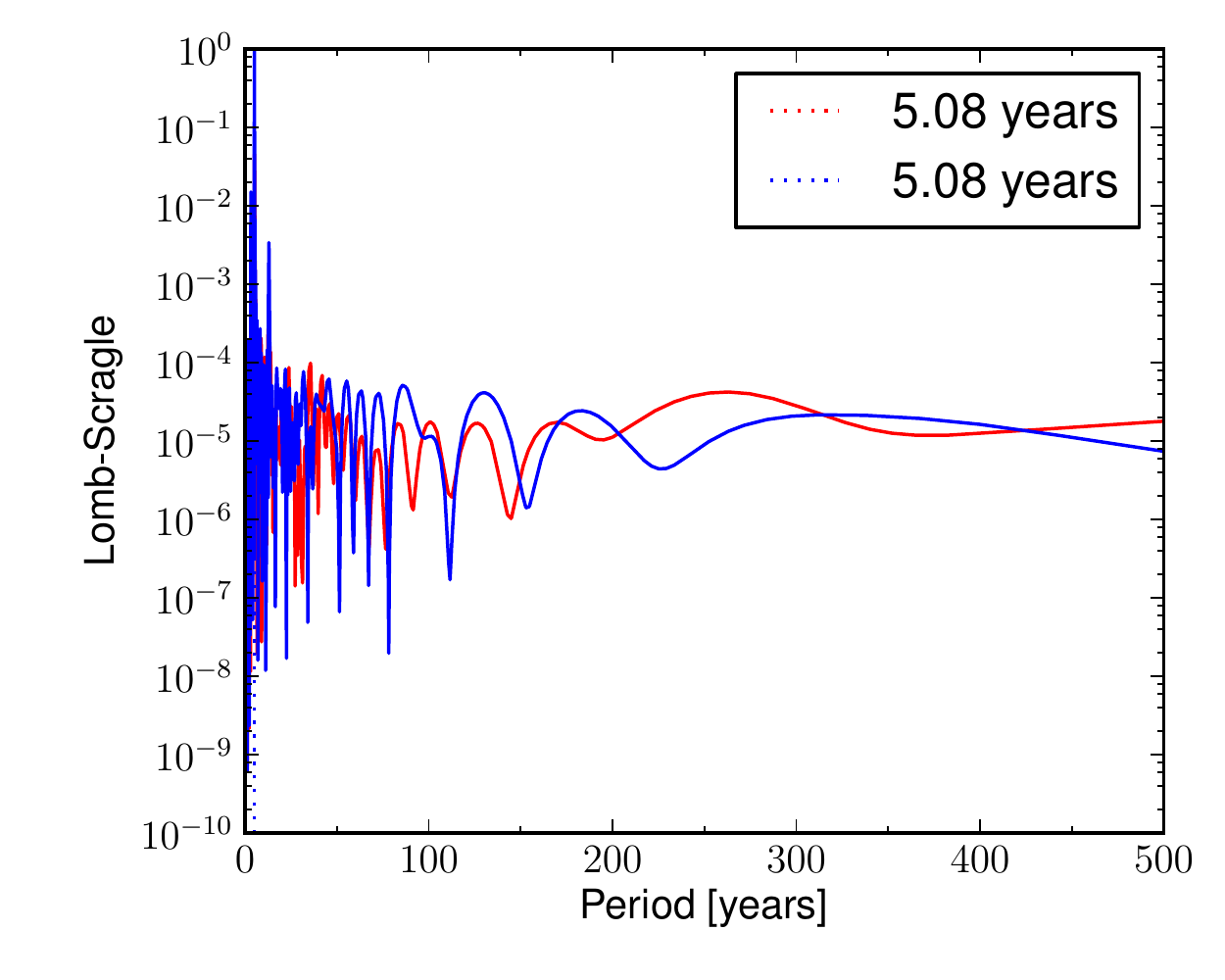} 
\includegraphics[height=3.2cm]{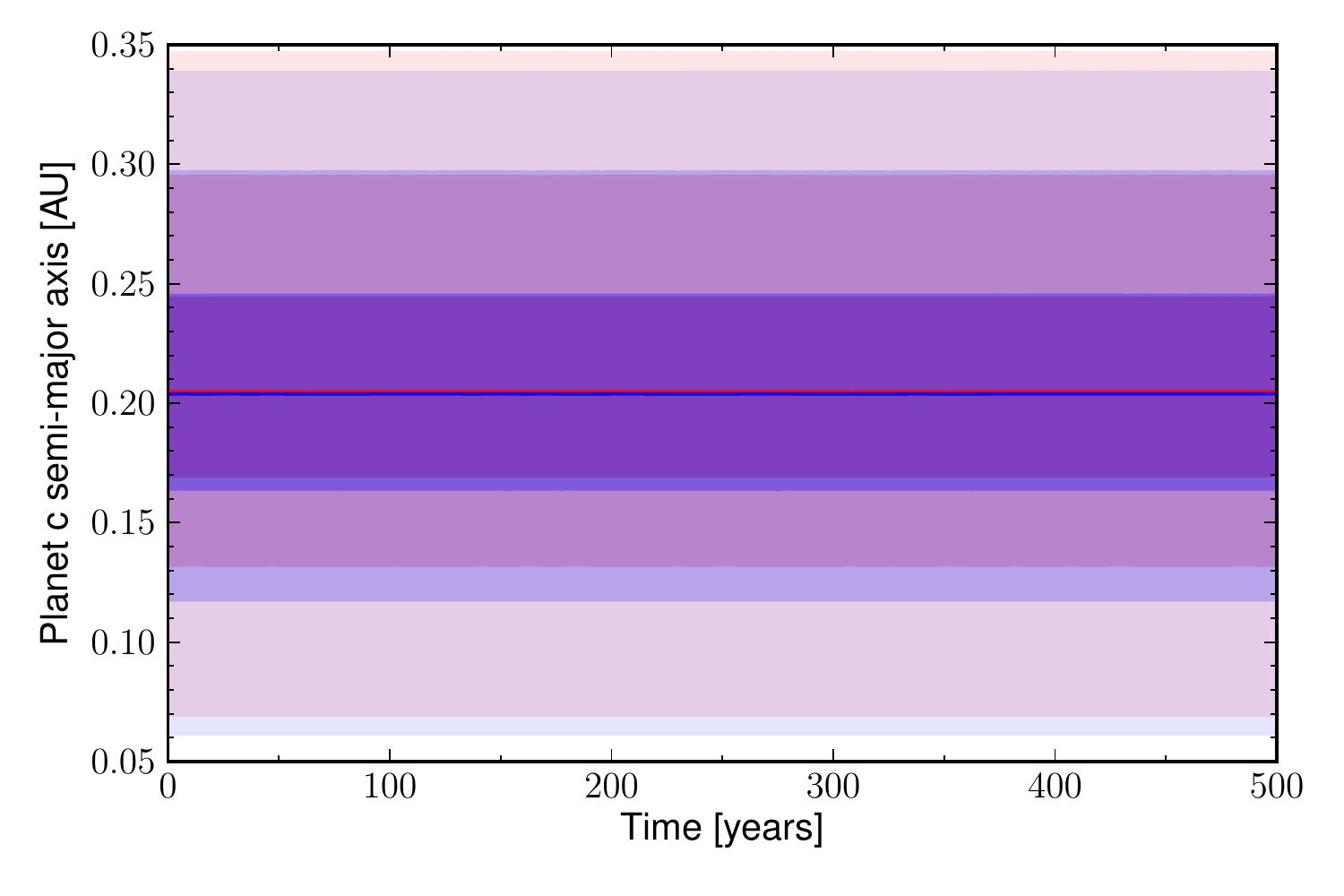}\includegraphics[height=3.2cm]{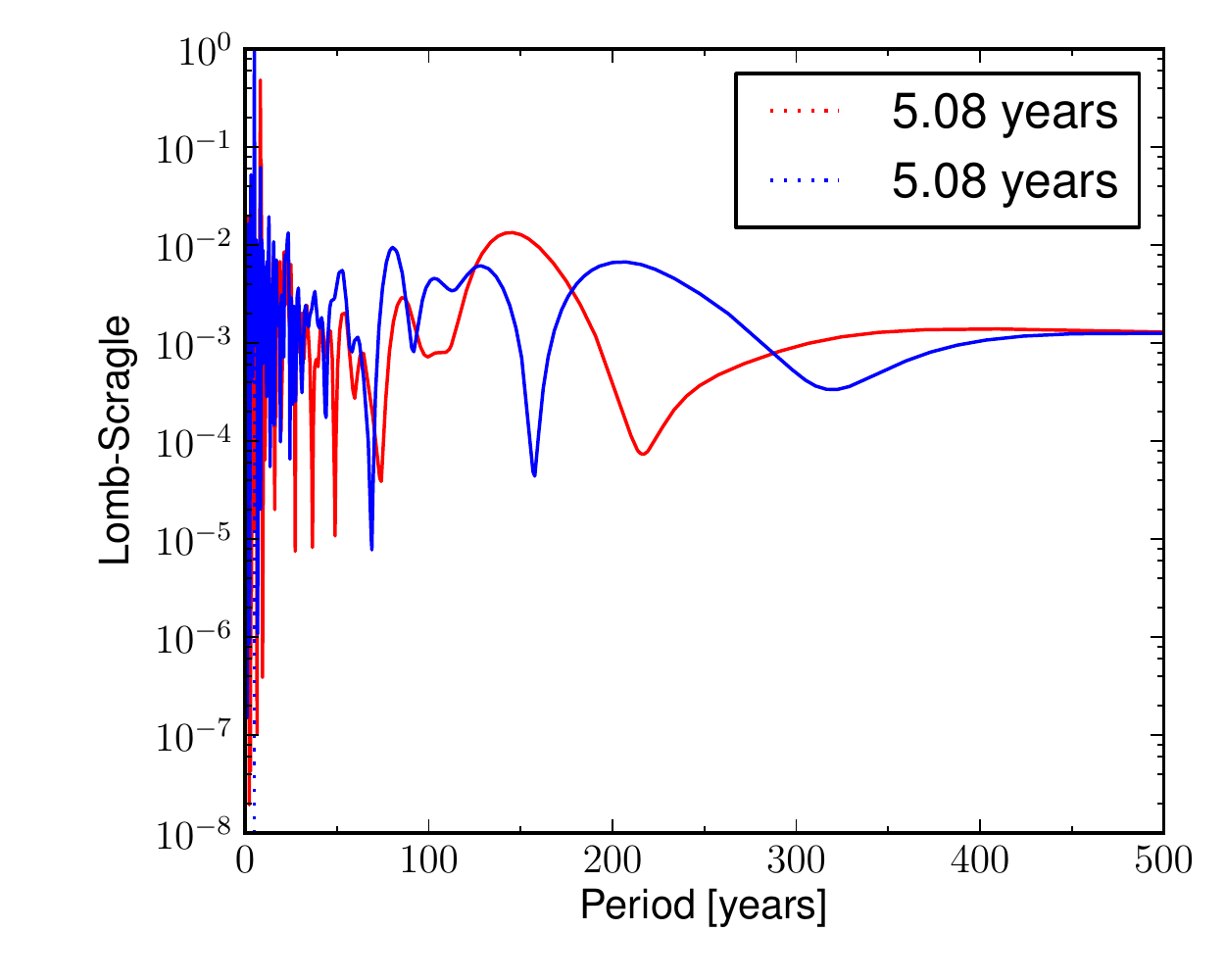}\\ 
\includegraphics[height=3.2cm]{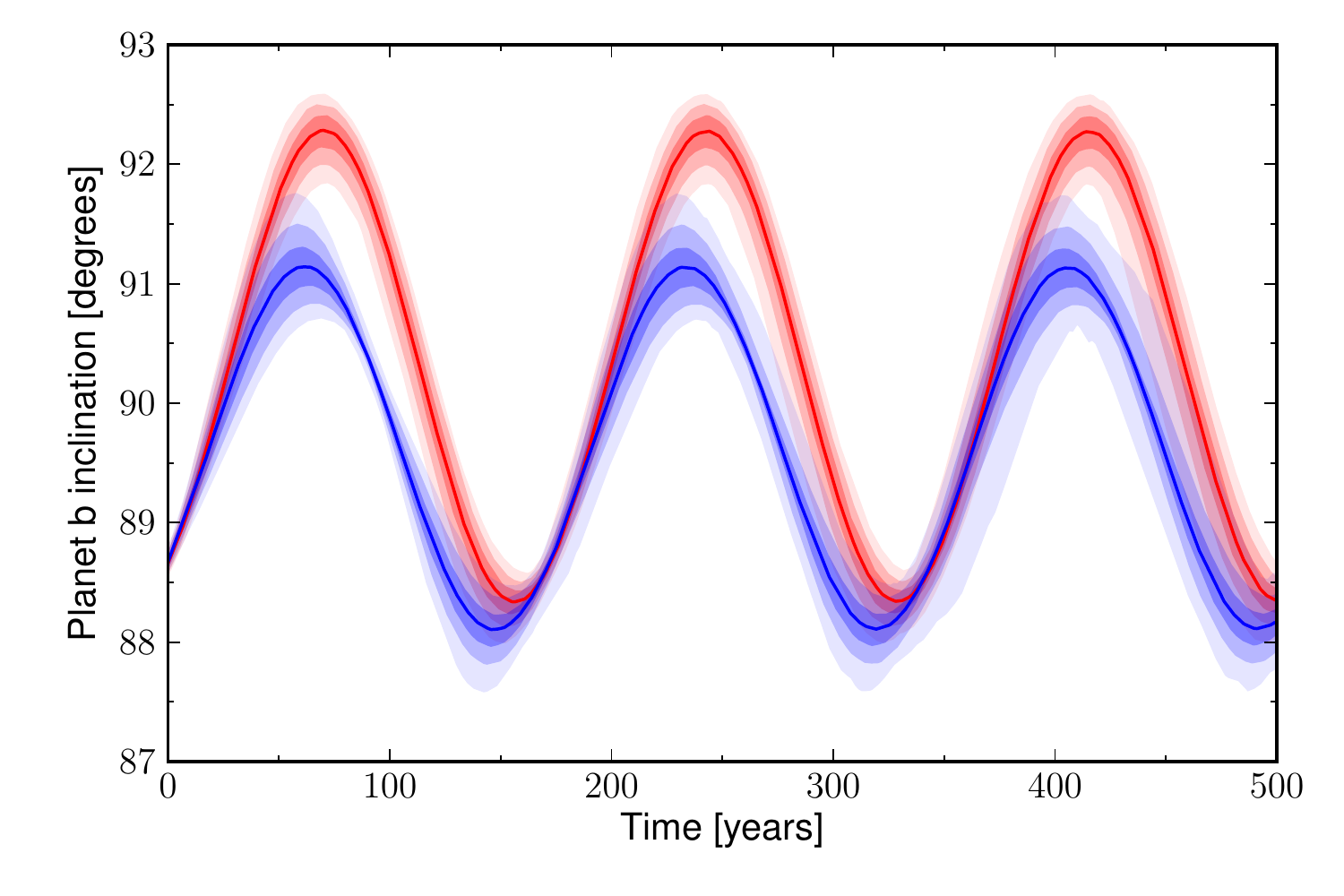}\includegraphics[height=3.2cm]{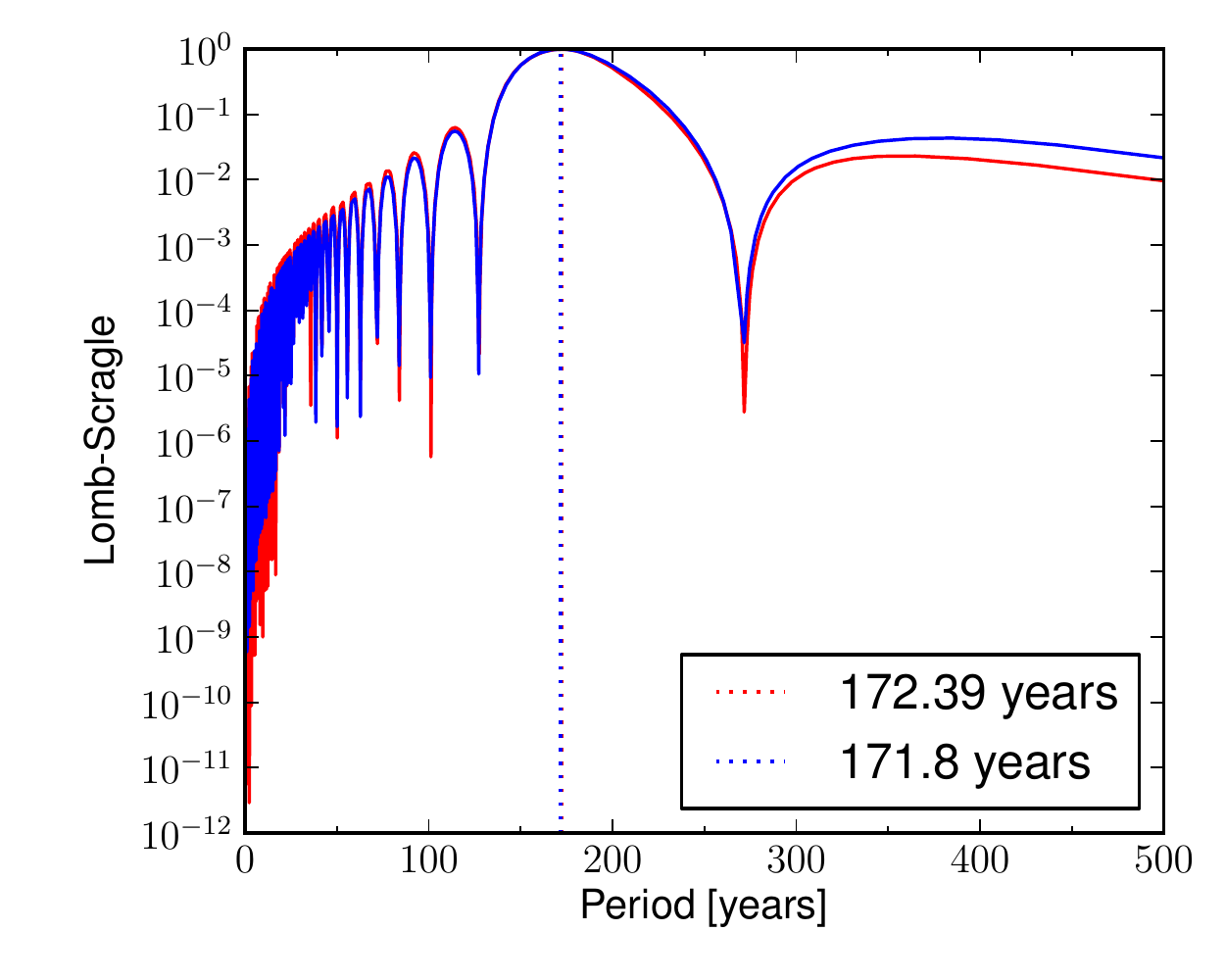}
\includegraphics[height=3.2cm]{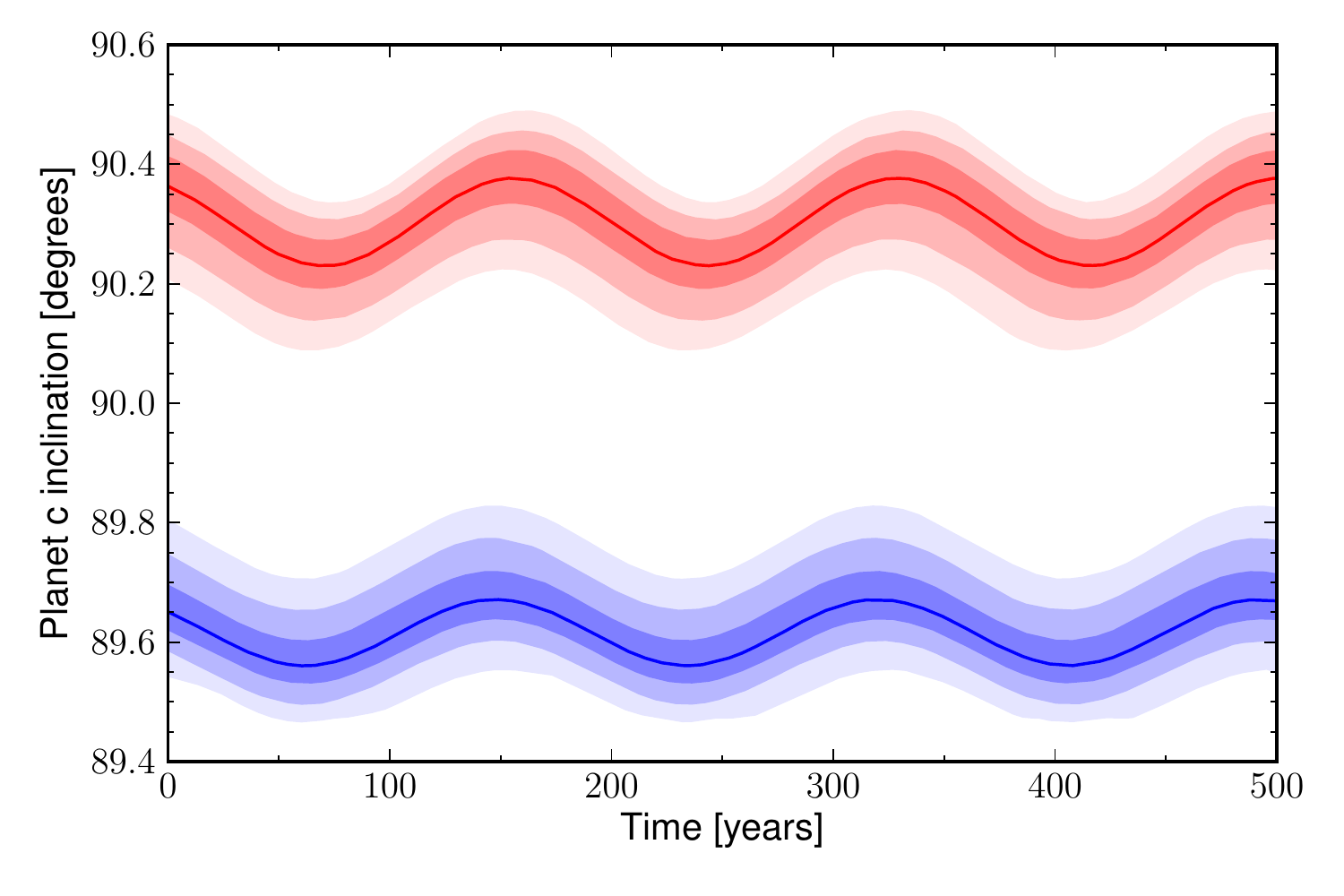}\includegraphics[height=3.2cm]{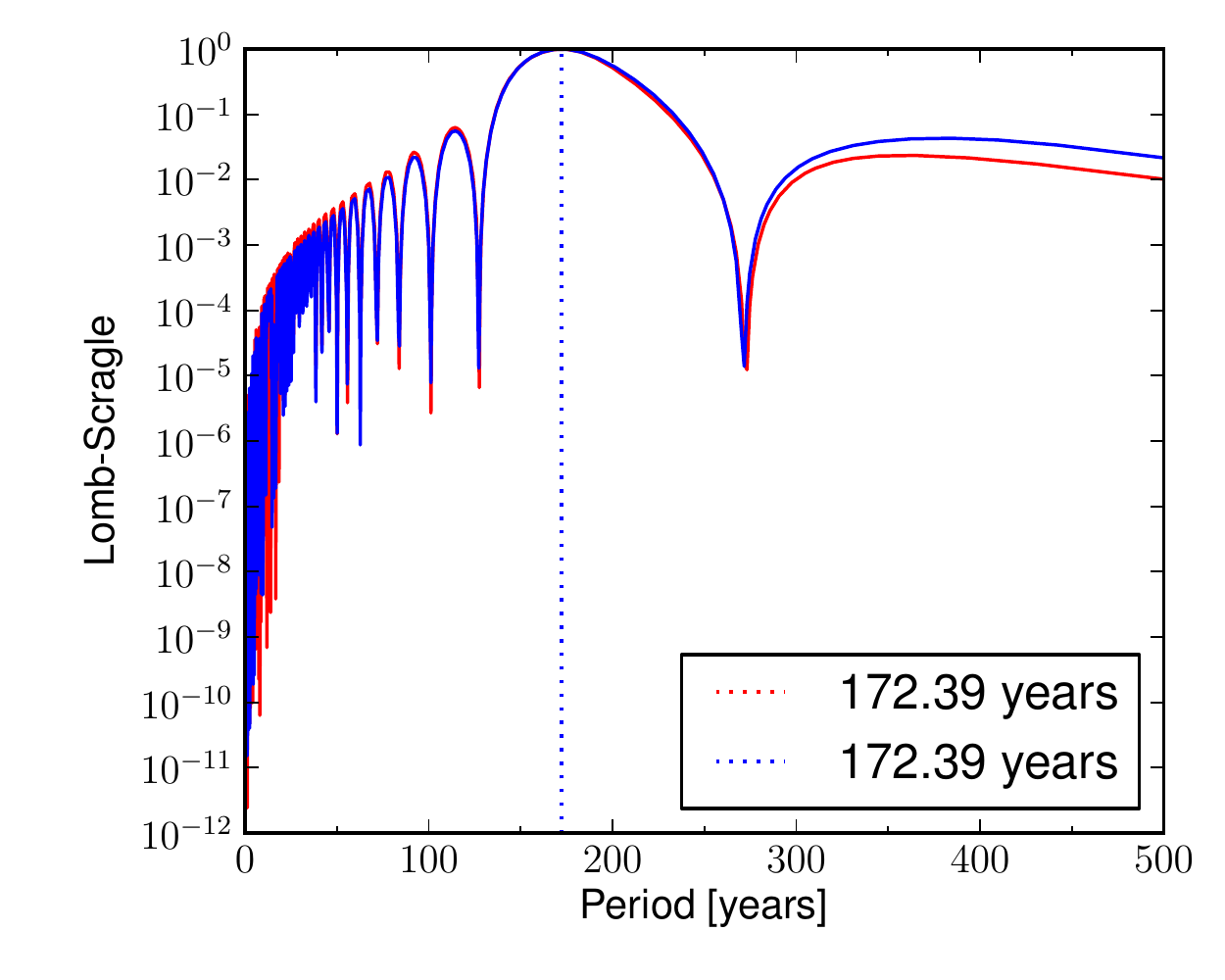}\\
\includegraphics[height=3.2cm]{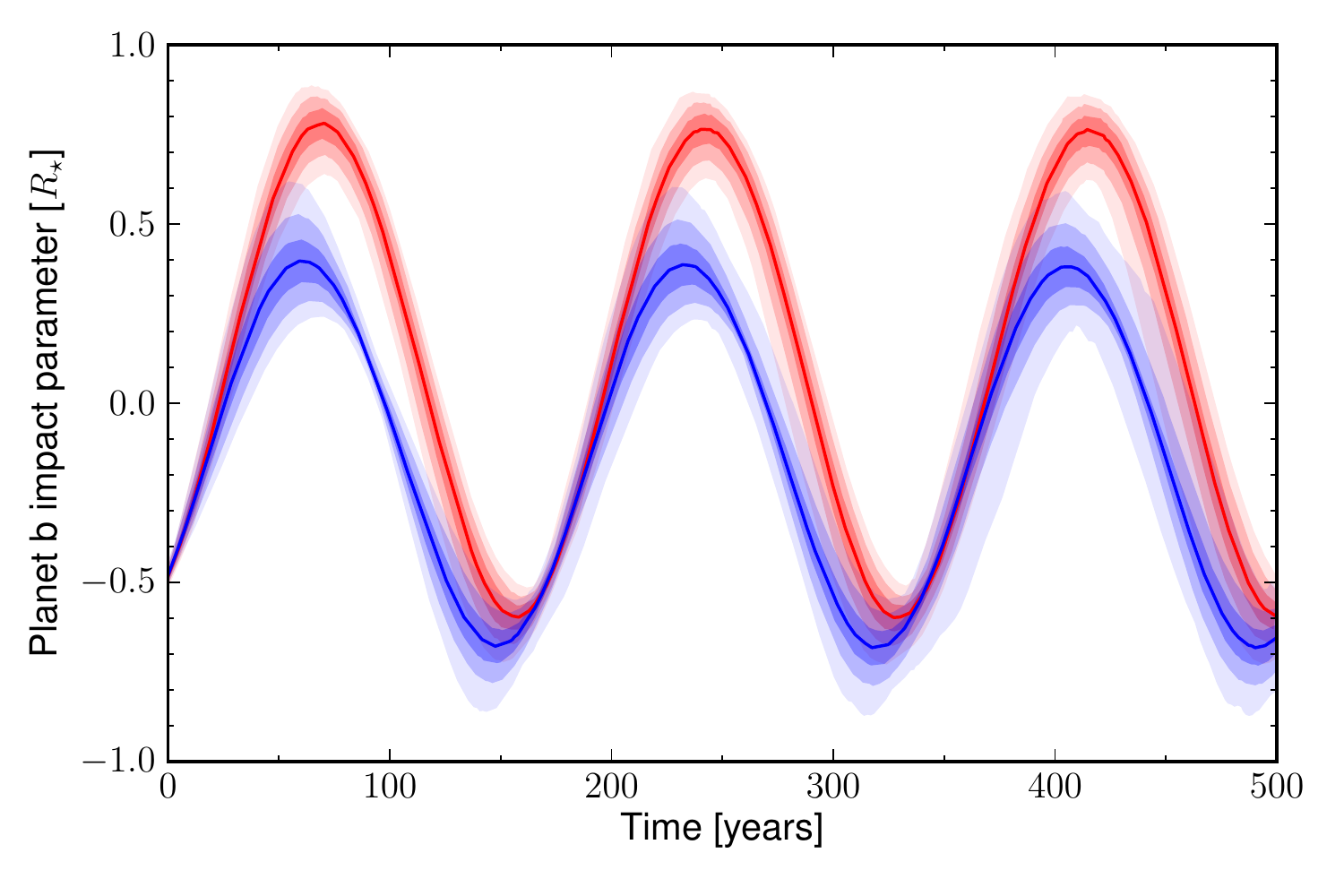}\includegraphics[height=3.2cm]{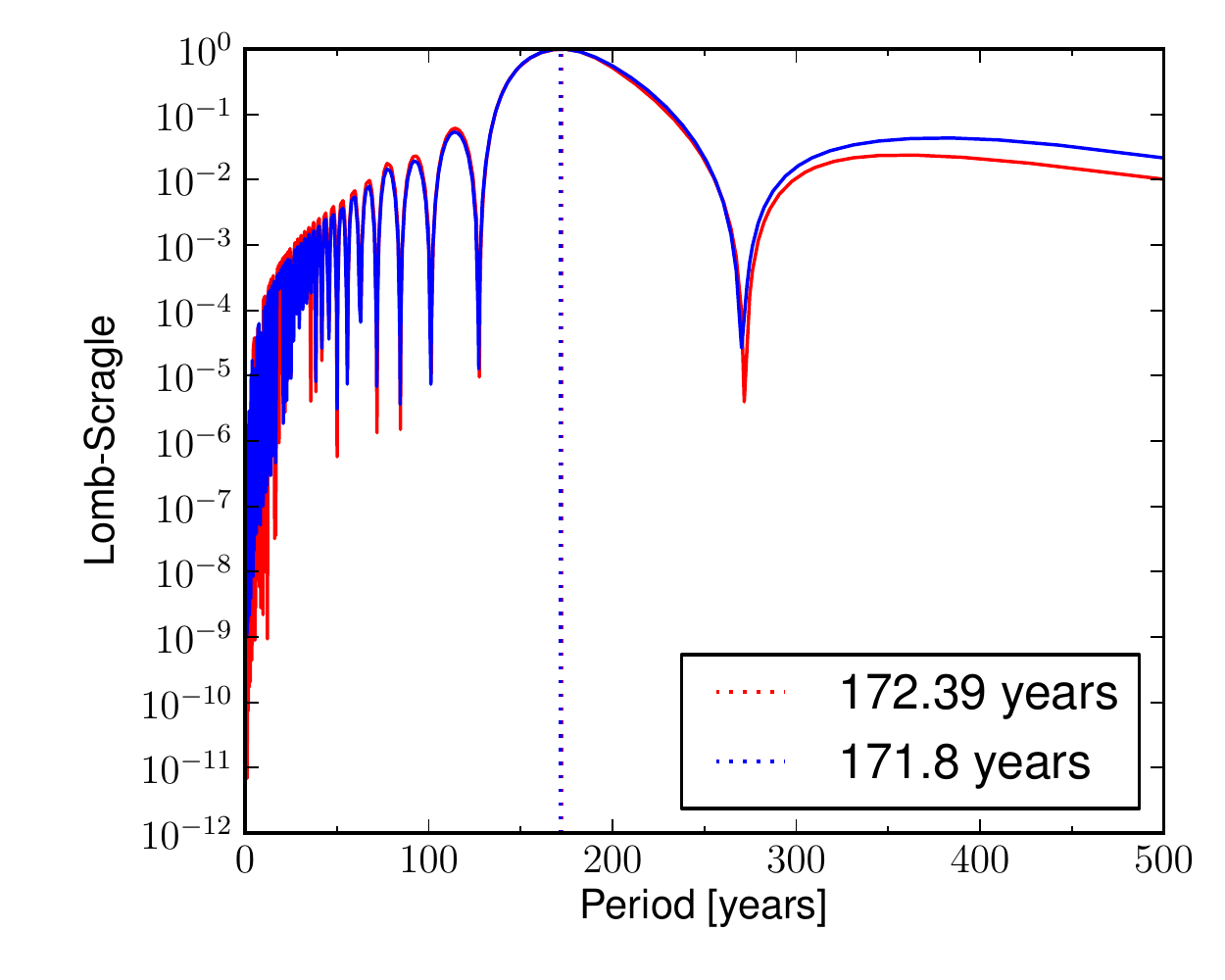}
\includegraphics[height=3.2cm]{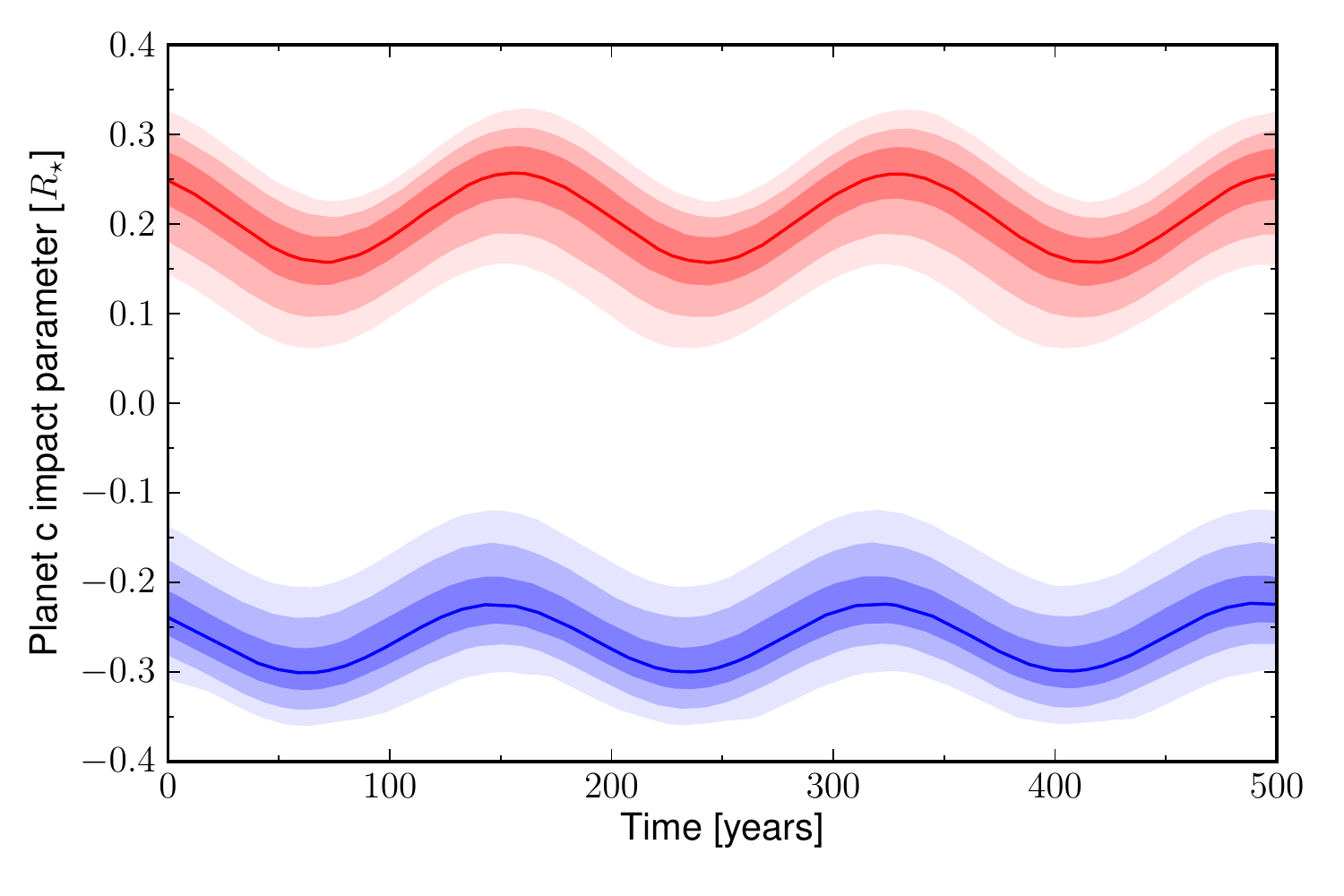}\includegraphics[height=3.2cm]{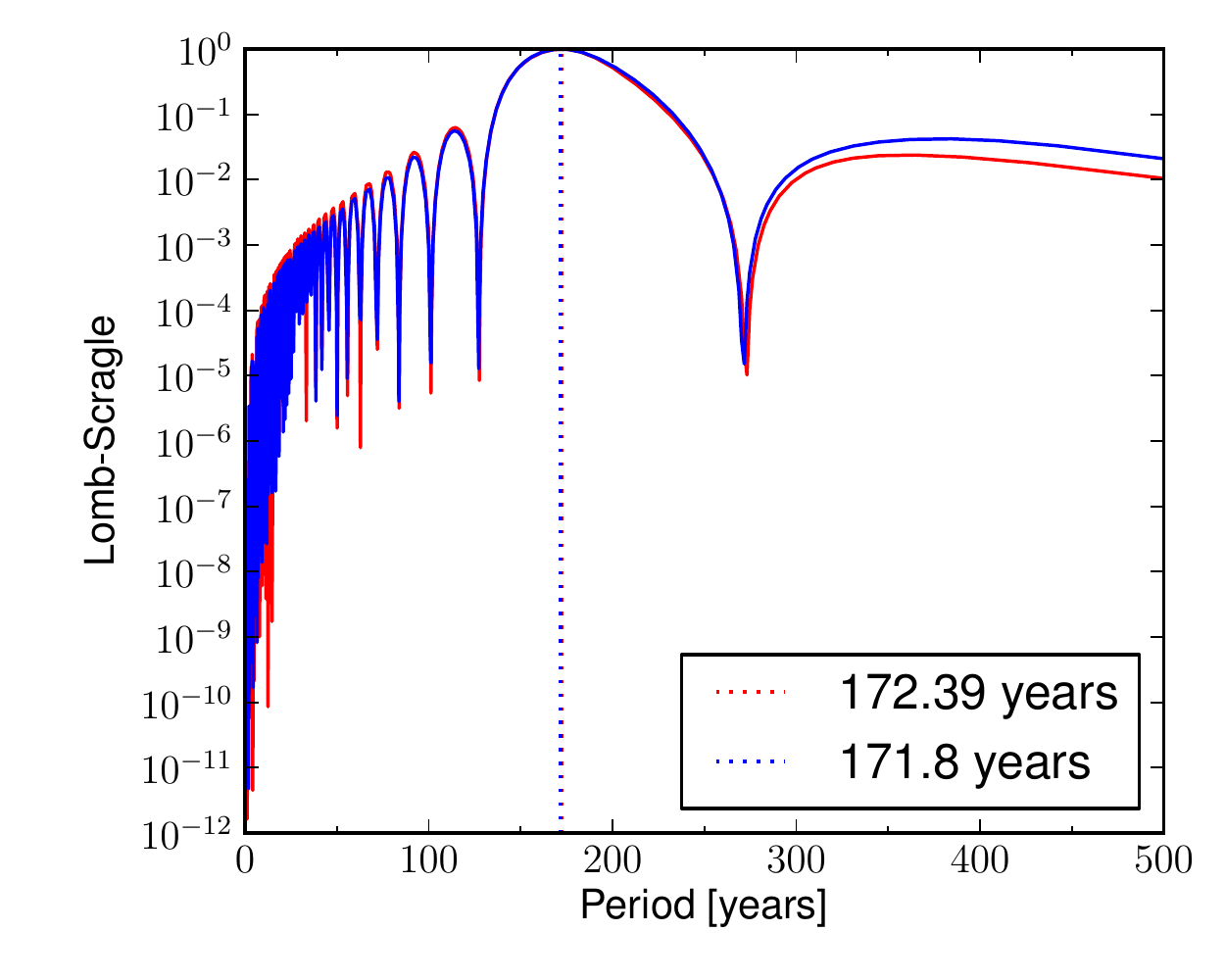}\\
\includegraphics[height=3.2cm]{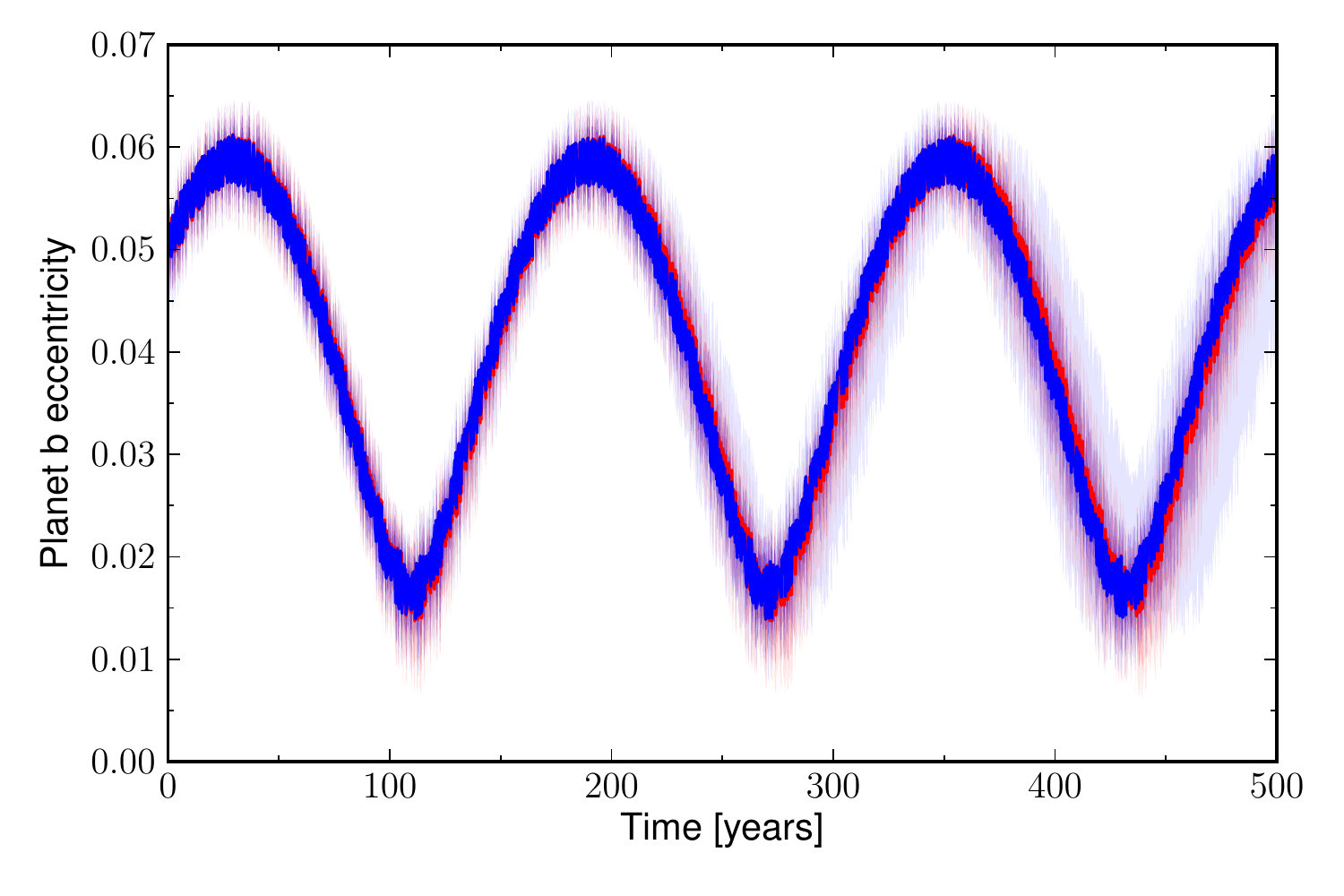}\includegraphics[height=3.2cm]{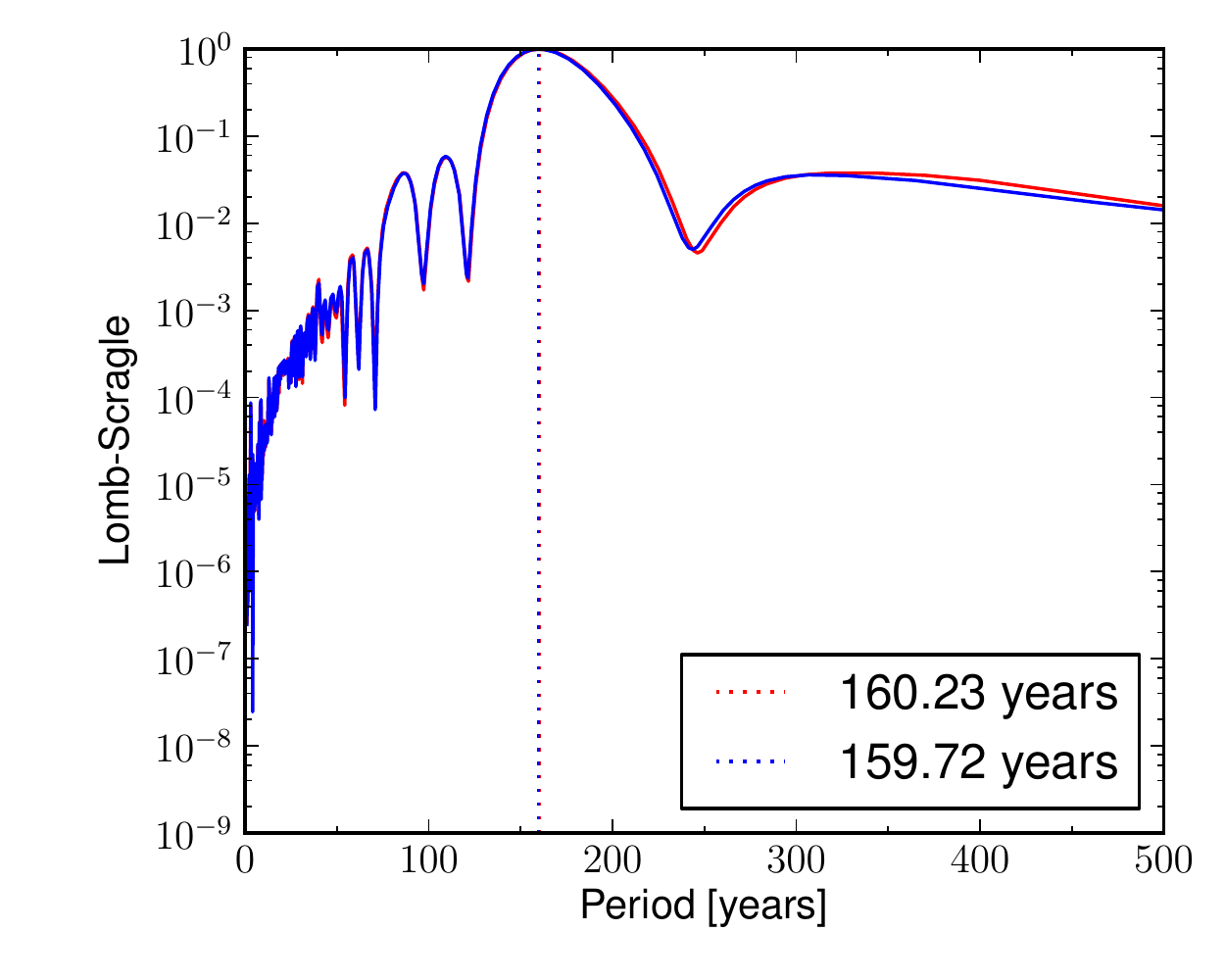}
\includegraphics[height=3.2cm]{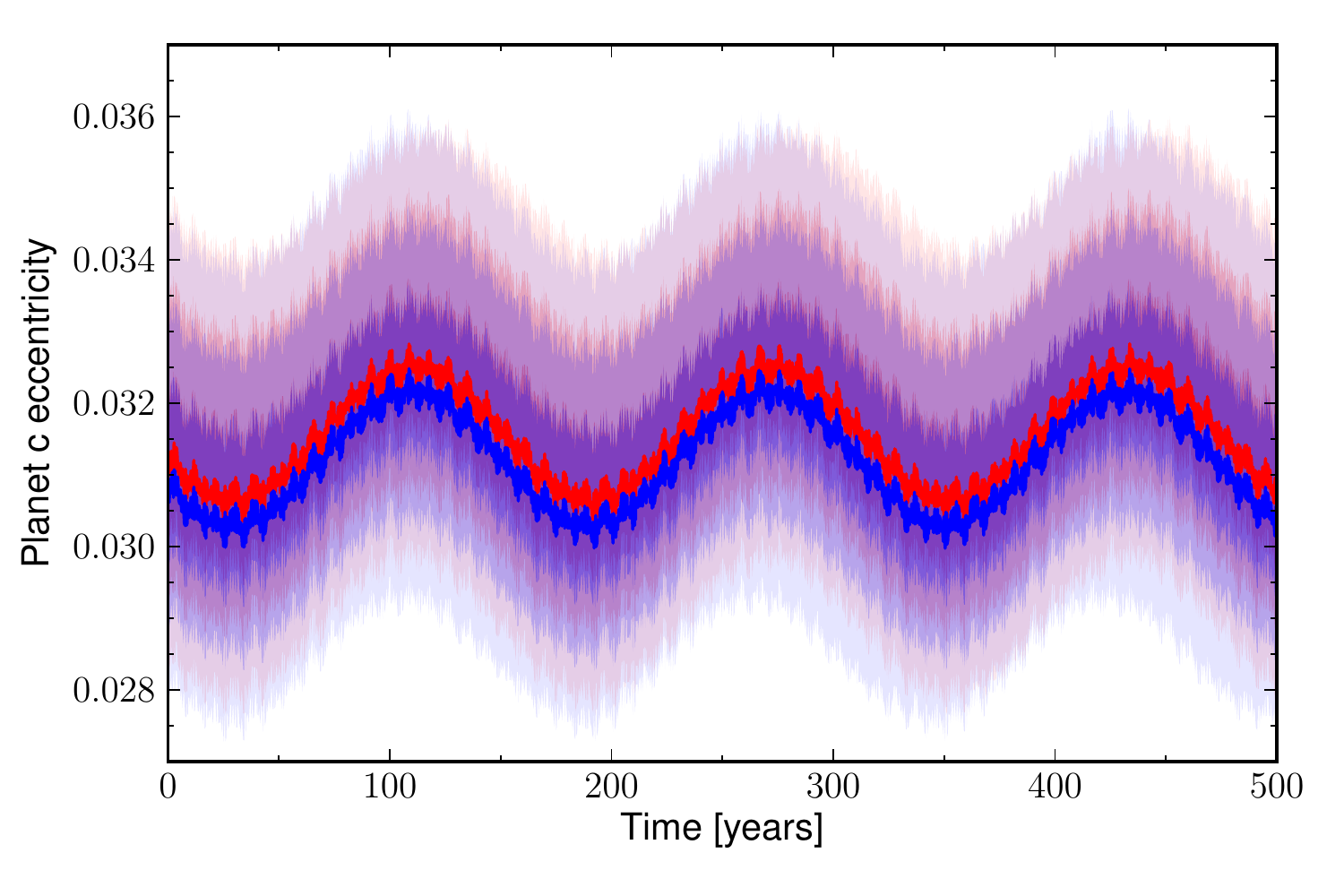}\includegraphics[height=3.2cm]{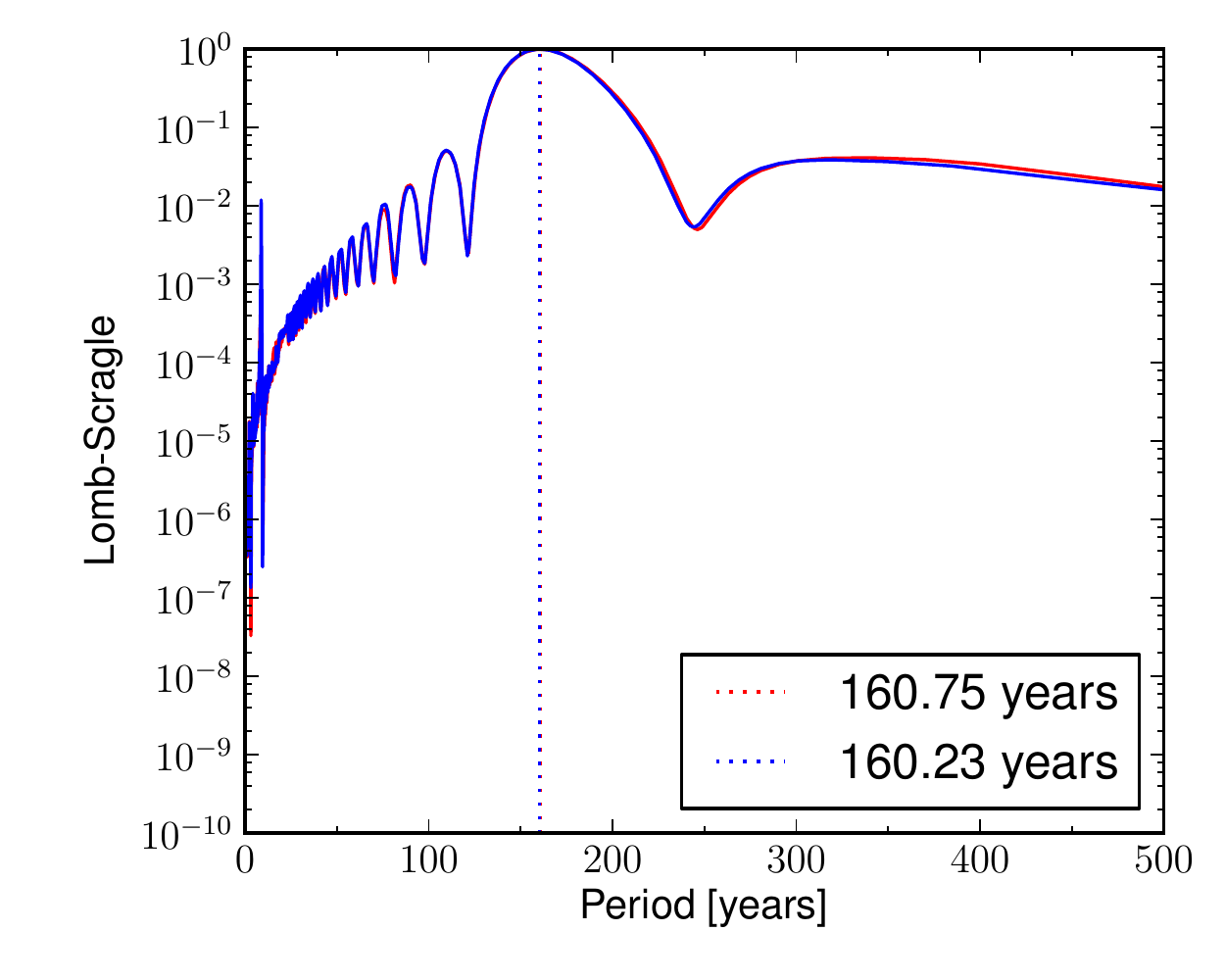}\\
\includegraphics[height=3.2cm]{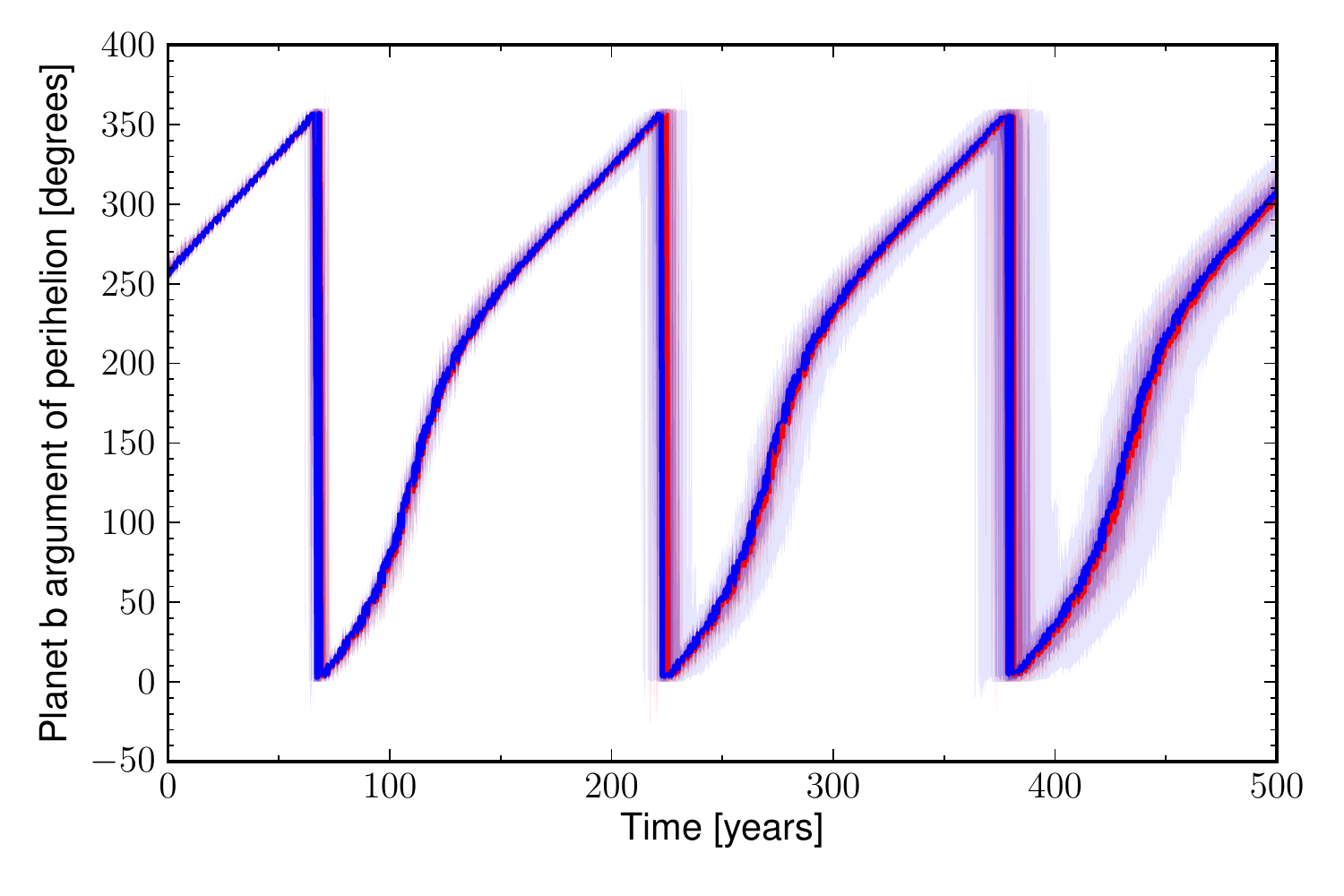}\includegraphics[height=3.2cm]{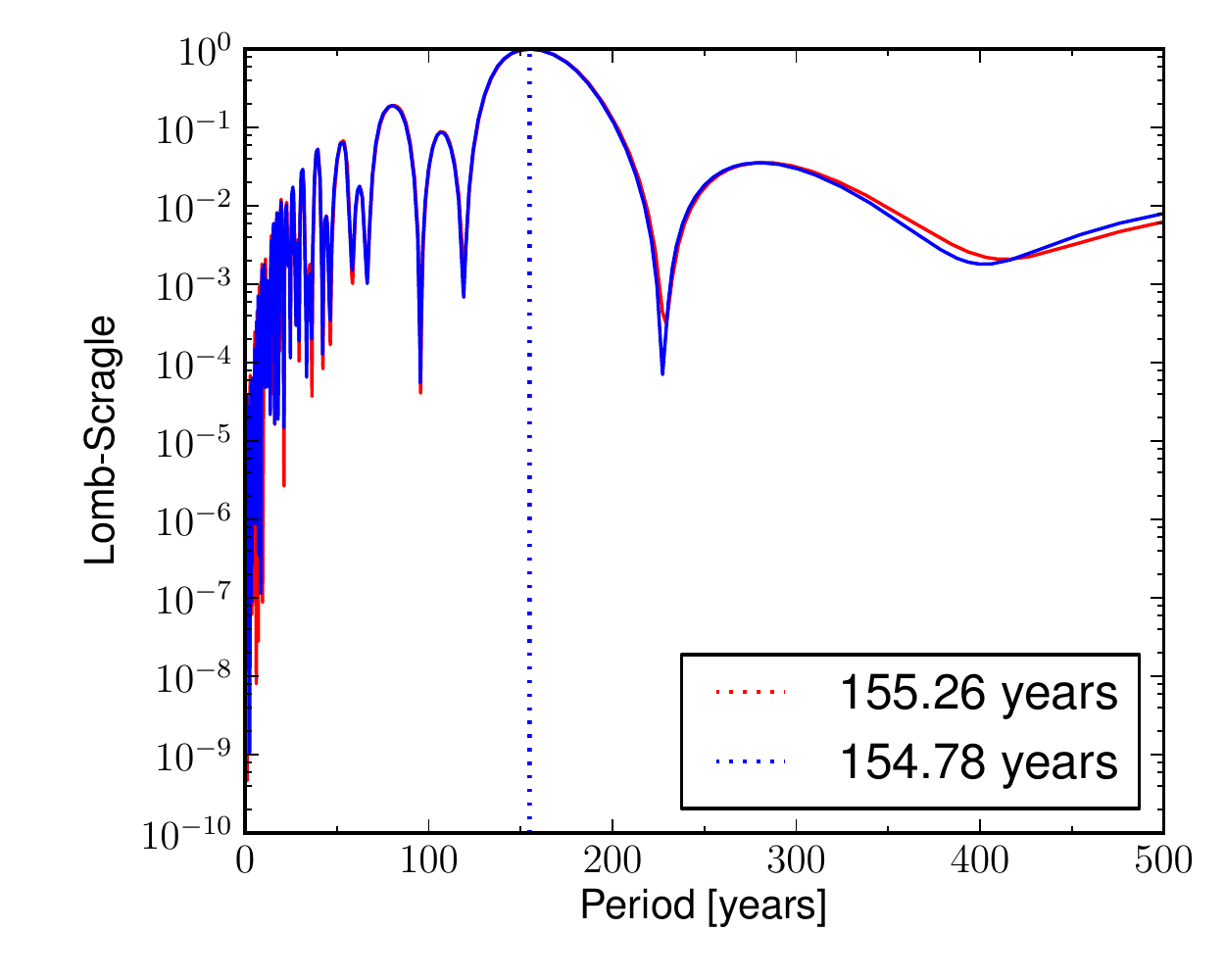}
\includegraphics[height=3.2cm]{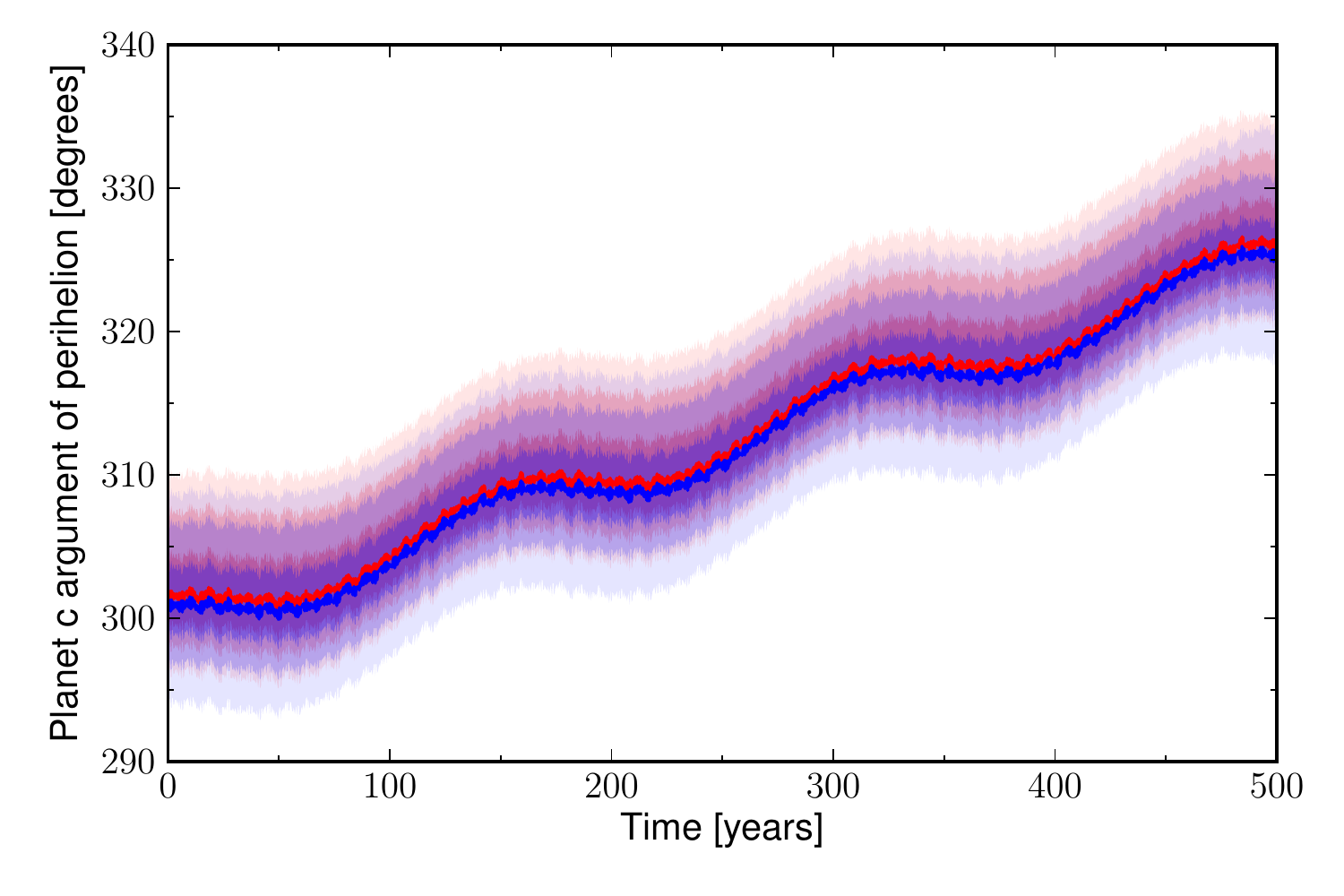}\includegraphics[height=3.2cm]{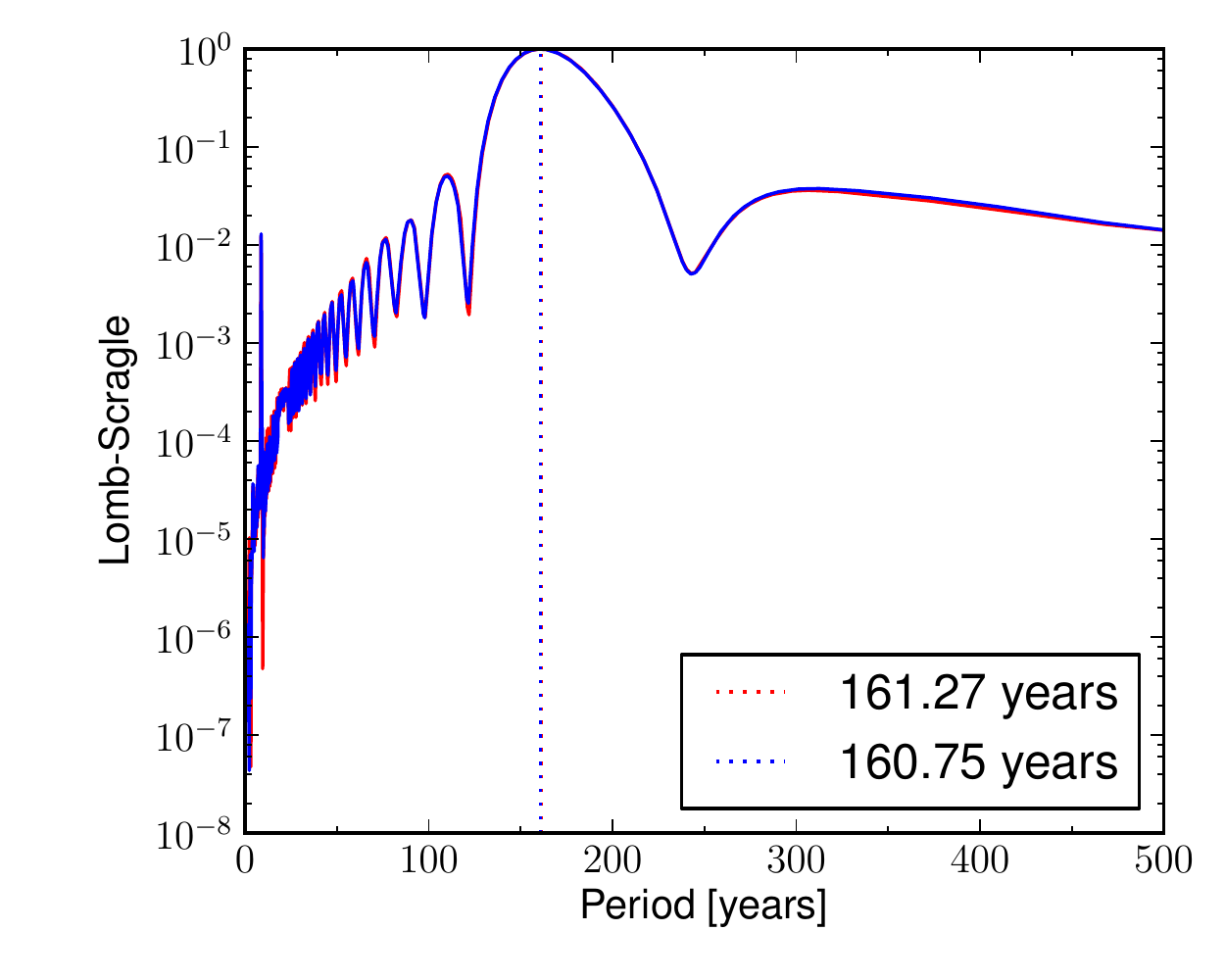}\\
\includegraphics[height=3.2cm]{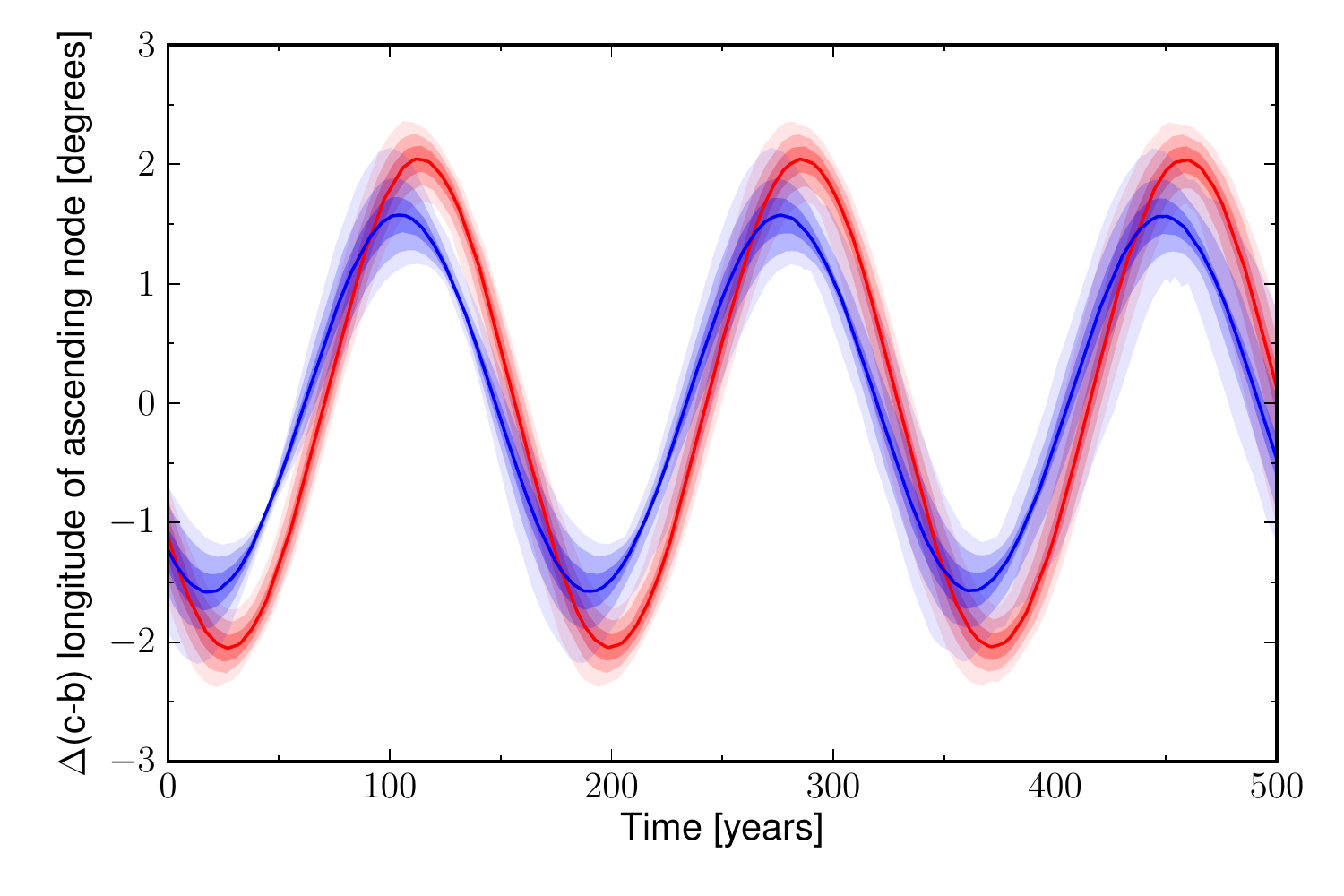}\includegraphics[height=3.2cm]{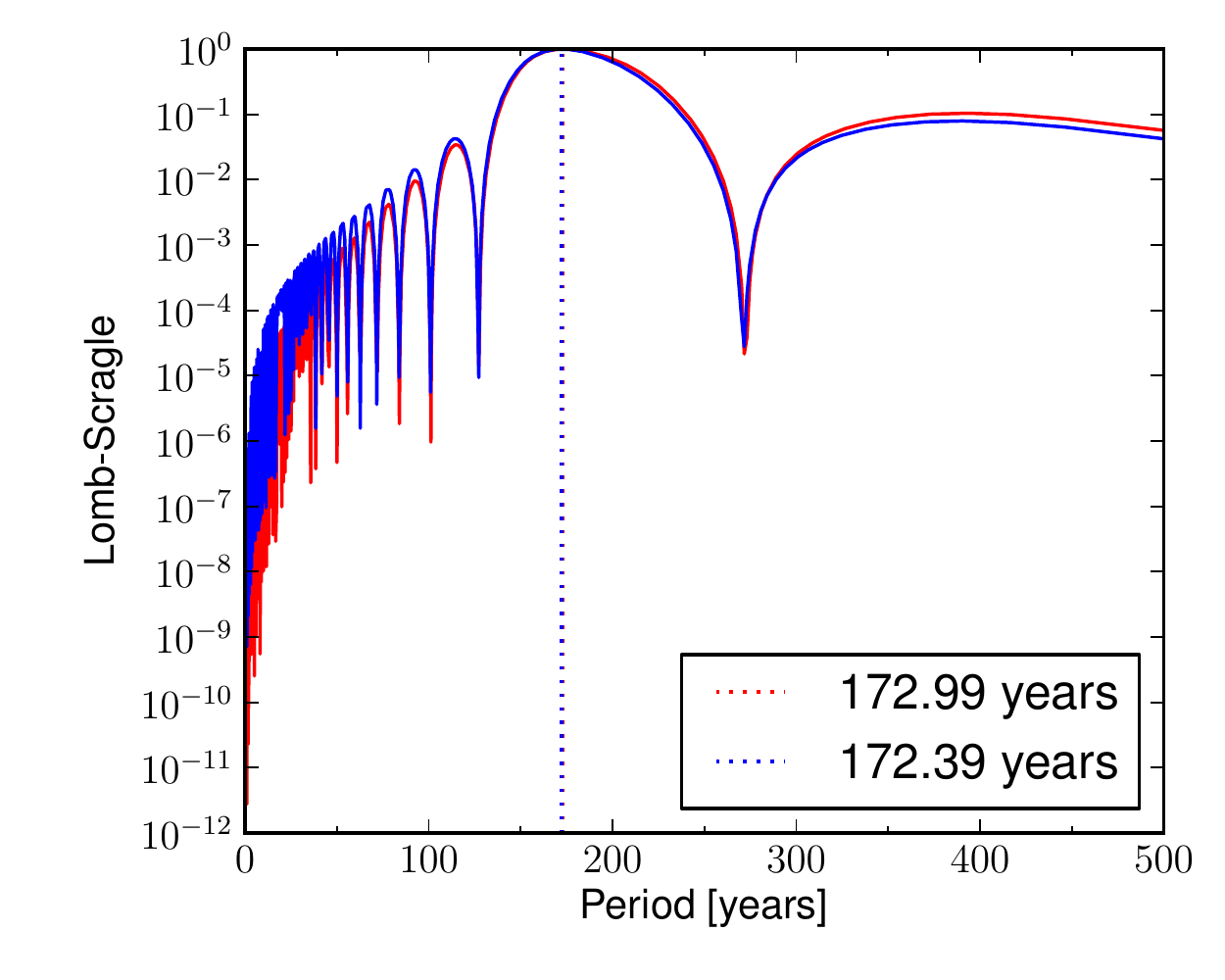}
\includegraphics[height=3.2cm]{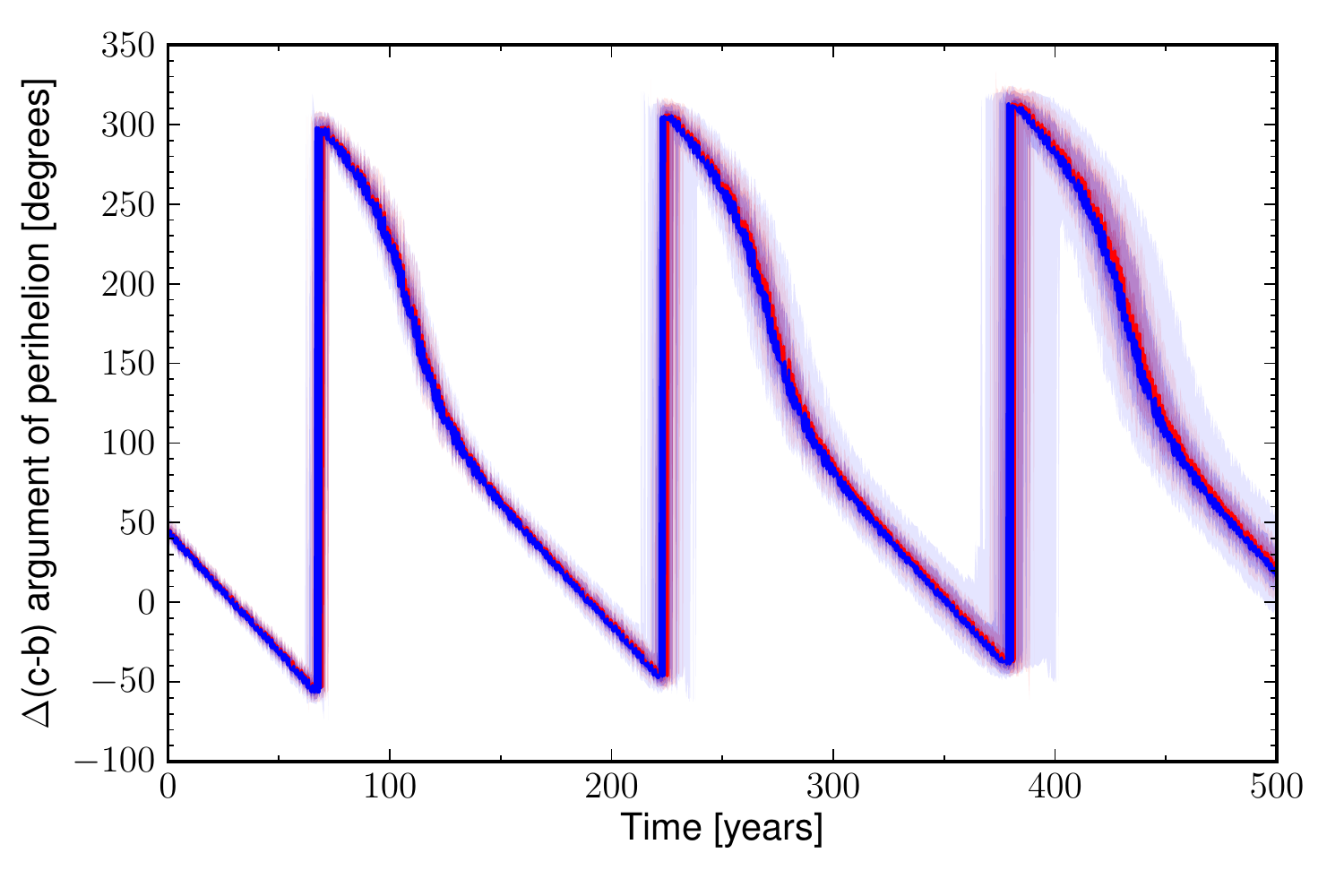}\includegraphics[height=3.2cm]{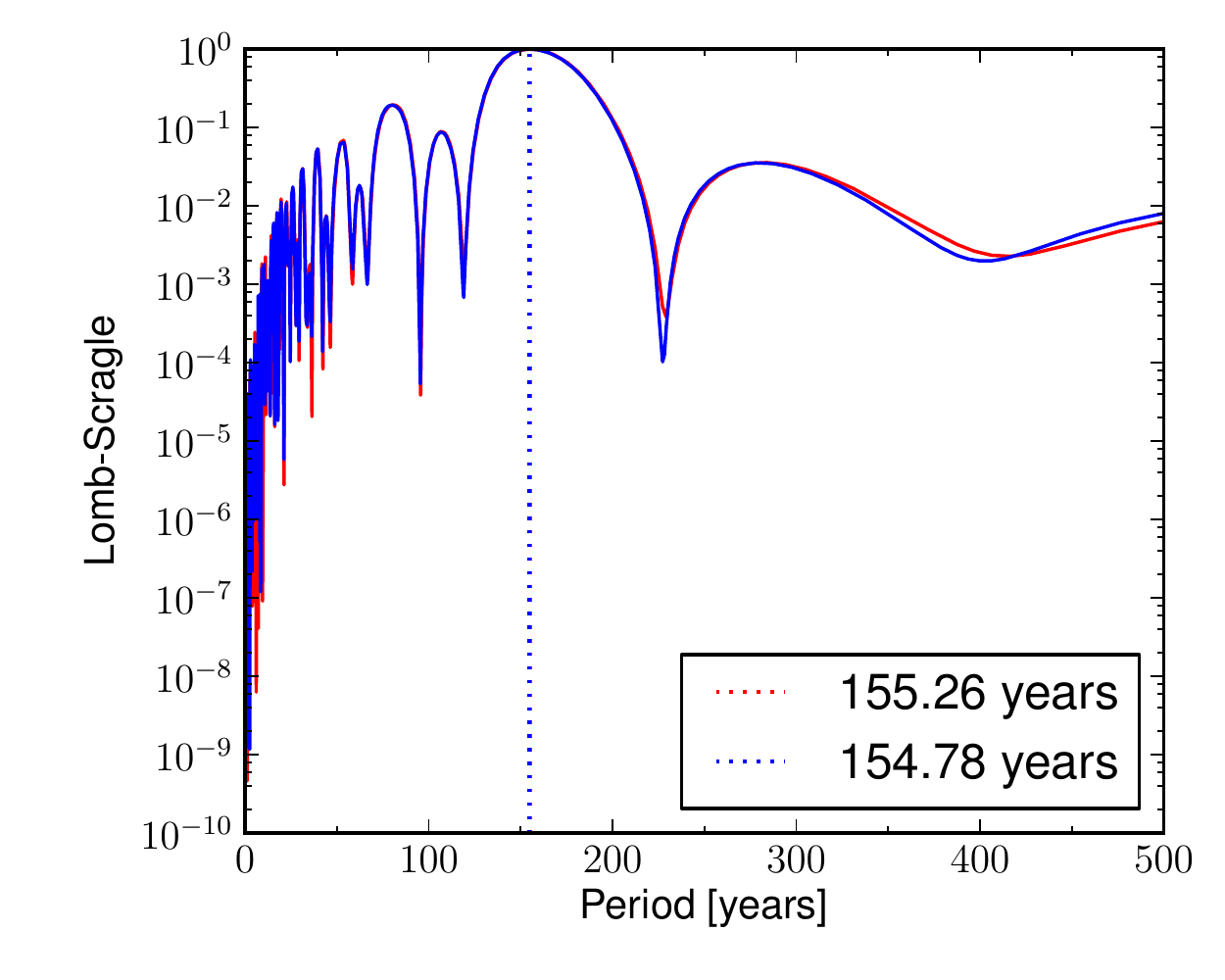}\\
\includegraphics[height=3.2cm]{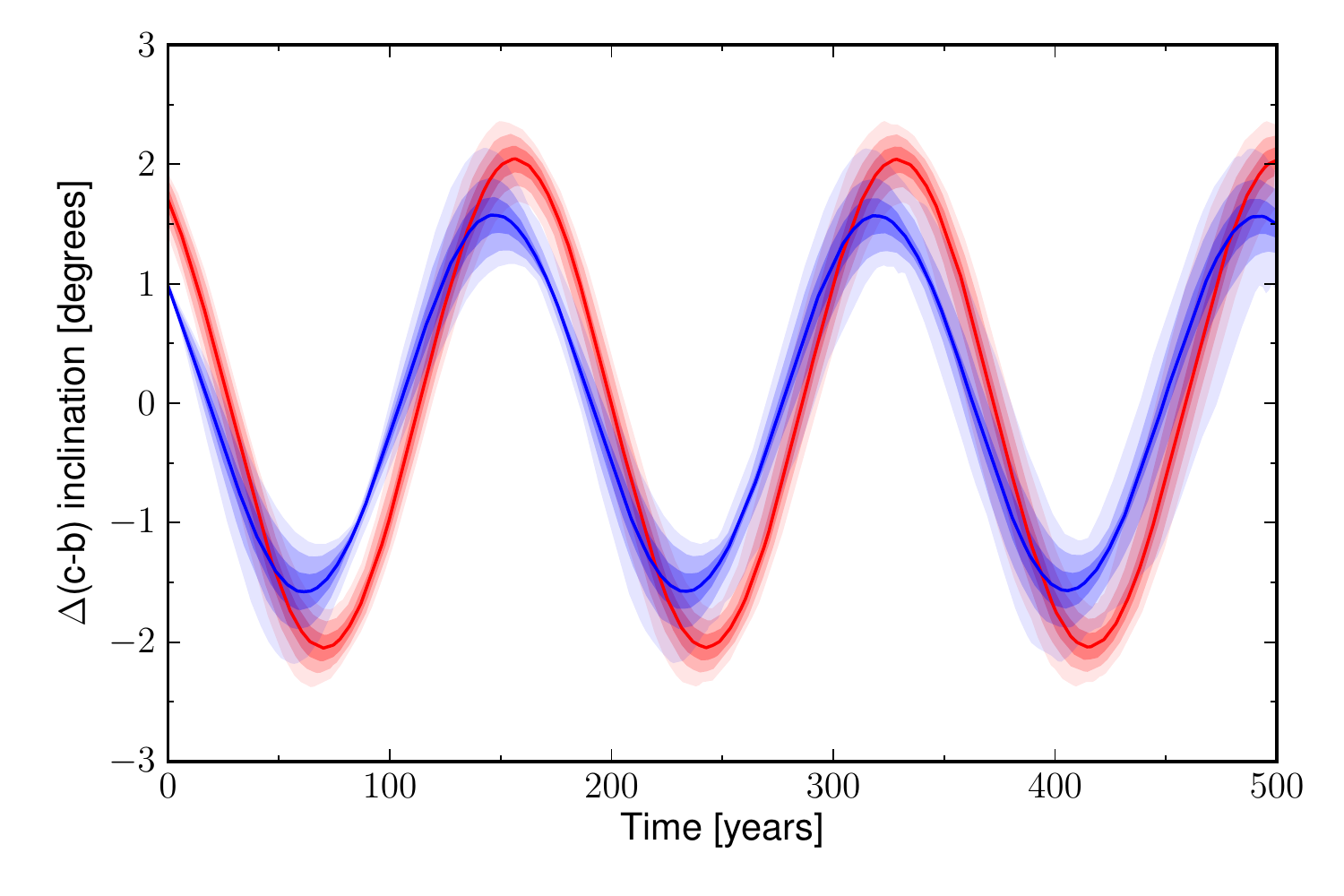}\includegraphics[height=3.2cm]{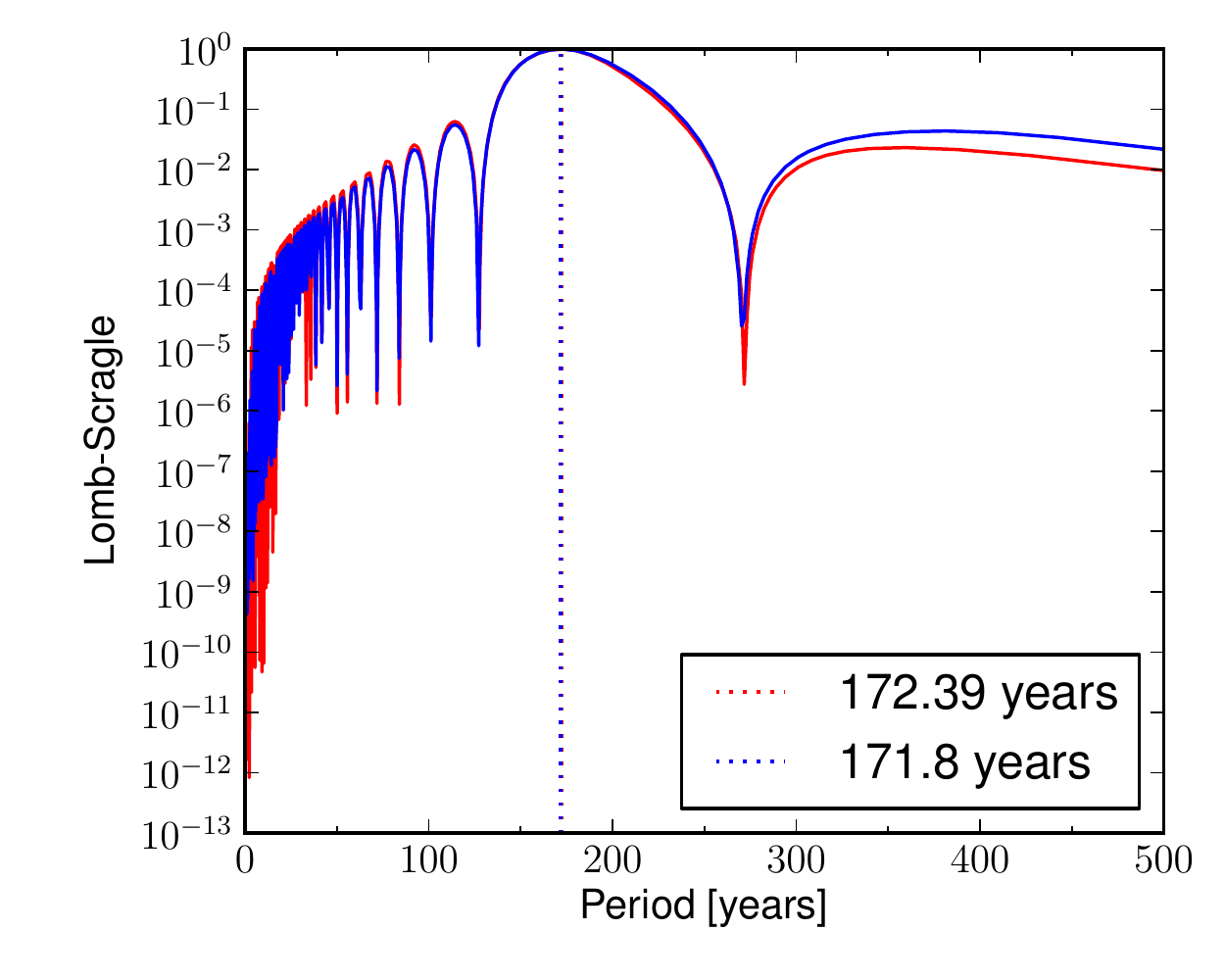}\hspace{3cm}\includegraphics[height=3.2cm]{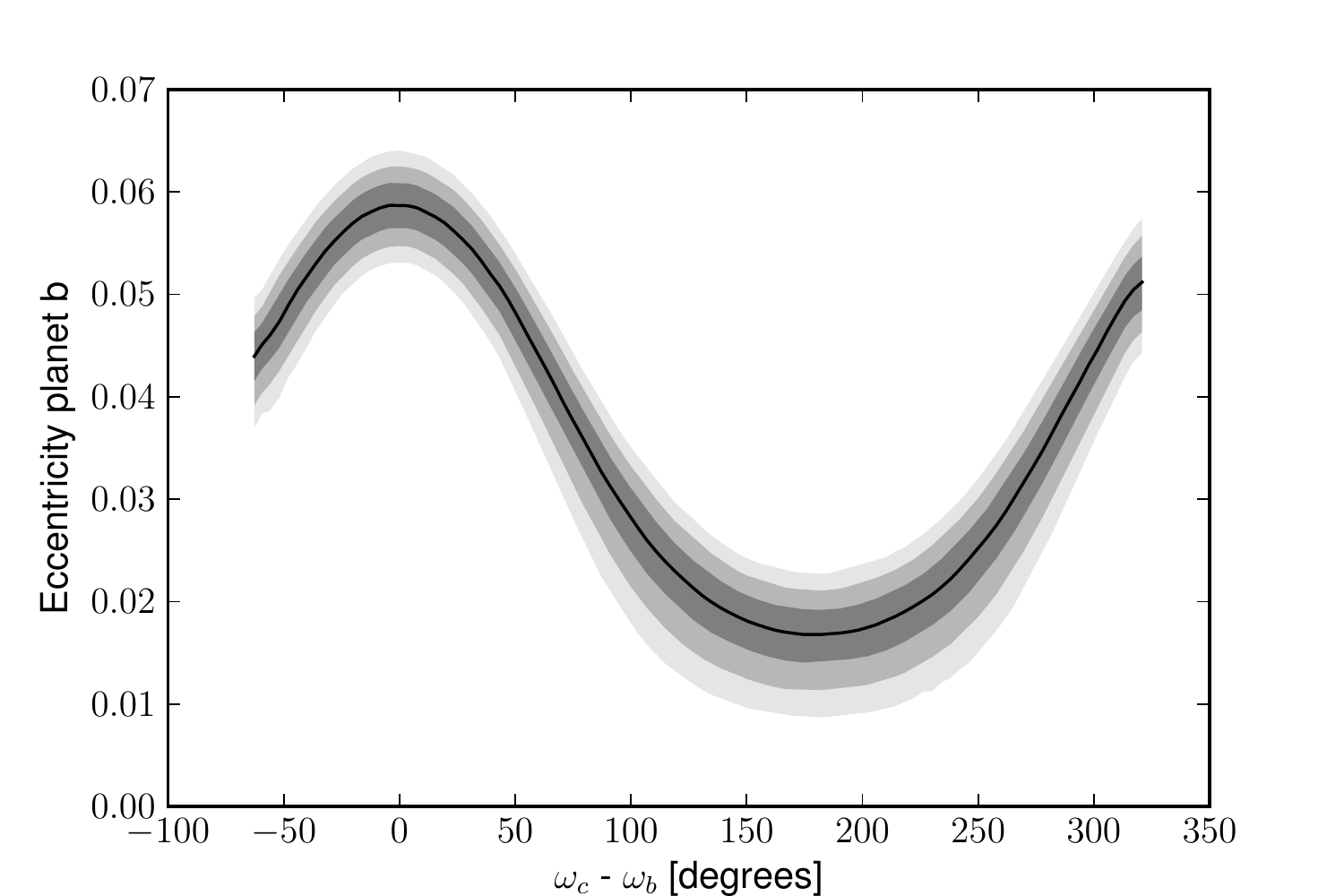}\\
\caption{Idem Fig.~\ref{ShortCen} for the long term evolution (500 years) of the orbital parameters since the beginning of \Kepler\ observations. The orbital parameters oscillate in the long term with a periodicity around 155-172 years.}
\label{LongCen}
\end{figure*}


\bsp

\label{lastpage}

\end{document}